\documentclass[12pt]{article}%
\usepackage{amssymb}
\usepackage{graphicx}
\usepackage{amsmath}%
\usepackage{amsfonts}
\newtheorem{theorem}{Theorem}
\newtheorem{acknowledgement}[theorem]{Acknowledgement}

\newtheorem{definition}[theorem]{Definition}

\newtheorem{lemma}[theorem]{Lemma}

\newtheorem{proposition}[theorem]{Proposition}

\newenvironment{proof}[1][Proof]{\textbf{#1.} }{\ \rule{0.5em}{0.5em}}
\begin{document}

\title{Poisson Hypothesis for information networks \\(A study in non-linear Markov processes)}
\author{Alexander Rybko\\Institute for the Information Transmission Problems,\\Russian Academy of Sciences, Moscow, Russia\\rybko@iitp.ru
\and Senya Shlosman\\Centre de Physique Theorique, CNRS, \\Luminy, 13288 Marseille, France\\shlosman@cpt.univ-mrs.fr}
\maketitle

\begin{abstract}
In this paper we prove the Poisson Hypothesis for the limiting behavior of the
large queueing systems in some simple cases. We show in particular that the
corresponding dynamical systems, defined by the non-linear Markov processes,
have a line of fixed points which are global attractors. To do this we derive
the corresponding non-linear integral equation and we explore its
self-averaging properties. 

MSC-class: 82C20 (Primary), 60J25 (Secondary)

\end{abstract}

\section{Introduction}

The Poisson Hypothesis deals with large queueing systems. For general systems
one can not compute exactly the quantities of interest, so various
approximations are used in practice. The Poisson Hypothesis was formulated
first by L. Kleinrock in \cite{K}. It is the statement that certain
approximation becomes exact in the appropriate limit. It concerns the
following situation. Suppose we have a large network of servers, through which
customers are travelling, being served at different nodes of the network. If
the node is busy, the customers wait in the queue. Customers are entering into
the systems via some nods, and the external flows of customers from the
outside are Poissonian. The service time at each node is random, with some
fixed distribution, depending on the node. We are interested in the stationary
distribution $\pi_{\mathcal{N}}$ at a given node $\mathcal{N} $: what is the
distribution of the queue at it, what is the average waiting time, etc. If the
service time distributions are different from the Poisson distribution, then
the distribution $\pi_{\mathcal{N}}$ in general can not be computed. The
recipe of the Poisson Hypothesis for approximate computation of $\pi
_{\mathcal{N}}$ is the following:

\begin{itemize}
\item consider the total flow $\mathcal{F}$ of customers to the node
$\mathcal{N}.$ (In general, $\mathcal{F}$ is not Poissonian, of course.)
Replace $\mathcal{F}$ with a constant rate Poisson flow $\mathcal{P},$ the
rate being equal to the average rate of $\mathcal{F}.$ Compute the stationary
distribution $\hat{\pi}_{\mathcal{N}}$ at $\mathcal{N},$ corresponding to the
inflow $\mathcal{P}.$ (This is an easy computation.) The claim is that
$\hat{\pi}_{\mathcal{N}}\approx\pi_{\mathcal{N}}.$
\end{itemize}

The Poisson Hypothesis is supposed to give a good estimate if the internal
flow to every node $\mathcal{N}$ is a sum of flows from many other nodes, and
each of these flows constitute only a small fraction of the total flow to
$\mathcal{N}.$

Clearly, the Poisson Hypothesis can not be literally true. It can hopefully
hold only after some kind of ``thermodynamic'' limit is taken.

In the present paper we prove the Poisson Hypothesis for the information
networks in some simple cases. Namely, we will consider the following closed
queueing network. Let there be $M$ servers and $N$ customers to be served. The
distribution of the service time is given by some fixed random variable
$\eta.$ Upon being served, the customer chooses one of $M$ servers with
probability $\frac{1}{M},$ and goes for the service there. Then in the limit
$M,N\rightarrow\infty,$ with $\frac{M}{N}\rightarrow\rho,$ the Poisson
Hypothesis holds, under certain general restrictions on $\eta$.

An important step in this problem was made in the paper \cite{KR1}. Namely, it
was shown there that the above mentioned flow $\mathcal{F}$ converges in our
limit to a Poisson random process with some rate function $\lambda\left(
t\right)  $. If one would be able to show additionally that $\lambda\left(
t\right)  \rightarrow c=const$ as $t\rightarrow\infty,$ that will be
sufficient to establish the Hypothesis. However, the technique of \cite{KR1}
was not enough to prove the \textbf{relaxation} property $\lambda\left(
t\right)  \rightarrow c.$ It was proven there that the situation at a given
single server is described by the so-called non-linear Markov process $\mu
_{t}$ with Poissonian input with rate $\lambda\left(  t\right)  ,$ and the
(non-Poissonian) output with the same rate. Another way of saying this is that
corresponding non-linear Markov process defines some complicated dynamical
system, and the problem was to study its invariant measures. Namely, this
system has one parameter family of fixed points, and the question is about
whether it has other invariant measures.

In the present paper we complete the picture, showing that the above
relaxation $\lambda\left(  t\right)  \rightarrow c$ indeed takes place, and so
$\mu_{t}\rightarrow\mu_{c},$ where $\mu_{c}$ is the stationary distribution of
the stationary Markov process with the Poisson input, corresponding to
constant rate $\lambda\left(  t\right)  =c$. In the language of dynamical
systems we show that there are no other invariant measures except these
defined by the fixed points.

The central discovery of the present paper, which seems to be the key to the
solution of the problem, is that, roughly speaking, the function
$\lambda\left(  t\right)  $ has to satisfy the following non-linear equation:
\begin{equation}
\lambda\left(  t\right)  =\left[  \lambda\left(  \cdot\right)  \ast
q_{\lambda\left(  \cdot\right)  ,t}\left(  \cdot\right)  \right]  \left(
t\right)  .\label{200}%
\end{equation}
Here $\ast$ stays for convolution: for two functions $a\left(  t\right)
,b\left(  t\right)  $ it is defined as
\[
\left[  a\left(  \cdot\right)  \ast b\left(  \cdot\right)  \right]  \left(
t\right)  =\int a\left(  t-x\right)  b\left(  x\right)  \,dx,
\]
while $q_{\lambda\left(  \cdot\right)  ,t}\left(  \cdot\right)  $ is a
one-parameter family of probability densities with $t$ real, which depends
also in an implicit way on the unknown function $\lambda\left(  \cdot\right)
$. We call $\left(  \ref{200}\right)  $ \textbf{the self-averaging property.
}The present paper consists therefore of two parts: we prove that indeed the
self-averaging relation holds, and we prove then that it implies relaxation.

It is amazing that the relation $\left(  \ref{200}\right)  $ depends crucially
on some purely combinatorial statement concerning certain problem of the
placement of the rods on the line $\mathbb{R}^{1},$ see Section 6.

To fix the terminology, we remind the reader here what we mean by the
\textbf{non-linear Markov process} (see \cite{M1}, \cite{M2}). We do this for
the simplest case of discrete time Markov chains, taking values in a finite
set $S,$ $\left|  S\right|  =k.$ In such a case the set of states of this
Markov chain is a simplex $\Delta_{k}$ of all probability measures on $S,$
$\Delta_{k}=\left\{  \mu=\left(  p_{1},...,p_{k}\right)  :p_{i}\geq
0,p_{1}+...+p_{k}=1\right\}  ,$ while the Markov evolution defines a map
$P:\Delta_{k}\rightarrow\Delta_{k}.$ In the case of usual Markov chain $P$ is
affine, and this is why we will call it \textbf{linear }chain. In this case
the matrix of transition probabilities coincides with $P.$ If $P$ is
non-linear, we will call such a process a non-linear Markov chain. It is
defined by a family of transition probability matrices $P_{\mu},$ $\mu
\in\Delta_{k},$ so that matrix element $P_{\mu}\left(  i,j\right)  $ is a
probability of going from $i$ to $j$ in one step, starting \textbf{in the
state }$\mu.$ The (non-linear) map $P$ is then defined by $P\left(
\mu\right)  =\mu P_{\mu}.$

The ergodic properties of the linear Markov chains are settled by the
Perron-Frobenius theorem. In particular, if the linear map $P$ is such that
the image $P\left(  \Delta_{k}\right)  $ belongs to the interior
$\mathrm{Int\,}\left(  \Delta_{k}\right)  $ of $\Delta_{k},$ then there is
precisely one point $\mu\in\mathrm{Int\,}\left(  \Delta_{k}\right)  ,$ such
that $P\left(  \mu\right)  =\mu,$ and for every $\nu\in\mathrm{\,}\Delta_{k}$
we have $P^{n}\left(  \nu\right)  \rightarrow\mu$ as $n\rightarrow\infty.$

In case $P$ is non-linear, we are dealing with more or less arbitrary
dynamical system, and the question about stationary states of the chain or
about measures on $\Delta_{k}$ invariant under $P$ can not be settled in general.

Therefore it is natural to ask about the specific features of our dynamical
system, which permit us to find all its invariant measures. We explain this in
the following subsection.

\textbf{Dynamical systems aspect. }Here we will use the notation of the paper,
though in fact the situation of the paper is more complicated; in particular
the underlying space is not a manifold, but a space of all measures over some
non-compact set.

Let $M$ be a manifold, supplied with the following structures:

\begin{itemize}
\item for every point $\mu\in M$ and every $\lambda>0$ a tangent vector
$X\left(  \mu,\lambda\right)  $ at $\mu$ is defined,

\item a function $b:M\rightarrow\mathbb{R}^{+}$ is fixed.
\end{itemize}

We want to study the dynamical system
\begin{equation}
\frac{d}{dt}\mu\left(  t\right)  =X\left(  \mu\left(  t\right)  ,b\left(
\mu\left(  t\right)  \right)  \right)  .\label{001}%
\end{equation}
Its flow conserves another given function, $N:M\rightarrow\mathbb{R}^{+},$ and
we want to prove that our dynamical system has one-parameter family of fixed
points - each corresponding to one value of $N$ - and no other invariant measures.

We have the following extra properties of our dynamical system:

Let $\lambda\left(  t\right)  >0;$ consider the differential equation
\begin{equation}
\frac{d}{dt}\mu\left(  t\right)  =X\left(  \mu\left(  t\right)  ,\lambda
\left(  t\right)  \right)  ,\;t\geq0,\label{002}%
\end{equation}
with $\mu\left(  0\right)  =\nu.$ We denote the solution to it by $\mu
_{\nu,\lambda\left(  \cdot\right)  }\left(  t\right)  .$ We know that

\begin{itemize}
\item for every $c>0$ and every initial data $\nu,$ the solution $\mu
_{\nu,\lambda\left(  \cdot\right)  }\left(  t\right)  $ to $\left(
\ref{002}\right)  $ converges to some stationary point $\nu_{c}\in M,$
\begin{equation}
\mu_{\nu,\lambda\left(  \cdot\right)  }\left(  t\right)  \rightarrow\nu
_{c},\text{ provided }\lambda\left(  t\right)  \rightarrow c\text{ as
}t\rightarrow\infty,\label{004}%
\end{equation}

\item for the function $N$ we have
\[
\frac{d}{dt}N\left(  \mu_{\nu,\lambda\left(  \cdot\right)  }\left(  t\right)
\right)  =\lambda\left(  t\right)  -b\left(  \mu_{\nu,\lambda\left(
\cdot\right)  }\left(  t\right)  \right)  .
\]
In particular, for every trajectory $\hat{\mu}_{\nu}\left(  t\right)  $ of
$\left(  \ref{001}\right)  $ (where $\hat{\mu}_{\nu}\left(  0\right)  =\nu$)
we have $N\left(  \hat{\mu}_{\nu}\left(  t\right)  \right)  =N\left(
\nu\right)  .$ Also, $N\left(  \nu_{c}\right)  $ is continuous and increasing
in $c$;

\item for every $\nu,\lambda\left(  \cdot\right)  $ and every $t>0$ there
exists a probability density $q_{\nu,\lambda,t}\left(  x\right)  ,\;x\geq0,$
such that
\[
b\left(  \mu_{\nu,\lambda\left(  \cdot\right)  }\left(  t\right)  \right)
=\left(  \lambda\ast q_{\nu,\lambda,t}\right)  \left(  t\right)  ,
\]
where
\[
\left(  \lambda\ast q_{\nu,\lambda,t}\right)  \left(  y\right)  =\int_{x\geq
0}q_{\nu,\lambda,t}\left(  x\right)  \lambda\left(  y-x\right)  \,dx.
\]
The family $q_{\nu,\lambda,t}\left(  x\right)  $ satisfies:
\[
\int_{0}^{1}q_{\nu,\lambda,t}\left(  x\right)  \,dx=1\text{ for all }%
\nu,\lambda,t,
\]
and
\[
\inf_{\substack{\nu,\lambda,t \\x\in\left[  0,1\right]  }}q_{\nu,\lambda
,t}\left(  x\right)  >0.
\]

\end{itemize}

Then for every initial state $\nu$
\begin{equation}
\hat{\mu}_{\nu}\left(  t\right)  \rightarrow\nu_{c},\label{003}%
\end{equation}
where $c$ satisfies $N\left(  \nu_{c}\right)  =N\left(  \nu\right)  .$\bigskip

Our statement follows from the fact that the self-averaging property,
\[
f\left(  t\right)  =\left(  f\ast q_{t}\right)  \left(  t\right)  ,
\]
with $q_{t}\left(  \cdot\right)  $ being a family of probability densities on
$\left[  0,1\right]  $, implies that $f\left(  t\right)  \rightarrow const$ as
$t\rightarrow\infty,$ so $\left(  \ref{003}\right)  $ follows from $\left(
\ref{004}\right)  .$

We feel that the relation $\left(  \ref{200}\right)  $ is an important feature
of the subject we are interested in. Therefore in the present paper we study
it and the related questions in some generality.

$i)$ We start with the equation
\begin{equation}
f\left(  t\right)  =\left[  f\left(  \cdot\right)  \ast q_{t}\left(
\cdot\right)  \right]  \left(  t\right)  .\label{201}%
\end{equation}
Here we suppose that $q_{t}\left(  \cdot\right)  $ is just some one-parameter
family of probability densities (without functional dependence), so $\left(
\ref{200}\right)  $ is a special case of $\left(  \ref{201}\right)  .$ On the
other hand, we suppose additionally that all the distributions $q_{t}\left(
\cdot\right)  $ are supported by some finite interval. We establish relaxation
in this case.

$ii)$ We then do the same for the case of distributions $q_{t}\left(
\cdot\right)  $ with unbounded support.

$iii)$ Last, we treat the true problem, where in addition to the infinite
support, an extra parameter $\mu$ appears and an extra perturbation is added
to convolution term in $\left(  \ref{201}\right)  :$%
\begin{equation}
\lambda\left(  t\right)  =\left(  1-\varepsilon_{\lambda,\mu}\left(  t\right)
\right)  \left[  \lambda\left(  \cdot\right)  \ast q_{\lambda,\mu,t}\left(
\cdot\right)  \right]  \left(  t\right)  +\varepsilon_{\lambda,\mu}\left(
t\right)  Q_{\lambda,\mu}\left(  t\right)  .\label{202}%
\end{equation}
Here the parameter $\varepsilon_{\lambda,\mu}\left(  t\right)  $ is small:
$\varepsilon_{\lambda,\mu}\left(  t\right)  \rightarrow0$ as $t\rightarrow
\infty,$ the term $Q_{\lambda,\mu}\left(  t\right)  $ is uniformly bounded,
while the meaning of $\mu$ will be explained later.

As we proceed from $i)$ to $iii),$ we will have to assume more about the class
of distributions $\left\{  q_{\cdot}\right\}  ,$ for which the self-averaging
implies relaxation.

We finish this introduction by a brief discussion of the previous work on the
subject, and their methods.

As we said before, part of the proof of the Poissonian Hypothesis -- the so
called Weak Poissonian Hypothesis -- was obtained in \cite{KR1}. By proving
that the Markov semigroups describing the Markov processes for finite $M,N,$
converge, after factorization by the symmetry group of the model, to the
semigroup, describing the non-linear Markov process, the authors have proven
that the limit flows to each node are independent Poisson flows with the same
rate function $\lambda\left(  t\right)  .$ This statement is fairly general,
and can be generalized to other models with the same kind of the symmetry --
the so-called mean-field models. The general theory -- see, for example,
\cite{L} -- implies then, that all the limit points of the stationary measures
of the Markov processes with finite $M,N$ are invariant measures of the
limiting non-linear Markov process. The remaining step -- the proof that the
limiting dynamical system has no other attractors except the one-parameter
family of the fixed points -- is done in the present paper. Apriori this fact
is not at all clear, and one can construct natural examples of the systems
with many complicated attractors, which are reflected in the complex behavior
of the Markov processes with finite $M,N.$ However, the self-averaging
property, explained above, rules out such a possibility. It seems that the
self-averaging property can also be generalized to other mean-field models.

The Poisson Hypothesis was fully established in a pioneer paper \cite{St} for
a special case when the service time is non-random. This is a much simpler
case, and the methods of the paper can not be extended to our situation. They
are sufficient for a simpler case of the Poissonian service times, which case
was studied in \cite{KR2}.

The paper \cite{DKV} deals with another mean-field model, describing some open
queueing network. Though the Poisson Hypothesis does not hold for it, the
spirit of the main statement there is the same as in the present paper: the
limiting dynamical system has precisely one global attractor, which
corresponds to the fixed point.

One of specific feature of the method of the paper \cite{DKV}, as well as
related paper \cite{DF}, is that the Markov processes have countable sets of
values. So one can in principle use monotonicity arguments and stochastic
domination. In our situation the phase space is (one-dimensional) real
manifold, and this technique does not seem to be applicable.

The importance of the Poisson Hypothesis as the central problem of the theory
of large queueing systems was emphasized, among others, by Roland Dobrushin
\cite{D} and Alexander Borovkov \cite{B}.

\section{Notation}

In this section we will fix the notation for the non-linear Markov process,
which takes place at a given server in the above described limit.

\textit{Server.} It is defined by specifying the distribution of the random
serving time $\eta$, i.e. by the function
\[
F\left(  t\right)  =\mathbf{\Pr}\left\{  \text{serving time }\eta\geq
t\right\}  .
\]

We suppose that $\eta$ is such that:

\begin{itemize}
\item the density function $p\left(  t\right)  $ of $\eta$ is positive on
$t\geq0$ and uniformly bounded from above;

\item $p\left(  t\right)  $ satisfies the following strong Lipschitz
condition: for some $C<\infty$, some positive function $u\left(  t\right)  $
and for all $t\geq0$
\begin{equation}
\left|  p\left(  t+\Delta t\right)  -p\left(  t\right)  \right|  \leq
Cp\left(  t\right)  \left|  \Delta t\right|  ,\label{02}%
\end{equation}
provided $t+\Delta t>0$ and $\left|  \Delta t\right|  <u\left(  t\right)  ;$

\item introducing the random variables
\[
\eta\Bigm|_{\tau}=\left(  \eta-\tau\Bigm|\eta>\tau\right)  ,\tau\geq0,
\]
we suppose that for some $\delta>0,$ $M_{\delta}<\infty$
\begin{equation}
\mathbb{E}\left(  \eta\Bigm|_{\tau}\right)  ^{2+\delta}<M_{\delta}\label{154}%
\end{equation}
for all $\tau;$ therefore, for the conditional expectations we have
\begin{equation}
\mathbb{E}\left(  \eta\Bigm|_{\tau}\right)  <\bar{C}\label{11}%
\end{equation}
uniformly in $\tau\geq0;$

\item Without loss of generality we can suppose that
\begin{equation}
\mathbb{E}\left(  \eta\right)  =1.\label{153}%
\end{equation}
In what follows, the function $p\left(  \cdot\right)  $ will be fixed.
\end{itemize}

\textit{Configurations. }By a configuration of a server at a given time moment
$t$ we mean the following data:

\begin{itemize}
\item The number $n\geq0$ of customers waiting to be served. The customer who
is served at $t$, is included in the total amount $n.$ Therefore by
definition, the length of the queue is $n-1$ for $n\geq1,$ and $0$ for $n=0. $

\item The duration $\tau$ of the elapsed service time of the customer under
the service at the moment $t.$
\end{itemize}

Therefore the set of all configurations $\Omega$ is the set of all pairs
$\left(  n,\tau\right)  ,$ with an integer $n>0$ and a real $\tau>0,$ plus the
point $\mathbf{0}$, describing the situation of the server being idle. For a
configuration $\omega=\left(  n,\tau\right)  \in\Omega$ we define $N\left(
\omega\right)  =n.$ We put $N\left(  \mathbf{0}\right)  =0.$

\textit{States. }By a state of the system we mean a probability measure $\mu$
on $\Omega.$ We denote by $\mathcal{M}\left(  \Omega\right)  $ the set of all
states on $\Omega.$

\textit{Observables. }There are some natural random variables associated with
our system. One is the queue length, $N_{\mu}=N_{\mu}\left(  \omega\right)  .$
We denote by $N\left(  \mu\right)  $ the mean queue length in the state $\mu:$%
\[
N\left(  \mu\right)  =\mathbb{E}\left(  N_{\mu}\right)  \equiv\left\langle
N_{\mu}\left(  \omega\right)  \right\rangle _{\mu},
\]
and we introduce the subsets $\mathcal{M}_{q}\left(  \Omega\right)
\subset\mathcal{M}\left(  \Omega\right)  ,\,q\geq0$ by
\[
\mathcal{M}_{q}\left(  \Omega\right)  =\left\{  \mu\in\mathcal{M}\left(
\Omega\right)  :N\left(  \mu\right)  =q\right\}  .
\]
Another one is the expected service time
\[
S\left(  \omega\right)  =\left\{
\begin{array}
[c]{ll}%
0 & \text{ for }\omega=\mathbf{0,}\\
\left(  n-1\right)  \mathbb{E}\left(  \eta\right)  +\mathbb{E}\left(
\eta\Bigm|_{\tau}\right)  & \text{for }\omega=\left(  n,\tau\right)
,\text{with }n>0.
\end{array}
\right.
\]
Again, we define
\[
S\left(  \mu\right)  =\left\langle S\left(  \omega\right)  \right\rangle
_{\mu}.
\]
Clearly,
\begin{equation}
S\left(  \mu\right)  \leq\bar{C}N\left(  \mu\right)  .\label{12}%
\end{equation}

\textit{Input flow. }Suppose that a function $\lambda\left(  t\right)  \geq0$
is given. We suppose that the input flow to our server is a Poisson process
with rate function $\lambda\left(  t\right)  ,$ which means in particular that
the probabilities $P_{k}\left(  t,s\right)  $ of the events that $k$ new
customers arrive during the time interval $\left[  t,s\right]  $ satisfy
\[
P_{k}\left(  t,t+\Delta t\right)  =\left\{
\begin{array}
[c]{ll}%
\lambda\left(  t\right)  \Delta t+o\left(  \Delta t\right)  & \text{ for
}k=1,\\
1-\lambda\left(  t\right)  \Delta t+o\left(  \Delta t\right)  & \text{ for
}k=0,\\
o\left(  \Delta t\right)  & \text{ for }k>1,
\end{array}
\right.
\]
as $\Delta t\rightarrow0,$ while for non-intersecting time segments $\left[
t_{1},s_{1}\right]  ,\,\,\left[  t_{2},s_{2}\right]  $ the flows are independent.

\textit{Output flow. }Suppose the initial state $\nu=\mu\left(  -T\right)  ,$
$T>0,$ as well as the rate function $\lambda\left(  t\right)  ,$ with
$\lambda\left(  t\right)  =0$ for $t\leq-T,$ of the input flow are given. Then
the system evolves in time, and its state at the moment $t$ is given by the
measure
\[
\mu\left(  t\right)  =\mu_{\nu,\lambda\left(  \cdot\right)  }\left(  t\right)
.
\]
In particular, the probabilities $Q_{k}\left(  t,s\right)  =Q_{k}\left(
t,s;\nu,\lambda\left(  \cdot\right)  ,p\left(  \cdot\right)  \right)  $ of the
events that $k$ customers have finished their service during the time interval
$\left[  t,s\right]  $ are defined. We suppose that the customer, once served,
leaves the system.

The resulting random point process $Q_{\cdot}\left(  \cdot,\cdot\right)  $
need not, of course, be Poissonian. However we still can define its rate
function $b\left(  t\right)  $ as the one satisfying
\[
Q_{k}\left(  t,t+\Delta t\right)  =\left\{
\begin{array}
[c]{ll}%
b\left(  t\right)  \Delta t+o\left(  \Delta t\right)  & \text{ for }k=1,\\
1-b\left(  t\right)  \Delta t+o\left(  \Delta t\right)  & \text{ for }k=0,\\
o\left(  \Delta t\right)  & \text{ for }k>1,
\end{array}
\right.
\]
as $\Delta t\rightarrow0.$ The rate function $b\left(  \cdot\right)  $ of the
output flow is determined once the initial state $\nu=\mu\left(  0\right)  $
and the rate function $\lambda\left(  \cdot\right)  $ of the input flow are
given. Therefore the following (non-linear) operator $A$ is well defined:
\[
b\left(  \cdot\right)  =A\left(  \nu,\lambda\left(  \cdot\right)  ,-T\right)
.
\]
We will call the general situation, described by the triple $\nu
,\lambda\left(  \cdot\right)  ,b\left(  \cdot\right)  =A\left(  \nu
,\lambda\left(  \cdot\right)  ,-T\right)  $, as a General Flow Process (GFP).

The following is known about the operator $A,$ see \cite{KR1}:

\begin{itemize}
\item For every initial state $\nu$ the equation
\[
A\left(  \nu,\lambda\left(  \cdot\right)  \right)  \equiv A\left(  \nu
,\lambda\left(  \cdot\right)  ,0\right)  =\lambda\left(  \cdot\right)
\]
has exactly one solution $\lambda\left(  \cdot\right)  =\lambda_{\nu}\left(
\cdot\right)  .$ Then the evolving state $\mu_{\nu,\lambda_{\nu}\left(
\cdot\right)  }\left(  t\right)  $ is what is called \textit{the non-linear
Markov process, }which we will abbreviate as NMP.

\item The non-linear Markov process has the following conservation property:
for all $t$%
\[
N\left(  \mu_{\nu,\lambda_{\nu}\left(  \cdot\right)  }\left(  t\right)
\right)  =N\left(  \nu\right)
\]
(because ``the rates of the input flow and the output flow coincide''). So one
can say that the spaces $\mathcal{M}_{q}\left(  \Omega\right)  $ are invariant
under non-linear Markov evolutions.

\item All the functions $\lambda_{\nu}\left(  \cdot\right)  $ are bounded:
\begin{equation}
\lambda_{\nu}\left(  t\right)  \leq C=C\left(  \eta\right) \label{13}%
\end{equation}
uniformly in $\nu$ and $t.$

\item For every constant $c\in\lbrack0,1)$ there exists the initial state
$\nu_{c},$ such that
\begin{equation}
A\left(  \nu_{c},c\right)  =c.\label{03}%
\end{equation}
(Here we identify the constant $c$ with the function taking just one value $c
$ everywhere.) Moreover, this measure $\nu_{c}$ is a stationary state:
$\mu_{\nu_{c},c}\left(  t\right)  =\nu_{c}$ for all $t>0.$ The function
$c\rightsquigarrow N\left(  \nu_{c}\right)  $ is continuous increasing, with
$N\left(  \nu_{0}\right)  =0,$ $N\left(  \nu_{c}\right)  \uparrow\infty$ as
$c\rightarrow1.$
\end{itemize}

The non-linear Markov process $\mu_{\nu,\lambda_{\nu}\left(  \cdot\right)
}\left(  t\right)  $ is the main object of the present paper. Therefore we
will give below another definition of this process, via jump rates of
transitions during the infinitesimal time, $\Delta t.$ So suppose that our
process is in the state $\mu\in\mathcal{M}\left(  \Omega\right)  ,$ and
assumes the value $\omega=\left(  n,\tau\right)  \in\Omega.$ During the time
increment $\Delta t$ the following two transitions can happen with
probabilities of order of $\Delta t:$

\begin{itemize}
\item the customer under the service will finish it and will leave the server,
so the value $\left(  n,\tau\right)  $ will become $\left(  n-1,\varsigma
\right)  ,$ with $\varsigma\leq\Delta t.$ The probability of this event is
\[
c_{1}\Delta t+o\left(  \Delta t\right)  ,
\]
where
\[
c_{1}=c_{1}\left(  \omega\right)  =\lim_{\Delta t\rightarrow0}\frac{1}{\Delta
t}\frac{\int_{\tau}^{\tau+\Delta t}p\left(  x\right)  \,dx}{\int_{\tau
}^{\infty}p\left(  x\right)  \,dx};
\]

\item a new customer will arrive to the server, so the value $\left(
n,\tau\right)  $ will become $\left(  n+1,\tau+\Delta t\right)  .$ The
probability of this event is given by
\[
c_{2}\Delta t+o\left(  \Delta t\right)  ,
\]
where
\[
c_{2}=c_{2}\left(  \mu\right)  =\mathbb{E}_{\mu}\left(  c_{1}\left(
\omega\right)  \right)  .
\]
In words, the input rate is the average rate of the output in the state $\mu.$
\end{itemize}

\bigskip

It is curious to note that while for the general nonlinear continuous time
Markov processes its rates depend both on the value and on the state of the
process, in our case the rate $c_{1}$ depends only on the value, while the
rate $c_{2}$ -- only on the state of the process.

\section{More facts from \cite{KR1}}

Consider the following continuous time Markov process $\mathfrak{M}$. Let
there be $M$ servers and $N$ customers. The serving time is $\eta.$ The
configuration of the system consists of specifying the numbers of customers
$n_{i},\;i=1,...,M,$ waiting at each server, plus the duration $\tau_{i}$ of
the service time for every customer under service. Therefore it is a point in
\[
\Theta_{M,N}=\left\{  \left(  \omega_{1},...,\omega_{M}\right)  \in\Pi
_{i=1}^{M}\Omega_{i}:n_{1}+...+n_{M}=N\right\}  .
\]
Upon being served, the customer goes to one of $M$ servers with equal
probability $1/M,$ and is there the last in the queue.

The permutation group $\mathcal{S}_{M}$ acts on $\Theta_{M,N},$ leaving the
transition probabilities invariant. Therefore we can consider the
factor-process. Its values are (unordered) finite subsets of $\Omega.$ It can
be equivalently described as a measure
\[
\nu=\frac{1}{M}\sum_{i=1}^{M}\delta_{\left(  n_{i},\tau_{i}\right)  }.
\]
We can identify such measures with the configurations of the symmetrized
factor-process. Note that
\[
\left\langle n\right\rangle _{\nu}=\frac{N}{M}.
\]
So if we introduce the notation $\mathcal{M}_{q}\left(  \Omega\right)
\subset\mathcal{M}\left(  \Omega\right)  $ for the measures $\mu$ on $\Omega$
for which $\left\langle n\right\rangle _{\mu}=q,$ then we have that $\nu
\in\mathcal{M}_{\frac{N}{M}}\left(  \Omega\right)  .\;$We also introduce the
notation $\mathcal{M}_{\frac{N}{M},M}\left(  \Omega\right)  \subset
\mathcal{M}_{\frac{N}{M}}\left(  \Omega\right)  $ for the family of atomic
measures, such that each atom has a weight $\frac{k}{M}$ for some integer $k.$

A state of our Markov process is a probability measure on the set of
configurations, i.e. an element of $\mathcal{M}\left(  \mathcal{M}\left(
\Omega\right)  \right)  .$ If the initial state of the process is supported by
$\mathcal{M}_{q}\left(  \Omega\right)  ,$ then at any positive time it is
still the element of $\mathcal{M}\left(  \mathcal{M}_{q}\left(  \Omega\right)
\right)  .$ A natural embedding $\mathcal{M}\left(  \Omega\right)
\subset\mathcal{M}\left(  \mathcal{M}\left(  \Omega\right)  \right)  ,$ which
to each configuration $\nu\in\mathcal{M}\left(  \Omega\right)  $ corresponds
the atomic measure $\delta_{\nu},$ will be denoted by $\delta.$

For $\mu_{0}=\delta_{\nu}\in\mathcal{M}\left(  \mathcal{M}_{\frac{N}{M}%
,M}\left(  \Omega\right)  \right)  $ to be the initial state of our Markov
process, we denote by $\mu_{t}$ the evolution of this state. Clearly, in
general $\mu_{t}\notin\delta\left(  \mathcal{M}\left(  \Omega\right)  \right)
$ for positive $t.$ This process is ergodic. We denote by $\pi_{M,N} $ the
stationary measure of this process.

Let now $\kappa\in\mathcal{M}_{q}\left(  \Omega\right)  $ be some measure, let
the sequences of integers $N_{j},M_{j}\rightarrow\infty$ be such that
$\frac{N_{j}}{M_{j}}\rightarrow q,$ and let the measures $\nu^{j}%
\in\mathcal{M}_{\frac{N_{j}}{M_{j}},M_{j}}\left(  \Omega\right)  $ be such
that $\nu_{j}\rightarrow\kappa$ weakly. Consider the Markov processes $\mu
_{t}^{j}\in\mathcal{M}\left(  \mathcal{M}_{\frac{N_{j}}{M_{j}},M_{j}}\left(
\Omega\right)  \right)  ,$ corresponding to the initial conditions
$\delta_{\nu^{j}}.$ As we just said, in general $\mu_{t}^{j}\notin
\delta\left(  \mathcal{M}_{\frac{N_{j}}{M_{j}},M_{j}}\left(  \Omega\right)
\right)  $ for any $j,$ once $t>0.$ However, for the limit $\mu_{t}%
=\lim_{j\rightarrow\infty}\mu_{t}^{j}$ we have that $\mu_{t}\in\mathcal{M}%
\left(  \mathcal{M}_{q}\left(  \Omega\right)  \right)  ,$ and moreover
$\mu_{t}\in\delta\left(  \mathcal{M}_{q}\left(  \Omega\right)  \right)  ,$ so
we can say that the random evolutions $\mu_{t}^{j}$ tend to the non-random
evolution $\kappa_{t}\equiv\mu_{t}$ as $N_{j},M_{j}\rightarrow\infty.$

Therefore we have a dynamical system
\begin{equation}
\mathcal{T}_{t}:\mathcal{M}_{q}\left(  \Omega\right)  \rightarrow
\mathcal{M}_{q}\left(  \Omega\right)  .\label{160}%
\end{equation}
This dynamical system $\mu_{t}$ is nothing else but the non-linear Markov
process, mentioned above.

Another way of obtaining the same dynamical system is to look on the behavior
of a given server. Here instead of taking the symmetrization of the initial
process $\mathfrak{M}$ on $\Theta_{M,N},$ we have to consider its projection
on the first coordinate, $\Omega_{1}$, say. To make the correspondence with
the above, we have to take for the initial state of this process a measure
$\tilde{\nu}^{j}$ on $\Theta_{M,N},$ which is $\mathcal{S}_{M}$-invariant, and
which symmetrization is the initial state $\nu^{j}$ of the preceding
paragraph. The projection of $\mathfrak{M}$ on $\Omega_{1}$ would not be, of
course, a Markov process. However, it becomes the very same non-linear Markov
process $\mu_{t}$ in the above limit $N_{j},M_{j}\rightarrow\infty$.

We can generalize further, and study the projection of $\mathfrak{M}$ to a
finite product, $\prod_{j=1}^{R}\Omega_{j}.$ Then in the limit $N_{j}%
,M_{j}\rightarrow\infty$ this projection converges to a process on
$\prod_{j=1}^{R}\Omega_{j},$ which factors into the product of $R$ independent
copies of the same non-linear Markov process $\mu_{t}.$ This statement is
known as the ``propagation of chaos'' property.

The main result of the present paper is that for every $q$ the dynamical
system $\left(  \ref{160}\right)  $ has exactly one fixed point $\nu_{c},$
$c=c\left(  q\right)  $, and that it is globally attractive. In particular
that means that $\pi_{N_{j},M_{j}}\rightarrow\nu_{c},$ provided $\frac{N_{j}%
}{M_{j}}\rightarrow q$ as $j\rightarrow\infty$ and $c=c\left(  q\right)  .$

\section{Main result}

Our goal is to show the following:

\begin{theorem}
For every initial state $\nu$ the solution $\lambda_{\nu}\left(  \cdot\right)
$ of the equation
\[
A\left(  \nu,\lambda\left(  \cdot\right)  \right)  =\lambda\left(
\cdot\right)
\]
has the \textbf{relaxation }property:
\[
\lambda_{\nu}\left(  t\right)  \rightarrow c\text{ as }t\rightarrow\infty,
\]
where the constant $c$ satisfies
\[
\mathbb{E}_{\nu}\left(  N\left(  \omega\right)  \right)  =\mathbb{E}_{\nu_{c}%
}\left(  N\left(  \omega\right)  \right)  .
\]
Moreover, $\mu_{\nu,\lambda_{\nu}\left(  \cdot\right)  }\left(  t\right)
\rightarrow\nu_{c}$ weakly, as $t\rightarrow\infty.$
\end{theorem}

A special case of the above theorem is the following

\begin{proposition}
\label{P} Let $T>0$ be some time moment, and suppose that the function
$\lambda\left(  \cdot\right)  $ satisfies
\begin{equation}
A\left(  \mathbf{0},\lambda\left(  \cdot\right)  ,-T\right)  =b\left(
\cdot\right) \label{30}%
\end{equation}
with
\begin{equation}
\lambda\left(  t\right)  =b\left(  t\right)  \text{ for all }t\geq0.\label{31}%
\end{equation}
Let also
\[
\int_{-T}^{0}\lambda\left(  t\right)  \,dt\leq C<\infty.
\]
Then for some $c\geq0$%
\begin{equation}
\lambda\left(  t\right)  \rightarrow c\text{ as }t\rightarrow\infty.\label{32}%
\end{equation}

\end{proposition}

Our theorem follows from the Proposition \ref{P} immediately in the special
case when the initial state $\nu$ is of the form $\nu=\mu_{\mathbf{0}%
,\lambda\left(  \cdot\right)  }\left(  t\right)  $ for some $\lambda$ and some
$t>0.$ These initial states are easier to handle, so we treat them separately.

The heuristics behind the Proposition \ref{P} is the following. One expects
that if
\[
b\left(  \cdot\right)  =A\left(  \nu,\lambda\left(  \cdot\right)  \right)  ,
\]
then the function $b$ for large times is ``closer to a constant'' than the
function $\lambda.$ More precisely, if $t\ $belongs to some segment $\left[
T_{1},T_{2}\right]  $, with $T_{1}\gg1,$ then the dependence of $b\left(
t\right)  $ on $\nu$ is very weak, so $b$ is determined mainly by $\lambda.$
One then argues that under that assumption $\sup_{t\in\left[  T_{1}%
,T_{2}\right]  }b\left(  t\right)  $ should be strictly less than $\sup
_{t\in\left[  T_{1},T_{2}\right]  }\lambda\left(  t\right)  .$ Indeed, one can
visualize the random configuration of the exit moments $y_{i}$-s as being
obtained from the input flow configuration of $x_{i}$-s by making it
\textit{sparser}. Namely, we have to consider a sequence $\eta_{i}$ of i.i.d.
random variables, having the same distribution as $\eta,$ and then to move the
particles $x_{i}$ to the right, positioning them at locations $y_{i},$ so that
in the result
\begin{equation}
y_{i+1}-y_{i}\geq\eta_{i}\label{10}%
\end{equation}
for all $i$-s, \textit{\ }see $\left(  \ref{07}\right)  $, $\left(
\ref{06}\right)  $ below for more details. However this is a very rough idea,
since some particles need not be moved, due to the fact that (\ref{10}) may
hold even prior to the sparsening step, in which case it will happen that
$y_{i+1}=x_{i+1},$ while $y_{i}>x_{i},$ and so the configuration becomes
locally denser. (And if $\lambda$ is a constant, then $b$ is this same
constant, so again the above argument is not literally true.)

To be more precise, we will show the following\textbf{\ self-averaging
property}. Let the functions $\lambda\left(  \cdot\right)  $ and $b\left(
\cdot\right)  $ are related by
\[
b\left(  \cdot\right)  =A\left(  \mathbf{0},\lambda\left(  \cdot\right)
,-T\right)  .
\]
One of the main points of the following would be to show that for every $x$
one can find a probability density $q_{\lambda,x}\left(  t\right)  ,$
vanishing for $t\leq0,$ such that
\begin{equation}
b\left(  x\right)  =\left[  \lambda\ast q_{\lambda,x}\right]  \left(
x\right)  .\label{34}%
\end{equation}
We then will show that this self-averaging property of the system implies
(\ref{32}), provided we know in advance certain regularity properties of the
family $\left\{  q_{\lambda,x}\right\}  $. Note that apriori the condition
(\ref{34}) is not evident at all for our FIFO system: one has to rule out the
situation that, say, the input rate function $\lambda$ is uniformly bounded
from above by $1,$ while the output rate $b$ is occasionally reaching the
level $2;$ this is clearly inconsistent with (\ref{34}).

In general case, when
\[
b\left(  \cdot\right)  =A\left(  \mathbf{\mu},\lambda\left(  \cdot\right)
\right)  ,
\]
we have
\begin{equation}
b\left(  x\right)  =\left(  1-\varepsilon_{\lambda,\mu}\left(  x\right)
\right)  \left[  \lambda\ast q_{\lambda,\mu,x}\right]  \left(  x\right)
+\varepsilon_{\lambda,\mu}\left(  x\right)  Q_{\lambda,\mu}\left(  x\right)
,\label{134}%
\end{equation}
where $\varepsilon_{\lambda,\mu}\left(  x\right)  >0,$ $\varepsilon
_{\lambda,\mu}\left(  x\right)  \rightarrow0$ as $x\rightarrow\infty,$ while
$Q_{\lambda,\mu}\left(  x\right)  $ is a bounded term, see Section 8 for details.

To get the above mentioned regularity property we will need few preparatory lemmas.

\begin{lemma}
Let $\mu_{\nu,\lambda_{\nu}\left(  \cdot\right)  }\left(  \cdot\right)  $ be
NMP, with $N\left(  \mu_{\nu,\lambda_{\nu}\left(  \cdot\right)  }\left(
t\right)  \right)  =N\left(  \nu\right)  =q.$ Then there exists a time moment
$T=T\left(  q\right)  $ and $\varepsilon=\varepsilon\left(  q\right)  >0,$
such that for all $t>T$%
\begin{equation}
\left\langle \omega=\mathbf{0}\right\rangle _{\mu_{\nu,\lambda_{\nu}\left(
\cdot\right)  }\left(  t\right)  }>\varepsilon.\label{302}%
\end{equation}
In words, the probability of observing the system $\mu_{\nu,\lambda_{\nu
}\left(  \cdot\right)  }\left(  t\right)  $ to be in the idle state is
uniformly positive, after some time $T\left(  q\right)  $.
\end{lemma}

\begin{proof}
Let the initial state $\kappa$ of the NMP be such that $N\left(
\kappa\right)  =q.$ Then by (\ref{11}), (\ref{12}), we have $S\left(
\kappa\right)  \leq\bar{C}q.$ Consider now the GFP, started in $\kappa$ and
having zero input flow, i.e. $\lambda\equiv0.$ We denote it by $\mu_{\kappa
,0}\left(  t\right)  .$ Consider the probability $\left\langle N\left(
\omega\right)  >0\right\rangle _{\mu_{\kappa,0}\left(  t\right)  }$ that such
a system is still occupied at the moment $t.$ Then clearly
\[
S\left(  \kappa\right)  \geq t\;\left\langle N\left(  \omega\right)
>0\right\rangle _{\mu_{\kappa,0}\left(  t\right)  }.
\]
Therefore
\[
\left\langle \omega=\mathbf{0}\right\rangle _{\mu_{\kappa,0}\left(  t\right)
}\geq1-\frac{\bar{C}q}{t}.
\]
In particular, if we put $T=2\bar{C}q,$ then for all $t\geq T$
\[
\left\langle \omega=\mathbf{0}\right\rangle _{\mu_{\kappa,0}\left(  t\right)
}\geq\frac{1}{2}.
\]

Consider now the NMP started at $\kappa.$ Let us introduce the event
\[
\mathcal{E}_{\kappa}\left(  t\right)  =\left\{
\begin{array}
[c]{l}%
\text{in the Poisson random flow, defined by the rate}\\
\lambda_{\kappa}\left(  \cdot\right)  ,\text{ no customer arrives before time
}t.
\end{array}
\right\}  \text{ }%
\]
Then
\begin{align}
\left\langle \omega=\mathbf{0}\right\rangle _{\mu_{\kappa,\lambda_{\kappa
}\left(  \cdot\right)  }\left(  T\right)  }  & \geq\mathbf{\Pr}\left(
\mathcal{E}_{T}\right)  \left\langle \omega=\mathbf{0}\Bigm|\mathcal{E}%
_{\kappa}\left(  T\right)  \right\rangle _{\mu_{\kappa,\lambda_{\kappa}\left(
\cdot\right)  }\left(  T\right)  }\label{14}\\
& =\mathbf{\Pr}\left(  \mathcal{E}_{\kappa}\left(  T\right)  \right)
\left\langle \omega=\mathbf{0}\right\rangle _{\mu_{\kappa,0}\left(  T\right)
}\geq\frac{1}{2}\mathbf{\Pr}\left(  \mathcal{E}_{\kappa}\left(  T\right)
\right)  ,\nonumber
\end{align}
so we need an estimate on the probability $\mathbf{\Pr}\left(  \mathcal{E}%
_{\kappa}\left(  T\right)  \right)  .$ This is easy, because of (\ref{13}):
\begin{equation}
\mathbf{\Pr}\left(  \mathcal{E}_{\kappa}\left(  T\right)  \right)
=\exp\left\{  -\int_{0}^{T}\lambda_{\kappa}\left(  t\right)  \,dt\right\}
\geq\exp\left\{  -TC\left(  \eta\right)  \right\}  .\label{15}%
\end{equation}
That proves our statement with $T=2\bar{C}q$ and $\varepsilon=\exp\left\{
-TC\left(  \eta\right)  \right\}  /2,$ though thus far only for $t=T.$

But in fact we are done! Indeed, for an arbitrary $t>T$ let us take the state
$\kappa=\kappa_{t-T}=\mu_{\nu,\lambda_{\nu}\left(  \cdot\right)  }\left(
t-T\right)  $ of the process $\mu_{\nu,\lambda_{\nu}\left(  \cdot\right)
}\left(  \cdot\right)  $ as the initial state of a new NMP, $\mu
_{\kappa,\lambda_{\kappa}\left(  \cdot\right)  }\left(  \cdot\right)  .$ Then
for every $\tau>T-t$ we have $\mu_{\nu,\lambda_{\nu}\left(  \cdot\right)
}\left(  \tau\right)  =\mu_{\kappa,\lambda_{\kappa}\left(  \cdot\right)
}\left(  \tau-\left(  t-T\right)  \right)  ,$ so in particular $\mu
_{\nu,\lambda_{\nu}\left(  \cdot\right)  }\left(  t\right)  =\mu
_{\kappa,\lambda_{\kappa}\left(  \cdot\right)  }\left(  T\right)  .$ Since
$\kappa_{t-T}\in\mathcal{M}_{q}\left(  \Omega\right)  ,$ we can apply
(\ref{14}), (\ref{15}) and thus complete the proof.
\end{proof}

\begin{lemma}
\label{la} Let $\mu_{\nu,\lambda_{\nu}\left(  \cdot\right)  }\left(
\cdot\right)  $ be NMP, with $N\left(  \mu_{\nu,\lambda_{\nu}\left(
\cdot\right)  }\left(  t\right)  \right)  =N\left(  \nu\right)  =q.$ Then
there exists a time moment $T^{\prime}=T^{\prime}\left(  q\right)  $ and
$\varepsilon^{\prime}=\varepsilon^{\prime}\left(  q\right)  >0,$ such that for
all $T\geq T^{\prime}$
\begin{equation}
\int_{0}^{T}\lambda_{\nu}\left(  t\right)  \,dt<T\left(  1-\varepsilon
^{\prime}\right)  .\label{01}%
\end{equation}

\end{lemma}

\begin{proof}
A configuration $\chi$ of our process in the segment $\left[  0,T^{\prime
}\right]  $ consists from

$i)$ the initial configuration $\left(  n,\tau\right)  ,$ drawn from the
distribution $\mu;$

$ii)$ the random set $0<x_{1}<...<x_{m}<T^{\prime},$ which is a realization of
the Poisson random field defined by the rate function $\lambda_{\nu}$
(restricted to the segment $\left[  0,T^{\prime}\right]  $), independent of
$\left(  n,\tau\right)  $;

$iii)$ one realization $\eta_{1}$ of the conditional random variable $\left(
\eta-\tau\Bigm|\eta>\tau\right)  $ and $n+m-1$ independent realizations
$\eta_{k},k=2,...,n+m$ of the random variable $\eta.$ We denote by
$\mathbb{P}_{\mu\otimes\lambda\otimes\eta}\left(  d\chi\right)  $ the
corresponding (product) distribution.

Let $\bar{A}\left(  \chi\right)  \subset\lbrack0,\infty)$ be the set on the
real line, covered by the rods of $\chi$ after the resolution of conflicts.
Let $B\left(  \chi\right)  =[0,\infty)\,\backslash\,\bar{A}\left(
\chi\right)  .$ Finally, let $A\left(  \chi\right)  \subset\bar{A}\left(
\chi\right)  $ be the set covered only by the last $m$ rods, while $C\left(
\chi\right)  =\bar{A}\left(  \chi\right)  \,\backslash\,A\left(  \chi\right)
$. A moment thought shows that
\begin{equation}
\mathbb{E}_{\chi}\left(  \int_{0}^{\infty}\mathbf{I}_{A\left(  \chi\right)
}\left(  x\right)  dx\right)  =\int\mathrm{mes}\left\{  A\left(  \chi\right)
\right\}  \mathbb{P}_{\mu\otimes\lambda\otimes\eta}\left(  d\chi\right)
=\int_{0}^{T^{\prime}}\lambda_{\nu}\left(  t\right)  \,dt.\label{114}%
\end{equation}
Also
\[
\int_{0}^{T^{\prime}}\left(  \mathbf{I}_{\bar{A}\left(  \chi\right)  }\left(
x\right)  +\mathbf{I}_{B\left(  \chi\right)  }\left(  x\right)  \right)
dx\equiv T^{\prime}.
\]
Evidently,
\[
\mathbb{E}_{\chi}\left(  \mathbf{I}_{B\left(  \chi\right)  }\left(  x\right)
\right)  =\mathbf{\Pr}\left\{  \text{the system is idle at the moment
}x\right\}  .
\]
From the pervious lemma we know that $\mathbb{E}_{\chi}\left(  \mathbf{I}%
_{B\left(  \chi\right)  }\left(  x\right)  \right)  >\varepsilon$ for all $x$
large enough. Therefore
\begin{equation}
\mathbb{E}_{\chi}\left(  \int_{0}^{T^{\prime}}\mathbf{I}_{\bar{A}\left(
\chi\right)  }\left(  x\right)  dx\right)  <T^{\prime}\left(  1-\varepsilon
/2\right) \label{115}%
\end{equation}
once $T^{\prime}$ is large enough. Finally,
\begin{align}
& \left|  \mathbb{E}_{\chi}\left(  \int_{0}^{T^{\prime}}\mathbf{I}_{\bar
{A}\left(  \chi\right)  }\left(  x\right)  dx-\int_{0}^{\infty}\mathbf{I}%
_{A\left(  \chi\right)  }\left(  x\right)  dx\right)  \right| \nonumber\\
& =\left|  \mathbb{E}_{\chi}\left(  \int_{0}^{\infty}\mathbf{I}_{C\left(
\chi\right)  }\left(  x\right)  dx\right)  -\mathbb{E}_{\chi}\left(
\int_{T^{\prime}}^{\infty}\mathbf{I}_{\bar{A}\left(  \chi\right)  }\left(
x\right)  dx\right)  \right| \label{116}%
\end{align}
Note that each of the last two expectations is the mean occupation time of our
system when it is initially in the states $\mu_{\nu,\lambda_{\nu}\left(
\cdot\right)  }\left(  0\right)  =\nu$ and $\mu_{\nu,\lambda_{\nu}\left(
\cdot\right)  }\left(  T^{\prime}\right)  ,$ while no extra input flows are
present. Since $N\left(  \nu\right)  =N\left(  \mu_{\nu,\lambda_{\nu}\left(
\cdot\right)  }\left(  T^{\prime}\right)  \right)  =q,$ the difference between
the expectations of these occupation times does not exceed $2\bar{C},$ see
$\left(  \ref{11}\right)  .$ This, together with (\ref{114}-\ref{116}) proves
our statement to hold for $T^{\prime}$ large, with $\varepsilon^{\prime
}=\varepsilon/4.$
\end{proof}

We finish this section with a statement about the regularity of the exit flow.

\begin{lemma}
Let the function $p\left(  t\right)  $ satisfies the strong Lipschitz
condition $\left(  \ref{02}\right)  $: for some $C$%
\[
\left|  p\left(  t+\Delta t\right)  -p\left(  t\right)  \right|  \leq
Cp\left(  t\right)  \Delta t.
\]
Then the function $b\left(  t\right)  $ is Lipschitz.
\end{lemma}

\begin{proof}
Let $t$ be fixed. The idea of the proof is to correspond to every elementary
event, which contribute to the output rate $b\left(  t\right)  ,$ the
elementary event, contributing to $b\left(  t+\Delta t\right)  ,$ by enlarging
by $\Delta t$ the service time of the customer, whose service ends at the
moment $t.$ This correspondence, however, does not ``cover'' all the events,
contributing to $b\left(  t+\Delta t\right)  .$ Namely, the elementary events
not covered by the above correspondence, are precisely those, for which the
customer, whose service terminated at $t+\Delta t,$ started his service after
the moment $t.$

Let us first estimate the probability $\pi\left(  t,\Delta t\right)  $ of the
event $\Pi\left(  t,\Delta t\right)  $ that some customer started to be served
after the moment $t,$ and was served before $t+\Delta t.$ Consider an
elementary event, contributing to $\Pi\left(  t,\Delta t\right)  .$ It is some
configuration $\left(  \bar{x}_{1},...,\bar{x}_{n};\bar{l}_{1},...,\bar{l}%
_{n}\right)  ,$ where a certain rod $\bar{l}_{k}$ satisfies $\bar{l}_{k}%
\leq\Delta t.$ Comparing the collection of events $\left\{  \left(  \bar
{x}_{1},...,\bar{x}_{n};\bar{l}_{1},...\bar{l}_{k-1},l_{k},...,\bar{l}%
_{n}\right)  :l_{k}\leq\Delta t\right\}  $ with the collection $\left\{
\left(  \bar{x}_{1},...,\bar{x}_{n};\bar{l}_{1},...\bar{l}_{k-1}%
,l_{k},...,\bar{l}_{n}\right)  :l_{k}>\Delta t\right\}  $ (Peierls
transformation), we conclude that
\[
\pi\left(  t,\Delta t\right)  \leq\frac{\int_{0}^{\Delta t}p\left(  t\right)
dt}{\int_{\Delta t}^{\infty}p\left(  t\right)  dt}\leq C_{p}\Delta t
\]
for some $C_{p}<\infty.$

Denote by $\zeta\left(  t\right)  $ the random moment of the beginning of the
service of the client, who happens to be the last one started to be served
before $t.$ Then one can define the rate $\gamma_{t}\left(  x\right)  $ for
all $x<t$ by
\[
\gamma_{t}\left(  x\right)  =\lim_{\Delta x\rightarrow0}\frac{\mathbf{\Pr
}\left\{  \zeta\left(  t\right)  \in\left[  x,x+\Delta x\right]  \text{
}\right\}  }{\Delta x}.
\]
Then
\[
b\left(  t\right)  =\int_{0}^{\infty}\gamma_{t}\left(  t-x\right)  p\left(
x\right)  dx.
\]
Clearly,
\[
b\left(  t+\Delta t\right)  =\int_{0}^{\infty}\gamma_{t}\left(  t-x\right)
p\left(  x+\Delta t\right)  dx+\pi\left(  t,\Delta t\right)  .
\]
Therefore
\begin{align*}
\left|  b\left(  t+\Delta t\right)  -b\left(  t\right)  \right|   & \leq
C_{p}\Delta t+\int_{0}^{\infty}\gamma_{t}\left(  t-x\right)  \left|  p\left(
x+\Delta t\right)  -p\left(  x\right)  \right|  dx\\
& \leq C_{p}\Delta t+C\Delta t\int_{0}^{\infty}\gamma_{t}\left(  t-x\right)
p\left(  x\right)  dx\\
& =C_{p}\Delta t+C\Delta tb\left(  t\right)  .
\end{align*}
Since $b\left(  \cdot\right)  $ is uniformly bounded, the proof follows.
\end{proof}

\section{The self-averaging relation}

Here we will derive a formula, expressing the function $b\left(  \cdot\right)
=A\left(  \mathbf{0},\lambda\left(  \cdot\right)  \right)  $ in terms of the
functions $\lambda\left(  \cdot\right)  $ and $p\left(  \cdot\right)  . $ This
will be the needed self-averaging relation (\ref{34}).

\begin{theorem}
Let the functions $b\left(  \cdot\right)  $ and $\lambda\left(  \cdot\right)
$ are related by
\[
b\left(  \cdot\right)  =A\left(  \mathbf{0},\lambda\left(  \cdot\right)
\right)  .
\]
Then there exists a family of probability densities $q_{\lambda,x}\left(
t\right)  ,$ such that
\[
b\left(  x\right)  =\int_{0}^{\infty}\lambda\left(  x-t\right)  q_{\lambda
,x}\left(  t\right)  \,dt.
\]

\end{theorem}

\begin{proof}
To see this we first introduce some new notions.

Let $l_{1},...,l_{n}>0$ be a collection of positive real numbers, which we
will interpret as the lengths of hard rods, placed in $\mathbb{R}^{1}.$ A
configuration of rods can be then given by specifying, say, their left-ends,
$x_{1}<x_{2}<...<x_{n},$ so that the rod $l_{i}$ occupies the segment $\left[
x_{i},x_{i}+l_{i}\right]  .$ This configuration will be denoted by $\sigma
_{n}\left(  x_{1},...,x_{n};l_{1},...,l_{n}\right)  .$

In case some of the rods are intersecting over a nondegenerate segments, we
say that such a configuration has conflicts. By a resolution of conflicts we
call another configuration of the rods $l_{1},...,l_{n},$ where these rods
have the following set $z_{1}<z_{2}<...<z_{n}$ of the left-ends:

it is defined inductively by
\[
z_{1}=x_{1},
\]
and
\begin{equation}
z_{i}=\max\left\{  z_{i-1}+l_{i-1},x_{i}\right\}  .\label{07}%
\end{equation}
We will denote by $y$-s the corresponding set of the right-ends:
\begin{equation}
y_{i}=z_{i}+l_{i}.\label{06}%
\end{equation}
Any configuration with no conflicts, and in particular any configuration
obtained by resolution of the conflicting one, will be called an
\textbf{r-configuration}. The operation of resolving the conflict will be
denoted by $R,$ so
\[
\sigma_{n}\left(  z_{1},...,z_{n};l_{1},...,l_{n}\right)  =R\sigma_{n}\left(
x_{1},...,x_{n};l_{1},...,l_{n}\right)  .
\]
For any configuration $\sigma$ of rods we will denote by $Y\left(
\sigma\right)  $ the set of their right-ends. So, in our notations
\[
\left(  y_{1},...,y_{n}\right)  =Y\left(  R\sigma_{n}\left(  x_{1}%
,...,x_{n};l_{1},...,l_{n}\right)  \right)  .
\]

Suppose now that the lengths $l_{1},...,l_{n},$ as well as the locations
$x_{1},...,x_{n-1}$ and $y$ are specified. We define the values $X\left(
y\right)  \equiv X\left(  y\Bigm|x_{1},...,x_{n-1};l_{1},...,l_{n}\right)  $
as the solutions of the equation
\begin{equation}
y\in Y\left(  R\sigma_{n}\left(  x_{1},...,x_{n-1},X\left(  y\right)
;l_{1},...,l_{n}\right)  \right)  .\label{21}%
\end{equation}
Note that for the general position data $\left(  x_{1},...,x_{n-1}%
;l_{1},...,l_{n}\right)  $ the function $X\left(  y\Bigm|x_{1},...,x_{n-1}%
;l_{1},...,l_{n}\right)  $ is not defined for some $y$-s of positive measure,
while for some other $y$-s it is multivalued, having several (finitely many)
branches, provided $n\geq2$. (The case $n=1$ is simple: $X\left(
y\Bigm|l_{1}\right)  =y-l_{1}.$)

Now we can write the desired formula:
\begin{align}
& b\left(  y\right)  =\exp\left\{  -I_{\lambda}\left(  y\right)  \right\}
\sum_{n=1}^{\infty}\frac{1}{\left(  n-1\right)  !}\times\label{129}\\
& \times\underset{n}{\underbrace{\int_{0}^{\infty}...\int_{0}^{\infty}}%
}\left[  \underset{n-1}{\underbrace{\int_{0}^{y}...\int_{0}^{y}}}%
\lambda\left(  X\left(  y\Bigm|x_{1},...,x_{n-1};l_{1},...,l_{n}\right)
\right)  \prod_{i=1}^{n-1}\lambda\left(  x_{i}\right)  \,dx_{i}\right]
\prod_{i=1}^{n}p\left(  l_{i}\right)  \,dl_{i},\nonumber
\end{align}
where
\[
I_{\lambda}\left(  y\right)  =\int_{0}^{y}\lambda\left(  x\right)  \,dx.
\]
The integral in (\ref{129}) should be understood as follows: the range of
integration coincides with the domain where the function $X\left(
y\Bigm|x_{1},...,x_{n-1};l_{1},...,l_{n}\right)  $ is defined, while over the
domains where the function $X$ is multivalued one should integrate each branch
separately and then take the sum of integrals.

In words, the meaning of the relation $\left(  \ref{129}\right)  $ is the
following: for every realization $x_{1},...,x_{n-1}$ of the Poisson random
field and every realization $l_{1},...,l_{n}$ of the sequence of the service
times, we look for time moments $X=X\left(  y\Bigm|x_{1},...,x_{n-1}%
;l_{1},...,l_{n}\right)  ,$ at which the $l_{n}$-customer has to arrive, so as
to ensure that at the moment $y$ some (other) customer will exit, after being
served. In some cases such moments might not exist, while in other cases there
might be more than one such moment. If $X_{i}$ are these moments, we then have
to add all the rate values, $\lambda\left(  X_{i}\right)  ,$ and to integrate
the sum $\sum_{i}\lambda\left(  X_{i}\right)  $ over all $n$ and all
$x_{1},...,x_{n-1};l_{1},...,l_{n},$ thus getting the exit rate $b\left(
y\right)  .$

The first term $\left(  n=1\right)  $ in (\ref{129}) is by definition the
convolution,
\begin{equation}
b_{1}\left(  y\right)  =\int_{0}^{y}\lambda\left(  y-l\right)  p\left(
l\right)  \,dl.\label{18}%
\end{equation}
Since $p\left(  l\right)  \geq0$ and
\begin{equation}
\int_{0}^{y}p\left(  l\right)  \,dl\leq1,\label{17}%
\end{equation}
we have indeed that $b_{1}\left(  y\right)  <\sup_{x\leq y}\lambda\left(
x\right)  $ in case when, say, the maxima of $\lambda$ are isolated, or when
$\lambda$ is not a constant and the support of the distribution $p$ is the
full semiaxis $\left\{  l>0\right\}  .$ We want to show that in some sense the
same is true for all the functions $b_{n},$ defined as
\begin{equation}
b_{n}\left(  y\right)  =\int\left[  \int\lambda\left(  X\left(  y\Bigm|x_{1}%
,...,x_{n-1};l_{1},...,l_{n}\right)  \right)  \prod_{i=1}^{n-1}\left(
\frac{\lambda\left(  x_{i}\right)  }{I_{\lambda}\left(  y\right)  }%
\,dx_{i}\right)  \right]  \prod_{i=1}^{n}p\left(  l_{i}\right)  \,dl_{i}%
.\label{20}%
\end{equation}
The crucial step will be the analog of (\ref{18}), (\ref{17}) for all $n>1,$
that is that
\[
b_{n}\left(  y\right)  =\int_{0}^{y}\lambda\left(  y-l\right)  p_{n}\left(
l\right)  \,dl,
\]
with $p_{n}\left(  l\right)  \geq0,\;\int_{0}^{y}p_{n}\left(  l\right)
\,dl\nearrow1$ for $y\rightarrow\infty.$ This turns out to be quite an
involved combinatorial statement.

Note that, evidently, the measure $\prod_{i=1}^{n}p\left(  l_{i}\right)
\,dl_{i}$ is invariant under the coordinate permutations in $\mathbb{R}^{n};$
therefore we can rewrite the expression (\ref{20}) for the function
$b_{n}\left(  y\right)  $ as
\begin{equation}
b_{n}\left(  y\right)  =\int\left[  \int\frac{1}{n!}\lambda\left(  \bar
{X}\left(  y\Bigm|x_{1},...,x_{n-1};\left\{  l_{1},...,l_{n}\right\}  \right)
\right)  \prod_{i=1}^{n-1}\left(  \frac{\lambda\left(  x_{i}\right)
}{I_{\lambda}\left(  y\right)  }\,dx_{i}\right)  \right]  \prod_{i=1}%
^{n}p\left(  l_{i}\right)  \,dl_{i},\label{23}%
\end{equation}
where the following notations and conventions are used:

\begin{itemize}
\item the (multivalued) function $\bar{X}\left(  y\Bigm|x_{1},...,x_{n-1}%
;\left\{  l_{1},...,l_{n}\right\}  \right)  $ by definition assigns to every
$y$ the union of the sets of solutions $X\left(  y\right)  $ of all the
equations
\begin{equation}
y\in Y\left(  R\sigma_{n}\left(  x_{1},...,x_{n-1},X\left(  y\right)
;l_{\pi\left(  1\right)  },...,l_{\pi\left(  n\right)  }\right)  \right)
,\label{22}%
\end{equation}
with $\pi$ running over all the permutation group $\mathcal{S}_{n}$ (the
notation $\left\{  l_{1},...,l_{n}\right\}  $ stresses the fact that the
function $\bar{X}$ does not depend on the order of $l_{i}$-s);

\item the entries of the set $\bar{X}\left(  y\Bigm|x_{1},...,x_{n-1};\left\{
l_{1},...,l_{n}\right\}  \right)  $ have to be counted with multiplicities,
which for a given $x\in\bar{X}\left(  y\Bigm|x_{1},...,x_{n-1};\left\{
l_{1},...,l_{n}\right\}  \right)  $ is by definition the number of equations
(\ref{22}) with different $\pi$-s, to which $x$ is a solution;

\item the integration in (\ref{23}) of the multivalued function means that
each sheet should be integrated and the results added. Moreover, each sheet
has to be taken as many times as its multiplicity is.
\end{itemize}

Since each contribution $\lambda\left(  X\left(  y\Bigm|x_{1},...,x_{n-1}%
;l_{1},...,l_{n}\right)  \right)  $ to $\left(  \ref{20}\right)  $ appears
$n!$ times in $\left(  \ref{23}\right)  ,$ we have to divide by $n!.$

We repeat that while for some $x$-s, $\pi$-s and $l$-s the equation (\ref{22})
might have no solutions, for other data it can have more than one solution.
Clearly, the set $\bar{X}\left(  y\Bigm|x_{1},...,x_{n-1};\left\{
l_{1},...,l_{n}\right\}  \right)  $, for almost every data $x_{1}%
,...,x_{n-1},$ can have no other entries than those of the form
\[
x_{A,y,\left\{  l_{i}\right\}  }=y-\sum_{i\in A\subset\left\{
1,2,...,n\right\}  }l_{i},
\]
where $A$ runs over all nonempty subsets of $\left\{  1,2,...,n\right\}  $
(i.e. at most $2^{n}-1$ different entries). So the function $\bar{X}\left(
y\Bigm|x_{1},...,x_{n-1};\left\{  l_{1},...,l_{n}\right\}  \right)  $, as a
function of $x_{1},...,x_{n-1},$ has to be piece-wise constant. It is not
ruled out apriori that for some data the set $\bar{X}\left(  y\Bigm|x_{1}%
,...,x_{n-1};\left\{  l_{1},...,l_{n}\right\}  \right)  $ can be empty. This
is not, however, the case. Moreover, as the Theorem \ref{T6} below states,

\begin{itemize}
\item the number of elements in the set $\bar{X}\left(  y\Bigm|x_{1}%
,...,x_{n-1};\left\{  l_{1},...,l_{n}\right\}  \right)  ,$ counted with
multiplicities, is \textbf{precisely }$n!$ for almost every value of the arguments.
\end{itemize}

Therefore we have for the inner integral in (\ref{23}):
\begin{align*}
& \int\frac{1}{n!}\lambda\left(  \bar{X}\left(  y\Bigm|x_{1},...,x_{n-1}%
;\left\{  l_{1},...,l_{n}\right\}  \right)  \right)  \prod_{i=1}^{n-1}\left(
\frac{\lambda\left(  x_{i}\right)  }{I_{\lambda}\left(  y\right)  }%
\,dx_{i}\right) \\
& =\int\frac{1}{n!}\sum_{\substack{A\subset\left\{  1,2,...,n\right\}  ,
\\A\neq\emptyset}}k\left(  A,y,x_{1},...,x_{n-1};\left\{  l_{1},...,l_{n}%
\right\}  \right)  \lambda\left(  x_{A,y,\left\{  l_{i}\right\}  }\right)
\prod_{i=1}^{n-1}\left(  \frac{\lambda\left(  x_{i}\right)  }{I_{\lambda
}\left(  y\right)  }\,dx_{i}\right)  ,
\end{align*}
where the integer $k\left(  A,y,x_{1},...,x_{n-1};\left\{  l_{1}%
,...,l_{n}\right\}  \right)  $ is the multiplicity of the value
$x_{A,y,\left\{  l_{i}\right\}  }$ of the function $\bar{X}$ at the point
$\left(  y,x_{1},...,x_{n-1};\left\{  l_{1},...,l_{n}\right\}  \right)  .$
Since
\[
\sum_{\substack{A\subset\left\{  1,2,...,n\right\}  , \\A\neq\emptyset
}}k\left(  A,y,x_{1},...,x_{n-1};\left\{  l_{1},...,l_{n}\right\}  \right)
=n!
\]
a.e., we have
\begin{align*}
& \int\frac{1}{n!}\lambda\left(  \bar{X}\left(  y\Bigm|x_{1},...,x_{n-1}%
;\left\{  l_{1},...,l_{n}\right\}  \right)  \right)  \prod_{i=1}^{n-1}\left(
\frac{\lambda\left(  x_{i}\right)  }{I_{\lambda}\left(  y\right)  }%
\,dx_{i}\right) \\
& =\sum_{\substack{A\subset\left\{  1,2,...,n\right\}  , \\A\neq\emptyset
}}q_{\lambda,y}\left(  A\Bigm|\left\{  l_{1},...,l_{n}\right\}  \right)
\lambda\left(  x_{A,y,\left\{  l_{i}\right\}  }\right)  ,
\end{align*}
where
\begin{align}
& q_{\lambda,y}\left(  A\Bigm|\left\{  l_{1},...,l_{n}\right\}  \right)
\label{25}\\
& =\int\frac{1}{n!}k\left(  A,y,x_{1},...,x_{n-1};\left\{  l_{1}%
,...,l_{n}\right\}  \right)  \prod_{i=1}^{n-1}\left(  \frac{\lambda\left(
x_{i}\right)  }{I_{\lambda}\left(  y\right)  }\,dx_{i}\right)  ,\nonumber
\end{align}
so
\begin{equation}
0\leq q_{\lambda,y}\left(  A\Bigm|\left\{  l_{1},...,l_{n}\right\}  \right)
\leq1,\text{ with }\sum_{\substack{A\subset\left\{  1,2,...,n\right\}  ,
\\A\neq\emptyset}}q_{\lambda,y}\left(  A\Bigm|\left\{  l_{1},...,l_{n}%
\right\}  \right)  =1,\label{26}%
\end{equation}
since the measures $\frac{\lambda\left(  x_{i}\right)  }{I_{\lambda}\left(
y\right)  }\,dx_{i}$ are probability measures on $\left[  0,y\right]  $. (Note
that the functions $k\left(  A,y,x_{1},...,x_{n-1};\left\{  l_{1}%
,...,l_{n}\right\}  \right)  $ do depend on the variables $x_{1},...,x_{n-1};$
hence the measures $q_{\lambda,y}\left(  \cdot\Bigm|\left\{  l_{1}%
,...,l_{n}\right\}  \right)  $ indeed depend on $\lambda,y.$) Therefore, for
the function $b_{n}\left(  y\right)  $ we obtain a sort of a convolution
expression:
\begin{equation}
b_{n}\left(  y\right)  =\int\sum_{\substack{A\subset\left\{
1,2,...,n\right\}  , \\A\neq\emptyset}}\left[  q_{\lambda,y}\left(
A\Bigm|\left\{  l_{1},...,l_{n}\right\}  \right)  \lambda\left(
x_{A,y,\left\{  l_{i}\right\}  }\right)  \right]  \prod_{i=1}^{n}p\left(
l_{i}\right)  \,dl_{i}.\label{24}%
\end{equation}
Be it the case that the probability measure $q_{\lambda,y}\left(
\cdot\Bigm|\left\{  l_{1},...,l_{n}\right\}  \right)  $ is concentrated on
just one subset $A=\left\{  1,2,...,n\right\}  ,$ we would obtain the usual
convolution
\[
b_{n}\left(  y\right)  =\int\lambda\left(  y-l_{1}-...-l_{n}\right)
\prod_{i=1}^{n}p\left(  l_{i}\right)  \,dl_{i}=\lambda\ast\underset
{n}{\underbrace{p\ast...\ast p}}\left(  y\right)  .
\]
Here the situation is more subtle, and in (\ref{24}) we have a stochastic
mixture of convolutions with random number of summands.

Taking into account the relations (\ref{129}), (\ref{20}), (\ref{24}), the
result can be summarized as follows. Let $\nu\equiv\nu_{\lambda,y}$ be the
integer valued random variable with the distribution
\[
\mathbf{\Pr}\left\{  \nu=n\right\}  =\exp\left\{  -I_{\lambda}\left(
y\right)  \right\}  \frac{\left[  I_{\lambda}\left(  y\right)  \right]  ^{n}%
}{n!},\,n=0,1,2,...,
\]
and $\eta_{1},\eta_{2},...$ be the i.i.d. random serving times. Consider the
random function $\xi_{\lambda,y}=\xi_{\lambda,y}\left(  \nu_{\lambda,y}%
;\eta_{1},\eta_{2},...\right)  ,$ such that its conditional distribution under
condition that the realization $\nu_{\lambda,y};\eta_{1},\eta_{2},...$ is
given, is supported by the finite set
\[
L\left(  \nu_{\lambda,y};\eta_{1},\eta_{2},...\right)  =\left\{  \sum_{i\in
A}\eta_{i}:A\subset\left\{  1,2,...,\nu_{\lambda,y}\mathbf{+}1\right\}
,A\neq\emptyset\right\}  \subset\mathbb{R}^{1},
\]
and is given by
\[
\mathbf{\Pr}\left\{  \xi_{\lambda,y}=\sum_{i\in A}\eta_{i}\Bigm|\nu
_{\lambda,y};\eta_{1},\eta_{2},...\right\}  =q_{\lambda,y}\left(
A\Bigm|\left\{  \eta_{1},...,\eta_{\nu_{\lambda,y}+1}\right\}  \right)
\]
(see (\ref{25})). Then the following holds:
\[
b\left(  y\right)  =\mathbb{E}\left(  \lambda\left(  y-\xi_{\lambda,y}\right)
\right)  .
\]
This is precisely the relation (\ref{34}), with $q_{\lambda,y}$ being the
distribution of $\xi_{\lambda,y}.$
\end{proof}

\section{Combinatorics of the rod placements}

In this section we will prove the Theorem \ref{T6}, which was used in the
previous section. We will use the notation of the previous section, introduced
at its beginning, up to relation $\left(  \ref{21}\right)  .$

By a cluster of the r-configuration $\sigma_{n}\left(  z_{1},...,z_{n}%
;l_{1},...,l_{n}\right)  $ we call any maximal subsequence $z_{i}%
<z_{i+1}<...<z_{j}$ such that $z_{j}=z_{i}+l_{i}+l_{i+1}+...+l_{j-1}.$ If
$z_{i}<z_{i+1}<...<z_{j}$ is a cluster of an r-configuration, then the point
$z_{i}$ will be called the root of the cluster, while the point $z_{j}$ will
be called the head of the cluster. Note that for a general position
configuration $\sigma_{n}\left(  x_{1},...,x_{n};l_{1},...,l_{n}\right)  $ the
point $z_{i}$ is a root of a cluster of the corresponding r-configuration if
and only if $z_{i}=x_{i}.$ The segment $\left[  z_{i},z_{j}+l_{j}\right]  $
will be called the body of the cluster $z_{i}<z_{i+1}<...<z_{j},$ and the
point $z_{j}+l_{j}$ will be called the end of the cluster.

The notation $\sigma_{n}\left(  x_{1},...,x_{n};l_{1},...,l_{n}\right)
\cup\sigma_{1}\left(  X,L\right)  $ has the obvious meaning of adding an extra
rod of the length $L$ at the location $X.$ Note though, that in general
\[
R\left[  \sigma_{n}\left(  x_{1},...,x_{n};l_{1},...,l_{n}\right)  \cup
\sigma_{1}\left(  X,L\right)  \right]  \neq R\left[  R\sigma_{n}\left(
x_{1},...,x_{n};l_{1},...,l_{n}\right)  \cup\sigma_{1}\left(  X,L\right)
\right]  .
\]
It is however the case, if the point $X$ is outside the union of all bodies of
clusters of $R\sigma_{n}\left(  x_{1},...,x_{n};l_{1},...,l_{n}\right)  .$
This will be used later.

In what follows we will need a marked point in $\mathbb{R}^{1}.$ For all our
purposes it is convenient to chose the origin, $0\in\mathbb{R}^{1},$ as such a point.

We will say that the resolution of conflicts in the configuration

\noindent$\sigma_{n}\left(  x_{1},...,x_{n};l_{1},...,l_{n}\right)  $ results
in a hit of the origin, iff for some $k$ we have
\begin{equation}
y_{k}\equiv z_{k}+l_{k}=0.\label{04}%
\end{equation}
Such a hit will be called an $x_{r}$-hit, iff the cluster of the point $z_{k}
$ has its root at $z_{r}=x_{r}.$ (Necessarily, we have that $r\leq k.$) An
$x_{r}$-hit will be called an $\left(  x_{r},x_{k}\right)  $-hit, if
(\ref{04}) holds.

Now we are ready to formulate our problem. Let $n$ be an integer, and
$\lambda_{1}<\lambda_{2}<...<\lambda_{n}$ be a fixed set of positive lengths
of rods. Let $x_{1}<x_{2}<...<x_{n-1}$ be a set of $\left(  n-1\right)  $
left-ends. We want to compute the number $N\left(  x_{1},x_{2},...,x_{n-1}%
;\lambda_{1},\lambda_{2},...,\lambda_{n}\right)  ,$ which is defined as
follows. For any permutation $\pi$ of $n$ elements and for any $X\in
\mathbb{R}^{1},\,X\neq x_{1},x_{2},...,x_{n-1}$ we can consider the
configuration $\sigma_{n-1}\left(  x_{1},...,x_{n-1};\lambda_{\pi\left(
1\right)  },...,\lambda_{\pi\left(  n-1\right)  }\right)  $ $\cup\sigma
_{1}\left(  X,\lambda_{\pi\left(  n\right)  }\right)  $ of rods, when the rods
$l_{i}=\lambda_{\pi\left(  i\right)  }$ are placed at $x_{i},i=1,...,n-1,$
while the free rod $l_{n}=\lambda_{\pi\left(  n\right)  } $ is placed at $X.$
Given $\pi,$ we count the number $N_{\pi}\left(  x_{1},...,x_{n-1};\lambda
_{1},...,\lambda_{n}\right)  $ of different locations $X,$ such that the
corresponding r-configuration $R\left[  \sigma_{n-1}\left(  x_{1}%
,...,x_{n-1};\lambda_{\pi\left(  1\right)  },...,\lambda_{\pi\left(
n-1\right)  }\right)  \cup\sigma_{1}\left(  X,\lambda_{\pi\left(  n\right)
}\right)  \right]  $ has a hit, and moreover this hit is an $X$-hit. (In
certain cases one cannot produce an $X$-hit by putting the rod $l_{n}%
=\lambda_{\pi\left(  n\right)  }$ anywhere on $\mathbb{R}^{1};$ then $N_{\pi
}\left(  x_{1},...,x_{n-1};\lambda_{1},...,\lambda_{n}\right)  =0.$ In certain
other cases there are more than one possibility to place the free rod so as to
produce an $X$-hit.) Then we define
\[
N\left(  x_{1},...,x_{n-1};\lambda_{1},...,\lambda_{n}\right)  =\sum_{\pi
\in\mathcal{S}_{n}}N_{\pi}\left(  x_{1},...,x_{n-1};\lambda_{1},...,\lambda
_{n}\right)  .
\]

\begin{theorem}
\label{T6} For almost every $x_{1},...,x_{n-1}$ and $\lambda_{1}%
,...,\lambda_{n},$%
\[
N\left(  x_{1},...,x_{n-1};\lambda_{1},...,\lambda_{n}\right)  =n!
\]

\end{theorem}

\begin{proof}
Let us explain why the result is plausible. Let the set $x_{1},...,x_{n-1}$ be
given. Then we can choose the positive numbers $\lambda_{1},...,\lambda_{n}$
so small that for any $\pi$ the configuration

\noindent$\sigma_{n-1}\left(  x_{1},...,x_{n-1};\lambda_{\pi\left(  1\right)
},...,\lambda_{\pi\left(  n-1\right)  }\right)  \cup\sigma_{1}\left(
X=-\lambda_{\pi\left(  n\right)  },\lambda_{\pi\left(  n\right)  }\right)  ,$
having the $\left(  X,X\right)  $-hit, has no conflicts, while no other choice
of $X$ results in a hit. Therefore in our case \noindent$N_{\pi}\left(
x_{1},...,x_{n-1};\lambda_{1},...,\lambda_{n}\right)  =1$ for every $\pi,$ so indeed

\noindent$N\left(  x_{1},...,x_{n-1};\lambda_{1},...,\lambda_{n}\right)  =n! $.

Now we explain why our result is non-trivial. To see it, take $n=2,$
$x_{1}=-3,$ $\lambda_{1}=1,$ $\lambda_{2}=10.$ Then
\[
N_{12}\left(  x_{1};\lambda_{1},\lambda_{2}\right)  =2
\]
-- one can place the rod $10$ at $-10$ or at $-11.$ On the other hand,
\[
N_{21}\left(  x_{1};\lambda_{1},\lambda_{2}\right)  =0
\]
-- the rod $10,$ placed at $-3,$ blocks the origin from being hit. Still,
$2+0=2!.$ Note that this example is a general position one.

We will derive our theorem from its special case, explained in the first
paragraph of the present proof. The idea of computing $N\left(  x_{1}%
,...,x_{n-1};\lambda_{1},...,\lambda_{n}\right)  $ for a general data is to
decrease one by one the numbers $\lambda_{1}<\lambda_{2}<...<\lambda_{n},$
starting from the smallest one, to the values very small, keeping track on the
quantities $N_{\pi}\left(  x_{1},...,x_{n-1};\lambda_{1},...,\lambda
_{n}\right)  .$ During this evolution some of these will jump, but the total
sum $N\left(  x_{1},...,x_{n-1};\lambda_{1},...,\lambda_{n}\right)  $ would
stay unchanged, as we will show. That will prove our theorem.

We begin by presenting a simple formula for the number $N_{\pi}\left(
x_{1},...,x_{n-1};\lambda_{1},..,\lambda_{n}\right)  .$ Consider the rod
configuration $R\left[  \sigma_{n-1}\left(  x_{1},...,x_{n-1};\lambda
_{\pi\left(  1\right)  },...,\lambda_{\pi\left(  n-1\right)  }\right)
\right]  ,$ which will be abbreviated as $R_{\pi}\left(  \lambda
_{1},..,\lambda_{n}\right)  \equiv R_{\pi}\left(  \mathbf{\lambda}\right)  .$
Let us compute the quantity $S_{\pi}\left(  x_{1},...,x_{n-1};\lambda
_{1},..,\lambda_{n}\right)  ,$ which is the number of points $y_{i}\in
Y\left(  R_{\pi}\left(  \lambda_{1},..,\lambda_{n}\right)  \right)  ,$ falling
into the segment $\left[  -\lambda_{\pi\left(  n\right)  },0\right]  .$ Then
\begin{equation}
N_{\pi}\left(  x_{1},...,x_{n-1};\lambda_{1},..,\lambda_{n}\right)  =\left\{
\begin{array}
[c]{ll}%
S_{\pi}\left(  x_{1},...,x_{n-1};\lambda_{1},..,\lambda_{n}\right)  &
\begin{array}
[c]{l}%
\text{ if the point }-\lambda_{\pi\left(  n\right)  }\text{ }\\
\text{belongs to a cluster }\\
\text{of }R_{\pi}\left(  \lambda_{1},..,\lambda_{n}\right)  ,
\end{array}
\\
& \\
S_{\pi}\left(  x_{1},...,x_{n-1};\lambda_{1},..,\lambda_{n}\right)  +1 &
\text{ otherwice.}%
\end{array}
\right. \label{011}%
\end{equation}
Indeed, for every $y_{i},$ falling inside $\left[  -\lambda_{\pi\left(
n\right)  },0\right]  $, there is a position

\noindent\ $X_{i}\left(  z_{1},...,z_{n-1},y_{1},...,y_{n-1}\right)  <0,$ such
that once the free rod $\lambda_{\pi\left(  n\right)  }$ is placed there, the
site $y_{i}$ is pushed to the right and hits the origin. In case the point
$-\lambda_{\pi\left(  n\right)  }$ is outside all clusters of $R_{\pi}\left(
\lambda_{1},..,\lambda_{n}\right)  ,$ placing the free rod $\lambda
_{\pi\left(  n\right)  }$ at $X_{0}=-\lambda_{\pi\left(  n\right)  }$ produces
an extra hit.

Now let $\Delta>0$ be such that
\[
\lambda_{1}<\lambda_{2}<...<\lambda_{i-1}<\lambda_{i}-\Delta<\lambda
_{i}+\Delta<\lambda_{i+1}<...<\lambda_{n},
\]
$i=1,...,n,$ and some of the functions $N_{\pi}$ exhibit jumps in $\lambda
_{i}$ as it goes down from $\lambda_{i}+\Delta$ to $\lambda_{i}-\Delta.$ We
denote by $\mathbf{\lambda}\left(  \delta\right)  $ the vector $\lambda
_{1},...,\lambda_{i}+\delta,...,\lambda_{n}.$ We suppose that $\Delta$ is
small enough, so that for any $\pi$ the difference
\[
\left|  N_{\pi}\left(  x_{1},...,x_{n-1};\mathbf{\lambda}\left(
\Delta\right)  \right)  -N_{\pi}\left(  x_{1},...,x_{n-1};\mathbf{\lambda
}\left(  -\Delta\right)  \right)  \right|
\]
is at most one. Moreover, we want $\Delta$ to be so small that on the segment
$\lambda\in\left[  \lambda_{i}-\Delta,\lambda_{i}+\Delta\right]  $ there is
precisely one point, say $\lambda_{i},$ at which some of the functions
$N_{\pi}\left(  x_{1},...,x_{n-1};\mathbf{\lambda}\right)  $ do jump. (In
general, there will be several permutations $\pi,$ for which such a jump will
happen at $\lambda=\lambda_{i}.$ Indeed, if we observe an $\left(
X,x_{k}\right)  $-hit in our rod configuration with $l_{i}=\lambda_{\pi\left(
i\right)  }$, while we have that $x_{1}<x_{2}<...x_{s-1}<X<x_{s}%
<...<x_{k}<...<x_{n-1},$ then in some cases we will have an $\left(
X,x_{k}\right)  $-hit for every rearrangement of the rods $l_{s},...,l_{k},$
i.e. for all permutations of the form $\pi\circ\rho,$ where $\rho$ permutes
the elements $s,...,k,$ leaving the other fixed.)

\textbf{\ }Let us begin with the case when
\begin{equation}
N_{\pi}\left(  x_{1},...,x_{n-1};\mathbf{\lambda}\left(  \Delta\right)
\right)  -N_{\pi}\left(  x_{1},...,x_{n-1};\mathbf{\lambda}\left(
-\Delta\right)  \right)  =1.\label{010}%
\end{equation}
That means that $N_{\pi}\left(  x_{1},...,x_{n-1};\mathbf{\lambda}\left(
\Delta\right)  \right)  \geq1.$ So the intersection $Y\left(  R_{\pi}\left(
\mathbf{\lambda}\left(  \Delta\right)  \right)  \right)  \cap\left[
-\mathbf{\lambda}\left(  \Delta\right)  _{\pi\left(  n\right)  },0\right]
\neq\emptyset.$ Let $y_{k}\left(  \mathbf{\lambda}\left(  \Delta\right)
,\pi\right)  <...<y_{r}\left(  \mathbf{\lambda}\left(  \Delta\right)
,\pi\right)  \ $are all the points of this intersection. The relation $\left(
\ref{010}\right)  $ implies via $\left(  \ref{011}\right)  $ that the point
$y_{k}\left(  \mathbf{\lambda}\left(  \delta\right)  ,\pi\right)  $ leaves the
segment $\left[  -\mathbf{\lambda}\left(  \Delta\right)  _{\pi\left(
n\right)  },0\right]  $ as $\delta$ passes the zero value:%

\begin{equation}
y_{k}\left(  \mathbf{\lambda}\left(  \delta\right)  ,\pi\right)
>-\mathbf{\lambda}\left(  \delta\right)  _{\pi\left(  n\right)  }\text{ for
}\delta>0,\label{012}%
\end{equation}

\begin{equation}
y_{k}\left(  \mathbf{\lambda}\left(  0\right)  ,\pi\right)  =-\mathbf{\lambda
}\left(  0\right)  _{\pi\left(  n\right)  },\label{013}%
\end{equation}

\begin{equation}
y_{k}\left(  \mathbf{\lambda}\left(  \delta\right)  ,\pi\right)
<-\mathbf{\lambda}\left(  \delta\right)  _{\pi\left(  n\right)  }\text{ for
}\delta<0.\label{014}%
\end{equation}

Moreover, the point $y_{k}\left(  \mathbf{\lambda}\left(  \delta\right)
,\pi\right)  $ is not the end of the cluster.

Therefore $y_{k}\left(  \mathbf{\lambda}\left(  \delta\right)  ,\pi\right)
=z_{k+1}\left(  \mathbf{\lambda}\left(  \delta\right)  ,\pi\right)  .$ We now
claim that if we assign the rod $\mathbf{\lambda}\left(  \delta\right)
_{\pi\left(  n\right)  }$ to $x_{k+1},$ and will take for the free rod the rod
$\mathbf{\lambda}\left(  \delta\right)  _{\pi\left(  k+1\right)  }$, then for
the corresponding permutation the opposite to $\left(  \ref{010}\right)  $
happens:
\begin{equation}
N_{\pi\left(  n\leftrightarrow k+1\right)  }\left(  x_{1},...,x_{n-1}%
;\mathbf{\lambda}\left(  \Delta\right)  \right)  -N_{\pi\left(
n\leftrightarrow k+1\right)  }\left(  x_{1},...,x_{n-1};\mathbf{\lambda
}\left(  -\Delta\right)  \right)  =-1.\label{015}%
\end{equation}
Here we denote by $\pi\left(  n\leftrightarrow k+1\right)  $ the permutation
which is the composition of the transposition $n\leftrightarrow k+1,$ followed
by $\pi.$ Indeed, after the above reassignment and the resolution of
conflicts, the rod $\mathbf{\lambda}\left(  \delta\right)  _{\pi\left(
n\right)  }$ will be positioned at the point $y_{k}\left(  \mathbf{\lambda
}\left(  \delta\right)  ,\pi\right)  .$ The relations $\left(  \ref{012}%
\right)  -\left(  \ref{014}\right)  $ then tell us, that during the $\delta
$-evolution the right endpoint of this rod will move from the positive
semiaxis to the negative one, thus adding one unit to the value $S_{\pi\left(
n\leftrightarrow k+1\right)  }\left(  x_{1},...,x_{n-1};\mathbf{\lambda
}\left(  \Delta\right)  \right)  .$

The above construction corresponds to every permutation $\pi,$ satisfying
$\left(  \ref{010}\right)  ,$ another permutation, $\pi^{\prime}=\Phi\left(
\pi\right)  ,$ which satisfy $\left(  \ref{015}\right)  .$ We will be done if
we show that $\Phi$ is one to one. We prove this by constructing the inverse
of $\Phi.$

So let $\pi^{\prime}$ be such that
\[
N_{\pi^{\prime}}\left(  x_{1},...,x_{n-1};\mathbf{\lambda}\left(
\Delta\right)  \right)  -N_{\pi^{\prime}}\left(  x_{1},...,x_{n-1}%
;\mathbf{\lambda}\left(  -\Delta\right)  \right)  =-1.
\]
According to the above that means that the intersection $Y\left(
R_{\pi^{\prime}}\left(  \mathbf{\lambda}\left(  \Delta\right)  \right)
\right)  \cap\left(  0,+\infty\right)  \neq\emptyset.$ Let $y_{k^{\prime}%
}\left(  \mathbf{\lambda}\left(  \Delta\right)  ,\pi^{\prime}\right)
<...<y_{r^{\prime}}\left(  \mathbf{\lambda}\left(  \Delta\right)  ,\pi
^{\prime}\right)  \ $are all the points of this intersection. The relation
$\left(  \ref{010}\right)  $ implies via $\left(  \ref{011}\right)  $ that the
point $y_{k^{\prime}}\left(  \mathbf{\lambda}\left(  \delta\right)
,\pi^{\prime}\right)  $ moves from the positive semiaxis to the negative one
as $\delta$ passes the zero value:
\[
y_{k^{\prime}}\left(  \mathbf{\lambda}\left(  \delta\right)  ,\pi^{\prime
}\right)  >0\text{ for }\delta>0,
\]
\[
y_{k^{\prime}}\left(  \mathbf{\lambda}\left(  0\right)  ,\pi^{\prime}\right)
=0,
\]
\[
y_{k^{\prime}}\left(  \mathbf{\lambda}\left(  \delta\right)  ,\pi^{\prime
}\right)  <0\text{ for }\delta<0.
\]
But that precisely means that the point $y_{k^{\prime}-1}\left(
\mathbf{\lambda}\left(  \delta\right)  ,\pi^{\prime}\right)  $ is inside the
segment $\left[  -\mathbf{\lambda}\left(  \delta\right)  _{\pi^{\prime}\left(
k^{\prime}\right)  },0\right]  $ for $\delta=\Delta,$ and outside it for
$\delta=-\Delta.$ So if we assign the free rod $\mathbf{\lambda}\left(
\delta\right)  _{\pi^{\prime}\left(  n\right)  }$ to the point $x_{k^{\prime}%
},$ making the rod $\mathbf{\lambda}\left(  \delta\right)  _{\pi^{\prime
}\left(  k^{\prime}\right)  }$ free, then we construct the permutation
$\pi^{\prime\prime}=\Phi^{\prime}\left(  \pi^{\prime}\right)  ,$ for which
$\left(  \ref{010}\right)  $ holds.

The statement that $\Phi^{\prime}$ is inverse to $\Phi$ is straightforward.
\end{proof}

Below we will need a version of the above theorem, which follows. Let $T,L$ be
positive reals, $L<T.$ Let again $n$ be an integer, and $\lambda_{1}%
<\lambda_{2}<...<\lambda_{n}$ be a fixed set of positive lengths of rods. Let
$-T<x_{1}<x_{2}<...<x_{n-1}<0$ be a set of $\left(  n-1\right)  $ left-ends.
We want to compute the number $\tilde{N}\left(  -T,x_{1},x_{2},...,x_{n-1}%
;L,\lambda_{1},\lambda_{2},...,\lambda_{n}\right)  ,$ which is defined as
follows. For any permutation $\pi$ of $n$ elements and for any $X\in\left(
\mathbb{-}T,0\right)  ,\,X\neq x_{1},x_{2},...,x_{n-1}$ we can consider the
configuration $\sigma_{n}\left(  -T,x_{1},...,x_{n-1};L,\lambda_{\pi\left(
1\right)  },...,\lambda_{\pi\left(  n-1\right)  }\right)  $ $\cup\sigma
_{1}\left(  X,\lambda_{\pi\left(  n\right)  }\right)  $ of rods, when the rod
$L$ is placed at $-T,$ the rods $l_{i}=\lambda_{\pi\left(  i\right)  }$ are
placed at $x_{i},i=1,...,n-1,$ while the free rod $l_{n}=\lambda_{\pi\left(
n\right)  }$ is placed at $X,$ $-T<X<0.$ Given $\pi,$ we count the number
$\tilde{N}_{\pi}\left(  -T,x_{1},x_{2},...,x_{n-1};L,\lambda_{1},\lambda
_{2},...,\lambda_{n}\right)  $ of different locations $X,$ such that the
corresponding r-configuration

\noindent$R\left[  \sigma_{n}\left(  -T,x_{1},...,x_{n-1};L,\lambda
_{\pi\left(  1\right)  },...,\lambda_{\pi\left(  n-1\right)  }\right)
\cup\sigma_{1}\left(  X,\lambda_{\pi\left(  n\right)  }\right)  \right]  $ has
a hit, and moreover this hit is an $X$-hit. Then we define
\[
\tilde{N}\left(  -T,x_{1},x_{2},...,x_{n-1};L,\lambda_{1},\lambda
_{2},...,\lambda_{n}\right)  =\sum_{\pi\in\mathcal{S}_{n}}\tilde{N}_{\pi
}\left(  -T,x_{1},x_{2},...,x_{n-1};L,\lambda_{1},\lambda_{2},...,\lambda
_{n}\right)  .
\]

\begin{theorem}
\label{R} Suppose that
\begin{equation}
L+\lambda_{1}+\lambda_{2}+...+\lambda_{n}<T.\label{080}%
\end{equation}
Then $\tilde{N}\left(  -T,x_{1},x_{2},...,x_{n-1};L,\lambda_{1},\lambda
_{2},...,\lambda_{n}\right)  =n!$ for almost every $x_{1},...,x_{n-1}$ and
$\lambda_{1},...,\lambda_{n}.$
\end{theorem}

The theorem \ref{R} differs from the theorem \ref{T6} by the presence of the
additional rod $L,$ which is placed at $-T,$ and by the restriction that all
points $X,x_{1},x_{2},...,x_{n-1}$ has to be within the segment $\left(
-T,0\right)  .$ Therefore the rod $L$ will not move under the resolution of
conflicts. Without the restriction (\ref{080}) the statement of the theorem is
not valid, as it is easy to see.

\begin{proof}
Let the numbers $0<\varepsilon_{1}<...<\varepsilon_{n-1}$ be so small that the
sum $\varepsilon_{1}+...+\varepsilon_{n-1}$ is less than any of the numbers
$\left|  \delta_{0}\left(  T-L\right)  +\delta_{1}\lambda_{1}+\delta
_{2}\lambda_{2}+...+\delta_{n}\lambda_{n}\right|  ,$ where $\delta_{i}$ are
taking any of three values $-1,0,1,$ with the only restriction that not all of
them vanish simultaneously. Let us replace the configuration $x_{1}%
,x_{2},...,x_{n-1}$ by the configuration $x_{1}^{\prime},x_{2}^{\prime
},...,x_{n-1}^{\prime},$ where
\[
x_{i}^{\prime}=\left\{
\begin{array}
[c]{ll}%
L-T+\varepsilon_{i} & \text{ if }x_{i}<L-T,\\
x_{i} & \text{ otherwice.}%
\end{array}
\right.
\]
Let $k$ be the largest integer for which $x_{i}^{\prime}>x_{i}.$ (The meaning
of the configuration $x_{1}^{\prime},x_{2}^{\prime},...,x_{n-1}^{\prime}$ is
the following: were all $\varepsilon_{i}$ zeroes, it is the result of
resolving the first conflict, between the first rod $L$ and the rods
intersecting it, which have to be pushed to the right-hand end of $L.$ We use
positive $\varepsilon$-s in order to have all the point $x_{i}^{\prime}$
different.) By the previous theorem we know that $N\left(  x_{1}^{\prime
},x_{2}^{\prime},...,x_{n-1}^{\prime};\lambda_{1},...,\lambda_{n}\right)
=n!\,.$ Let the location $X$ be such that for some permutation $\pi$ the
corresponding r-configuration

\noindent$R\left[  \sigma_{n-1}\left(  x_{1}^{\prime},...,x_{n-1}^{\prime
};\lambda_{\pi\left(  1\right)  },...,\lambda_{\pi\left(  n-1\right)
}\right)  \cup\sigma_{1}\left(  X,\lambda_{\pi\left(  n\right)  }\right)
\right]  $ has an $X$-hit. The condition (\ref{080}) implies that the cluster
of the r-configuration

\noindent$R\left[  \sigma_{n-1}\left(  x_{1}^{\prime},...,x_{n-1}^{\prime
};\lambda_{\pi\left(  1\right)  },...,\lambda_{\pi\left(  n-1\right)
}\right)  \cup\sigma_{1}\left(  X,\lambda_{\pi\left(  n\right)  }\right)
\right]  ,$ rooted at $X,$ does not contain any of the points $z_{1}^{\prime
}=x_{1}^{\prime},z_{2}^{\prime},...,z_{k}^{\prime}$ (see (\ref{07}) for the
notation), so $X>L-T,$ and the r-configuration

\noindent$R\left[  \sigma_{n}\left(  -T,x_{1},...,x_{n-1};L,\lambda
_{\pi\left(  1\right)  },...,\lambda_{\pi\left(  n-1\right)  }\right)
\cup\sigma_{1}\left(  X,\lambda_{\pi\left(  n\right)  }\right)  \right]  $ has
an $X$-hit as well. Therefore

\noindent$\tilde{N}\left(  -T,x_{1},x_{2},...,x_{n-1};L,\lambda_{1}%
,\lambda_{2},...,\lambda_{n}\right)  \geq n!\,.$ On the other hand, if the r-configuration

\noindent$R\left[  \sigma_{n}\left(  -T,x_{1},...,x_{n-1};L,\lambda
_{\pi\left(  1\right)  },...,\lambda_{\pi\left(  n-1\right)  }\right)
\cup\sigma_{1}\left(  X,\lambda_{\pi\left(  n\right)  }\right)  \right]  $ has
an $X$-hit, then by the same reasoning $X$ has to be to the right of the
location $L-T,$ and moreover the cluster of this configuration, rooted at $X,$
does not contain any of the points $z_{1}=-T,z_{2}=-T+L,...,z_{k+1};$
therefore the r-configuration

\noindent$R\left[  \sigma_{n-1}\left(  x_{1}^{\prime},...,x_{n-1}^{\prime
};\lambda_{\pi\left(  1\right)  },...,\lambda_{\pi\left(  n-1\right)
}\right)  \cup\sigma_{1}\left(  X,\lambda_{\pi\left(  n\right)  }\right)
\right]  $ has an $X$-hit. Hence

\noindent$\tilde{N}\left(  -T,x_{1},x_{2},...,x_{n-1};L,\lambda_{1}%
,\lambda_{2},...,\lambda_{n}\right)  \leq n!\,,$ and the proof follows.
\end{proof}

\section{Estimates on dissipators}

For the future use we have to estimate the densities $q_{\lambda,x}\left(
t\right)  ,$ entering into the relation $b\left(  x\right)  =\left[
\lambda\ast q_{\lambda,x}\right]  \left(  x\right)  .$

\begin{lemma}%
\begin{equation}
q_{\lambda,y}\left(  t\right)  \geq p\left(  t\right)  \mathbf{\Pr}\left\{
\text{server is idle at the moment }y-t\right\}  .\label{301}%
\end{equation}

\end{lemma}

\begin{proof}
We will obtain this estimate by invoking the initial relation $\left(
\ref{129}\right)  $ for $b:$%
\begin{align*}
& b\left(  y\right)  =\exp\left\{  -I_{\lambda}\left(  y\right)  \right\}
\sum_{n=1}^{\infty}\frac{1}{\left(  n-1\right)  !}\underset{n}{\underbrace
{\int_{0}^{\infty}...\int_{0}^{\infty}}}\\
& \left[  \underset{n-1}{\underbrace{\int_{0}^{y}...\int_{0}^{y}}}%
\lambda\left(  X\left(  y\Bigm|x_{1},...,x_{n-1};l_{1},...,l_{n}\right)
\right)  \prod_{i=1}^{n-1}\lambda\left(  x_{i}\right)  \,dx_{i}\right]
\prod_{i=1}^{n}p\left(  l_{i}\right)  \,dl_{i}%
\end{align*}
Namely, the contribution to the value $q_{\lambda,y}\left(  t\right)  $ comes
from all realizations $\left(  x_{1},...,x_{n-1};l_{1},...,l_{n}\right)  ,$
for which $y-t\in X\left(  y\Bigm|x_{1},...,x_{n-1};l_{1},...,l_{n}\right)  .$
Among such realizations let us pick the following class: the rod $l_{n}=t,$
while the rods of the configuration $R\sigma_{n}\left(  x_{1},...,x_{n-1}%
;l_{1},...,l_{n-1}\right)  $ does not cover the point $y-t.$ Let us denote the
indicator of the complement to the union of rods forming the set $R\sigma
_{n}\left(  x_{1},...,x_{n-1};l_{1},...,l_{n-1}\right)  $ by $\mathbf{I}%
_{\mathbf{x,l}}.$ Then we have
\begin{align*}
& q_{\lambda,y}\left(  t\right)  \geq p\left(  t\right)  \exp\left\{
-I_{\lambda}\left(  y\right)  \right\}  \sum_{n=1}^{\infty}\frac{1}{\left(
n-1\right)  !}\\
& \times\underset{n-1}{\underbrace{\int_{0}^{\infty}...\int_{0}^{\infty}}%
}\left[  \underset{n-1}{\underbrace{\int_{0}^{y}...\int_{0}^{y}}}%
\mathbf{I}_{\mathbf{x,l}}\left(  y-t\right)  \prod_{i=1}^{n-1}\lambda\left(
x_{i}\right)  \,dx_{i}\right]  \prod_{i=1}^{n-1}p\left(  l_{i}\right)
\,dl_{i}\\
& =p\left(  t\right)  \mathbf{\Pr}\left\{  \text{server is idle at the moment
}y-t\right\}  .
\end{align*}

\end{proof}

Next we establish the upper bound on $q_{\lambda,y}.$

\begin{lemma}
\label{oc}
\begin{equation}
q_{\lambda,y}\left(  t\right)  \leq\sum_{n=1}^{\infty}p^{\ast n}\left(
t\right)  \mathbf{\Pr}\left\{  N_{t}^{\lambda,y}\geq n-1\right\}  ,\label{152}%
\end{equation}
where $N_{t}^{\lambda,y}$ is the random number of $\lambda$-Poisson points in
the segment $\left[  y-t,y\right]  .$

In particular, for $t\leq C$ there exists a constant $\tilde{C}=\tilde
{C}\left(  C,p\right)  ,$ such that for all $\lambda,y$%
\begin{equation}
q_{\lambda,y}\left(  t\right)  \leq\tilde{C}.\label{157}%
\end{equation}

\end{lemma}

\begin{proof}
As above, we have
\begin{align*}
&  q_{\lambda,y}\left(  t\right)  =\exp\left\{  -I_{\lambda}\left(  y\right)
\right\}  \sum_{n=1}^{\infty}\underset{n-1}{\underbrace{\int_{0}^{\infty
}...\int_{0}^{\infty}}}\\
&  \left[  \underset{n-1}{\underbrace{\int_{0}^{y}\int_{x_{1}}^{y}%
...\int_{x_{n-2}}^{y}}}\mathbf{I}_{\mathbf{x,l}}\left(  y-t\right)  \left(
\sum_{j}p\left(  \beta_{j}\right)  \right)  \prod_{i=1}^{n-1}\lambda\left(
x_{i}\right)  \,dx_{i}\right]  \prod_{i=1}^{n-1}p\left(  l_{i}\right)
\,dl_{i},
\end{align*}
where the values $\beta_{j}$ are all solutions $\beta$ of the equation
\begin{equation}
X\left(  y\Bigm|x_{1},...,x_{n-1};l_{1},...,l_{n-1},\beta\right)
=y-t,\label{120}%
\end{equation}
and where we integrate only over the set $0<x_{1}<...<x_{n-1}<y,$ so we do not
have the factorials any more. Note that the equation $\left(  \ref{120}%
\right)  $ has solutions only if the configuration $\left(  x_{1}%
,...,x_{n-1};l_{1},...,l_{n-1}\right)  $ satisfies the condition
$\mathbf{I}_{\mathbf{x,l}}\left(  y-t\right)  =1$; in that case the solutions
do not depend on all these $x_{i},l_{i},$ for which $x_{i}<y-t.$ Let us define
the values $\beta_{j}$ outside the event $\mathbf{I}_{\mathbf{x,l}}\left(
y-t\right)  =1$ as the solutions of
\[
X\left(  y\Bigm|x_{1},...,x_{n-1};\tilde{l}_{1},...,\tilde{l}_{n-1}%
,\beta\right)  =y-t,
\]
where
\[
\tilde{l}_{i}=\left\{
\begin{array}
[c]{ll}%
0 & \text{ if }x_{i}<y-t,\\
l_{i} & \text{ otherwice.}%
\end{array}
\right.
\]
Replacing the indicator by $1,$ we have
\begin{align}
&  q_{\lambda,y}\left(  t\right)  \leq\exp\left\{  -I_{\lambda}\left(
y\right)  \right\}  \sum_{n=1}^{\infty}\nonumber\\
&  \underset{n-1}{\underbrace{\int_{0}^{\infty}...\int_{0}^{\infty}}}\left[
\underset{n-1}{\underbrace{\int_{x_{1}=0}^{y}\int_{x_{2}}^{y}...\int_{x_{n-2}%
}^{y}}}\left(  \sum_{j}p\left(  \beta_{j}\right)  \right)  \prod_{i=1}%
^{n-1}\lambda\left(  x_{i}\right)  \,dx_{i}\right]  \prod_{i=1}^{n-1}p\left(
l_{i}\right)  \,dl_{i}\nonumber\\
&  =\exp\left\{  -I_{\lambda}\left(  y\right)  +I_{\lambda}\left(  y-t\right)
\right\}  \sum_{n=1}^{\infty}\label{121}\\
&  \underset{n-1}{\underbrace{\int_{0}^{\infty}...\int_{0}^{\infty}}}\left[
\underset{n-1}{\underbrace{\int_{y-t}^{y}\int_{x_{2}}^{y}...\int_{x_{n-2}}%
^{y}}}\left(  \sum_{j}p\left(  \beta_{j}\right)  \right)  \prod_{i=1}%
^{n-1}\lambda\left(  x_{i}\right)  \,dx_{i}\right]  \prod_{i=1}^{n-1}p\left(
l_{i}\right)  \,dl_{i},\nonumber
\end{align}
where the $x$-integration is now taken over $y-t<x_{1}<...<x_{n-1}<y.$ Let us
enumerate $\beta_{j}$ in decreasing order. Then clearly $\beta_{1}=t,$ and the
corresponding term in $\left(  \ref{121}\right)  $ equals $p\left(  t\right)
.$ The next solution, $\beta_{2},$ of the equation $\left(  \ref{120}\right)
,$ exists only if $l_{1}<t$ and $x_{1}<y-l_{1}.$ Then $\beta_{2}=t-l_{1}.$ The
second term is therefore
\begin{align*}
&  \exp\left\{  -I_{\lambda}\left(  y\right)  +I_{\lambda}\left(  y-t\right)
\right\}  \sum_{n=2}^{\infty}\underset{n-2}{\underbrace{\int_{0}^{\infty
}...\int_{0}^{\infty}}}\\
&  \left\{  \underset{n-2}{\underbrace{\int_{x_{2}}^{y}...\int_{x_{n-2}}^{y}}%
}\left[  \int_{0}^{\infty}\left(  \int_{y-t}^{y-l_{1}}\lambda\left(
x_{1}\right)  \,dx_{1}\right)  p\left(  t-l_{1}\right)  p\left(  l_{1}\right)
dl_{1}\right]  \prod_{i=2}^{n-1}\lambda\left(  x_{i}\right)  \,dx_{i}\right\}
\prod_{i=2}^{n-1}p\left(  l_{i}\right)  \,dl_{i}\\
&  \leq\exp\left\{  -I_{\lambda}\left(  y\right)  +I_{\lambda}\left(
y-t\right)  \right\}  \sum_{n=2}^{\infty}\underset{n-2}{\underbrace{\int
_{0}^{\infty}...\int_{0}^{\infty}}}\\
&  \left\{  \underset{n-2}{\underbrace{\int_{x_{2}}^{y}...\int_{x_{n-2}}^{y}}%
}\left[  \int_{0}^{\infty}\left(  \int_{y-t}^{y}\lambda\left(  x_{1}\right)
\,dx_{1}\right)  p\left(  y-t-l_{1}\right)  p\left(  l_{1}\right)
dl_{1}\right]  \prod_{i=2}^{n-1}\lambda\left(  x_{i}\right)  \,dx_{i}\right\}
\prod_{i=2}^{n-1}p\left(  l_{i}\right)  \,dl_{i}\\
&  =p^{\ast2}\left(  t\right)  \mathbf{\Pr}\left\{  \text{the }\lambda
\text{-Poisson field has at least }1\text{ point in the segment }\left[
y-t,y\right]  \right\}  .
\end{align*}
So by induction we arrive to the bound $\left(  \ref{152}\right)  $:
\[
q_{\lambda,y}\left(  t\right)  \leq\sum_{n=1}^{\infty}p^{\ast n}\left(
t\right)  \mathbf{\Pr}\left\{  N_{t}^{\lambda,y}\geq n-1\right\}
\equiv\mathcal{Q}_{\lambda,y}\left(  t\right)  ,
\]
where $N_{t}^{\lambda,y}$ is the random number of $\lambda$-Poisson points in
the segment $\left[  y-t,y\right]  .$

To see $\left(  \ref{157}\right)  ,$ we use a rough form of $\left(
\ref{152}\right)  :$%
\begin{equation}
q_{\lambda,y}\left(  t\right)  \leq\sum_{n=1}^{\infty}p^{\ast n}\left(
t\right)  .\label{8158}%
\end{equation}
Let $A=\sup_{t}p\left(  t\right)  .$ Then it is immediate from $\left(
\ref{8158}\right)  $ that for all $t\leq C$%
\[
q_{\lambda,y}\left(  t\right)  \leq A\left(  1+\sum_{n=1}^{\infty}\mathbf{\Pr
}\left\{  \eta_{1}+...+\eta_{n}\leq C\right\}  \right)  ,
\]
where $\eta_{i}$ are i.i.d. random variables, distributed as $\eta.$ But the
probabilities $\mathbf{\Pr}\left\{  \eta_{1}+...+\eta_{n}\leq C\right\}  $
decay exponentially in $n.$
\end{proof}

We will need the compactness estimate on the distributions $q_{\lambda
,y}\left(  t\right)  .$ We will obtain them using the estimate $\left(
\ref{152}\right)  .$ As the following statement show, the estimate $\left(
\ref{152}\right)  $ is rather rough; we believe that all the moments of the
distribution $q_{\lambda,y}\left(  t\right)  $ of order less than $1+\delta$
are finite.

\begin{lemma}
\label{Qu} Suppose that $\lambda$ is such that for some $T^{\prime}$ and
$\varepsilon^{\prime}>0$ and for all $T\geq T^{\prime}$
\[
\int_{0}^{T}\lambda_{\nu}\left(  t\right)  \,dt<T\left(  1-\varepsilon
^{\prime}\right)
\]
(see $\left(  \ref{01}\right)  $). Then for any $b<\frac{\delta}{2}$
\begin{equation}
\int_{0}^{\infty}t^{b}q_{\lambda,y}\left(  t\right)  \,dt<C\left(
\lambda,b\right)  <\infty,\label{303}%
\end{equation}
where $C\left(  \lambda,b\right)  $ depends on $\lambda$ only via $T^{\prime}$
and $\varepsilon^{\prime}.$
\end{lemma}

\textbf{Proof of Lemma \ref{Qu}. }We are going to use the simple estimate: for
every random variable $\zeta$ and every $\varkappa>0$
\begin{equation}
\tilde{Q}\left(  T\right)  \equiv\mathbf{\Pr}\left\{  \zeta>T\right\}  \leq
T^{-\varkappa}\mathbb{E}\left(  \left|  \zeta\right|  ^{\varkappa}\right)
.\label{151}%
\end{equation}

We also will need an estimate on $\int_{A}^{\infty}t^{a}\tilde{q}\left(
t\right)  \,dt,$ $a<\varkappa,$ where $\tilde{q}$ is the density of $\zeta.$
We have:
\begin{align*}
\int_{A}^{\infty}t^{a}\tilde{q}\left(  t\right)  \,dt  & =-\int_{A}^{\infty
}t^{a}\,d\left(  \tilde{Q}\left(  t\right)  \right) \\
& =A^{a}\tilde{Q}\left(  A\right)  +a\int_{A}^{\infty}t^{a-1}\tilde{Q}\left(
t\right)  \,dt.
\end{align*}
To apply $\left(  \ref{151}\right)  $ to $\left(  \ref{152}\right)  $ we will
use the Dharmadhikari-Yogdeo estimate (see, e.g. \cite{P}, p.79): if $\xi_{i}$
are independent centered random variables, then
\[
\mathbb{E}\left(  \left|  \xi_{1}+...+\xi_{n}\right|  ^{2+\delta}\right)  \leq
Rn^{\delta/2}\sum_{1}^{n}\mathbb{E}\left(  \left|  \xi_{i}\right|  ^{2+\delta
}\right)  .
\]
Here $R=R\left(  \delta\right)  $ is some universal constant. Introducing
$\xi_{i}=\eta_{i}-1$ (see $\left(  \ref{153}\right)  $), we have (see $\left(
\ref{154}\right)  $)
\begin{align}
Q_{n}\left(  t\right)   & \equiv\mathbf{\Pr}\left\{  \eta_{1}+...+\eta
_{n}>t\right\}  =\mathbf{\Pr}\left\{  \xi_{1}+...+\xi_{n}>t-n\right\}
\label{156}\\
& \leq RM_{\delta}\left(  t-n\right)  ^{-\left(  2+\delta\right)  }%
n^{1+\delta/2}.\nonumber
\end{align}
To proceed, we use $\left(  \ref{152}\right)  $ to write
\begin{align}
\int_{0}^{\infty}t^{b}q_{\lambda,y}\left(  t\right)  \,dt  & \leq\int
_{0}^{\infty}t^{b}\mathcal{Q}_{\lambda,y}\left(  t\right)  \,dt\label{155}\\
& =\sum_{n=1}^{\infty}\left[  \int_{0}^{\infty}t^{b}p^{\ast n}\left(
t\right)  \mathbf{\Pr}\left\{  N_{t}^{\lambda,y}\geq n-1\right\}  \,dt\right]
.\nonumber
\end{align}

Note that $\mathbb{E}\left(  N_{t}^{\lambda,y}\right)  \leq\left(
1-\alpha\right)  t$ once $t$ is large enough. The first step is to write
\begin{align}
& \int_{0}^{\infty}t^{b}p^{\ast n}\left(  t\right)  \mathbf{\Pr}\left\{
N_{t}^{\lambda,y}\geq n-1\right\}  \,dt\label{300}\\
& \leq\int_{0}^{n\left(  1+\frac{\alpha}{2}\right)  }t^{b}p^{\ast n}\left(
t\right)  \mathbf{\Pr}\left\{  N_{t}^{\lambda,y}\geq n-1\right\}
\,dt+\int_{n\left(  1+\frac{\alpha}{2}\right)  }^{\infty}t^{b}p^{\ast
n}\left(  t\right)  \,dt.\nonumber
\end{align}
Now, using $\left(  \ref{151}\right)  $ and $\left(  \ref{156}\right)  ,$ we
have
\begin{align*}
\int_{n\left(  1+\frac{\alpha}{2}\right)  }^{\infty}t^{b}p^{\ast n}\left(
t\right)  \,dt  & \leq\left[  n\left(  1+\frac{\alpha}{2}\right)  \right]
^{b}RM_{\delta}\left(  \frac{\alpha}{2}n\right)  ^{-\left(  2+\delta\right)
}n^{1+\delta/2}\\
& +bRM_{\delta}n^{1+\delta/2}\int_{n\left(  1+\frac{\alpha}{2}\right)
}^{\infty}t^{b-1}\left(  t-n\right)  ^{-\left(  2+\delta\right)  }\,dt\\
& \leq Cn^{b-1-\delta/2},
\end{align*}
where $C=C\left(  \alpha,\delta,M_{\delta}\right)  .$

The first term in $\left(  \ref{300}\right)  $ is negligible. To see that, we
first observe:

\begin{lemma}
Let $0<\nu<1,$ and $N^{\nu}$ be a Poisson random variable:
\[
\mathbf{\Pr}\left\{  N^{\nu}=k\right\}  =e^{-\nu n}\frac{\left(  \nu n\right)
^{k}}{k!}.
\]
Then
\[
\mathbf{\Pr}\left\{  N^{\nu}\geq n\right\}  \leq\frac{1}{1-\nu}e^{-\frac
{\left(  1-\nu\right)  ^{2}}{2}n},
\]
provided $n$ is large enough.
\end{lemma}

\begin{proof}
Note first of all, that if $\chi>0$ and $n>\chi,$ then
\[
e^{-\chi}\sum_{k\geq n}\frac{\chi^{k}}{k!}\leq e^{-\chi}\frac{\chi^{n}}%
{n!}\sum_{k\geq0}\left(  \frac{\chi}{n+1}\right)  ^{k}=e^{-\chi}\frac{\chi
^{n}}{n!}\frac{1}{1-\frac{\chi}{n+1}}.
\]
In our case we thus have
\[
\sum_{k\geq n}\mathbf{\Pr}\left\{  N^{\nu}=k\right\}  \leq e^{-\nu n}%
\frac{\left(  \nu n\right)  ^{n}}{n!}\frac{1}{1-\nu}.
\]
By Stirling, for $n$ large
\begin{align*}
\sum_{k\geq n}\mathbf{\Pr}\left\{  N^{\nu}=k\right\}   & \leq\frac{1}{1-\nu
}e^{-\nu n}\frac{\nu^{n}n^{n}}{n^{n}e^{-n}}\\
& =\frac{1}{1-\nu}e^{\left(  1-\nu+\ln\nu\right)  n}\\
& \leq\frac{1}{1-\nu}e^{-\frac{\left(  1-\nu\right)  ^{2}}{2}n}.
\end{align*}

\end{proof}

Therefore,
\begin{align*}
& \int_{0}^{n\left(  1+\frac{\alpha}{2}\right)  }t^{b}p^{\ast n}\left(
t\right)  \mathbf{\Pr}\left\{  N_{t}^{\lambda,y}\geq n-1\right\}  \,dt\\
& \leq\frac{1}{\alpha}e^{-\frac{\alpha^{2}}{2}n}\int_{0}^{n\left(
1+\frac{\alpha}{2}\right)  }t^{b}p^{\ast n}\left(  t\right)  \,dt\\
& \leq\frac{1}{\alpha}e^{-\frac{\alpha^{2}}{2}n}\left[  n\left(
1+\frac{\alpha}{2}\right)  \right]  ^{b}.
\end{align*}
Hence, the moment $\int_{0}^{\infty}t^{b}q_{\lambda,y}\left(  t\right)  \,dt$
is finite as soon as the series $\sum_{n}n^{b-1-\delta/2}$ converges, which
happens when $b<\frac{\delta}{2}.$ That proves Lemma \ref{Qu}. $\blacksquare$

\section{The self-averaging relation: general case}

Here we derive a formula, expressing the function $b\left(  \cdot\right)
=A\left(  \mathbf{\mu},\lambda\left(  \cdot\right)  \right)  $ in terms of the
functions $\lambda\left(  \cdot\right)  $, $p\left(  \cdot\right)  $ and the
measure $\mathbf{\mu.}$ This will be the needed self-averaging relation
(\ref{134}). We remind the reader that $\mathbf{\mu}$ is a probability measure
on the set of pairs $\left\{  \left(  n,\tau\right)  \right\}  \cup
\mathbf{0.}$

\begin{theorem}
Let $N\left(  \mu\right)  =q,$ and the rate function $\lambda\left(
\cdot\right)  $ satisfies the conclusions of the Lemma \ref{la} and the
relation $\left(  \ref{01}\right)  :$%
\[
\int_{0}^{T}\lambda_{\nu}\left(  t\right)  \,dt<T\left(  1-\varepsilon
^{\prime}\right)  \text{ }\,\text{for all }T\geq T^{\prime}>0.
\]
Then there exists the family of probability densities $q_{\lambda,\mu
,x}\left(  \cdot\right)  ,$ $x>0,$ and the functionals $\varepsilon
_{\lambda,\mu}\left(  x\right)  $ and $Q_{\lambda,\mu}\left(  x\right)  ,$
such that
\begin{equation}
b\left(  x\right)  =\left(  1-\varepsilon_{\lambda,\mu}\left(  x\right)
\right)  \left[  \lambda\ast q_{\lambda,\mu,x}\right]  \left(  x\right)
+\varepsilon_{\lambda,\mu}\left(  x\right)  Q_{\lambda,\mu}\left(  x\right)
.\label{103}%
\end{equation}

Moreover,
\begin{equation}
\varepsilon_{\lambda,\mu}\left(  x\right)  \rightarrow0\text{ as }%
x\rightarrow\infty,\label{102}%
\end{equation}
while $Q_{\lambda,\mu}\left(  x\right)  \leq C,$ uniformly in $\lambda,\mu$
and $x,$ once $q,T^{\prime}$ and $\varepsilon^{\prime}$ are fixed.
\end{theorem}

\begin{proof}
We start by defining the functional $\varepsilon_{\lambda,\mu}\left(
x\right)  .$ Note that the description of the realization of our process up to
the moment $x$ consists of the following data:

$i)$ the initial configuration $\left(  n,\tau\right)  ,$ drawn from the
distribution $\mu;$

$ii)$ the random set $0<x_{1}<...<x_{m}<x$ (with random number $m$ of points),
which is a realization of the Poisson random field defined by the rate
function $\lambda$ (restricted to the segment $\left[  0,x\right]  $),
independent of $\left(  n,\tau\right)  $;

$iii)$ one realization $\eta_{1}$ of the conditional random variable $\left(
\eta-\tau\Bigm|\eta>\tau\right)  $ and $n+m-1$ independent realizations
$\eta_{k},k=2,...,n+m$ of the random variable $\eta.$ We denote by
$\mathbb{P}_{\mu\otimes\lambda\otimes\eta}$ the corresponding (product)
distribution. The difference $1-\varepsilon_{\lambda,\mu}\left(  x\right)  $
is then just the $\mathbb{P}_{\mu\otimes\lambda\otimes\eta}$-probability of
the event
\begin{equation}
\sum_{1}^{n+m}\eta_{k}<x.\label{140}%
\end{equation}
(If $n=0,$ then by definition we put $\tau=0;$ we put also $\sum_{1}^{0}%
\equiv0.$)

The meaning of the decomposition (\ref{103}) can be explained now: the first
term corresponds to the exit flow computed over those realizations where the
relation (\ref{140}) holds, while the second term represents the rest of the flow.

Let us show $\left(  \ref{102}\right)  ,$ that is that
\[
\mathbf{\Pr}\left\{  \sum_{1}^{n+m}\eta_{k}>x\right\}  \rightarrow0\,\text{as
}x\rightarrow\infty.
\]
To do this, we introduce two independent random variables:
\[
S_{\mu}=\sum_{1}^{n}\eta_{k},\;S_{\lambda}=\sum_{n+1}^{n+m}\eta_{k}.
\]
Then for every $\alpha\in\left(  0,1\right)  $ we have
\begin{align*}
\mathbf{\Pr}\left\{  \sum_{1}^{n+m}\eta_{k}>x\right\}   & =\mathbf{\Pr
}\left\{  S_{\mu}+S_{\lambda}>x\right\} \\
& \leq\mathbf{\Pr}\left\{  S_{\mu}>\alpha x\right\}  +\mathbf{\Pr}\left\{
S_{\lambda}>\left(  1-\alpha\right)  x\right\}  .
\end{align*}
Indeed, if $S_{\mu}+S_{\lambda}>x,$ then either $S_{\mu}>\alpha x,$ or else
$S_{\lambda}>\left(  1-\alpha\right)  x.$ Since $\mathbb{E}\left(  S_{\mu
}\right)  \leq\bar{C}+q,$ the probability $\mathbf{\Pr}\left\{  S_{\mu}>\alpha
x\right\}  $ goes to zero for every $\alpha$ positive, as $x\rightarrow
\infty,$ uniformly in $\mu.$ For the second term we have
\begin{align*}
& \mathbf{\Pr}\left\{  S_{\lambda}>\left(  1-\alpha\right)  x\right\} \\
& =\sum_{m=1}^{\infty}\left(  \int_{\left(  1-\alpha\right)  x}^{\infty
}p^{\ast m}\left(  t\right)  \,dt\right)  \mathbf{\Pr}\left\{  N^{\lambda
,x}=m\right\}  .
\end{align*}
Here $N^{\lambda,x}$ is the random number of points in the realization of the
Poisson field in $\left[  0,x\right]  .$ Note that $\mathbb{E}\left(
N^{\lambda,x}\right)  <x\left(  1-\varepsilon^{\prime}\right)  $ once
$x>T^{\prime}.$ Therefore we can apply the same argument which was used in the
proof of Lemma \ref{Qu} when showing that the integral $\int_{T}^{\infty
}\mathcal{Q}_{\lambda,y}\left(  t\right)  \rightarrow0$ as $T\rightarrow
\infty,$ see $\left(  \ref{152}\right)  .$ It implies that $\mathbf{\Pr
}\left\{  S_{\lambda}>\left(  1-\alpha\right)  x\right\}  \rightarrow0$ once
$\alpha$ is small enough, uniformly in $\lambda$.

Next we define the distributions $q_{\lambda,\mu,x}.$ They are constructed
from the random field of the rods $\left\{  \eta_{k},k=1,...,n+m\right\}  ,$
defined above, placed at locations $\left\{  \underset{n}{\underbrace
{0,...,0}},x_{1},...,x_{m}\right\}  ,$ via the procedure of resolution of
conflicts, defined in the previous section. To do it we first introduce the
rate $b_{L}\left(  x\right)  $ to be the exit rate of the conditional service
process under the conditions that
\begin{equation}
\sum_{1}^{n}\eta_{k}=L,\;\;\sum_{n+1}^{n+m}\eta_{k}<x-L.\label{141}%
\end{equation}
We claim that for some probability distributions $q_{\lambda,L,x}$ we have
\[
b_{L}\left(  x\right)  =\left[  \lambda\ast q_{\lambda,L,x}\right]  \left(
x\right)  .
\]
The distribution $q_{\lambda,\mu,x}$ is then obtained by integration:
\[
q_{\lambda,\mu,x}=\int q_{\lambda,L,x}\mathbb{P}_{\mu\otimes\lambda\otimes
\eta}\left(  \sum_{1}^{n}\eta_{k}\in dL\right)  .
\]
(The random variable $\sum_{1}^{n}\eta_{k}$ is of course independent of the
Poisson $\lambda$-field.) The output rate $b_{L}\left(  x\right)  $
corresponds to the situation when we have customers arriving at the moments
$0,x_{1},...,x_{m},$ which have serving times $L,\eta_{n+1},...,\eta_{n+m},$
and which satisfy the relation
\[
L+\sum_{n+1}^{n+m}\eta_{k}<x.
\]
So we have to repeat the construction of the Section 5 in the present
situation. Few steps require some comments. The transition from the relation
(\ref{20}) to (\ref{23}) uses the fact that for any $s$ the measure
$\prod_{i=1}^{s}p\left(  l_{i}\right)  \,dl_{i}$ is invariant under the
coordinate permutations $S_{s}$ in $\mathbb{R}^{s}.$ But the same $S_{m}$
symmetry evidently holds for the conditional distribution of the random vector
$\left\{  \left(  \eta_{k},k=n+1,...,n+m\right)  \Bigm|\sum_{n+1}^{n+m}%
\eta_{k}<x-L\right\}  ,$ since both the unconditional distribution and the
distribution of the condition are $S_{m}$-invariant. The next crucial step was
the relation (\ref{26}), stating that the functions $q_{\lambda,y}$ are
probability distributions. It was based on the theorem \ref{T6}. The situation
at hand is somewhat more delicate, since the rods we are dealing now with, are
of two kinds: the first one has a non-random length $L,$ produced by the
initial state $\mu,$ while others are situated at the Poissonian locations
$\left\{  x_{i}\right\}  ,$ defined by the rate function $\lambda.$ However,
under condition $\sum_{n+1}^{n+m}\eta_{k}<x-L$ the needed combinatorial
statement (about $m!$) still holds, and is the content of the theorem \ref{R}.
These remarks allow one to carry over the construction of the previous
section, and so to establish the existence of the probability densities
$q_{\lambda,L,x},$ and thus also $q_{\lambda,\mu,x}.$ The upper and lower
estimates on $q_{\lambda,\mu,x}$ are obtained in the same way as were the
estimates for $q_{\lambda,x}$ in the preceding section.

The function $Q_{\lambda,\mu}\left(  x\right)  $ is the rate of exit flow of
our process, conditioned by the event
\[
\sum_{1}^{n+m}\eta_{k}\geq x.
\]
The boundedness of the $Q_{\lambda,\mu}\left(  x\right)  $ follows from the
following property of the service time distribution $p\left(  x\right)  $: for
every $x,\tau,\,x>\tau>0,$ $1>t>0$
\begin{equation}
\frac{p\left(  x\right)  }{p\left(  x+t\right)  }\leq C^{\prime}%
,\;\;\frac{p\left(  x-\tau\Bigm|\eta>\tau\right)  }{p\left(  x-\tau
+t\Bigm|\eta>\tau\right)  }\leq C^{\prime}.\label{104}%
\end{equation}
The relation (\ref{104}) follows easily from the condition $\left(
\ref{02}\right)  .$ To explain the boundedness, consider the elementary event
\[
\left(  n,\tau\right)  \times\left\{  x_{1},...,x_{m}:0<x_{1}<...<x_{m}%
<x\right\}  \times\left\{  \eta_{1},...,\eta_{n+m}\right\}  ,
\]
which contributes to the output flow inside the segment $\left[  x,x+\Delta
x\right]  ,$ which flow is accounted by the second term of (\ref{103}). That
means that our rod configuration produces after resolution of conflicts a hit
inside $\left[  x,x+\Delta x\right]  ,$ and also that
\begin{equation}
\sum_{1}^{n+m}\eta_{k}>x.\label{105}%
\end{equation}
In the notation of the Section 6 it means that after resolution of conflicts
the endpoint $y_{k}$ of some (shifted) rod fits within $\left[  x,x+\Delta
x\right]  ,$ for some $k\in\left\{  1,...,n+m\right\}  .$ Let $\bar{k}$ be the
smallest such index. But then the elementary events
\[
\left(  n,\tau\right)  \times\left\{  x_{1},...,x_{m}:0<x_{1}<...<x_{m}%
<x\right\}  \times\left\{  \eta_{1},...,\eta_{\bar{k}-1},\eta_{\bar{k}}%
+t,\eta_{\bar{k}+1},...,\eta_{n+m}\right\}  ,
\]
with any $t\in\left(  \Delta x,1\right)  ,$ do not contribute to the output
flow inside the segment $\left[  x,x+\Delta x\right]  ,$ while still
satisfying (\ref{105}). Therefore, due to (\ref{104}), the probability that
the customer would finish his service during the period $\left[  x,x+\Delta
x\right]  ,$ is of the order of $\Delta x,$ and, moreover,
\[
Q_{\lambda,\mu}\left(  x\right)  \leq\frac{1}{C^{\prime}}\text{.}%
\]

\end{proof}

Let now $\mathfrak{M\in}\mathcal{M}\left(  \mathcal{M}_{q}\left(
\Omega\right)  \right)  $ be some invariant measure of the dynamical system
$\left(  \ref{160}\right)  .$ Then $\mathfrak{M}$-almost every state
$\tilde{\mu}_{0}\in\mathcal{M}_{q}\left(  \Omega\right)  $ belongs to the
family $\left\{  \tilde{\mu}_{t}:-\infty<t<+\infty\right\}  ,$ such that for
all $\tau>0,$ all $t$
\[
\mathcal{T}_{\tau}\left(  \tilde{\mu}_{t}\right)  =\tilde{\mu}_{t+\tau}.
\]
Let us fix one such family $\left\{  \tilde{\mu}_{t}\right\}  $. Then the
function $\lambda\left(  t\right)  $, $-\infty<t<+\infty,$ which for every
$-\infty<\tau<+\infty$ satisfies on $[\tau,+\infty)$ the equation
\[
\lambda\left(  \cdot\right)  =A\left(  \tilde{\mu}_{\tau},\lambda\left(
\cdot\right)  ,\tau\right)  ,
\]
is well defined. Then, according to the equation $\left(  \ref{103}\right)  ,$
for every $\tau,$ $-\infty<\tau<+\infty,$ and for all $x\geq\tau$%

\begin{equation}
\lambda\left(  x\right)  =\left(  1-\varepsilon_{\lambda,\tilde{\mu}_{\tau}%
}\left(  x\right)  \right)  \left[  \lambda\ast q_{\lambda,\tilde{\mu}_{\tau
},x}\right]  \left(  x\right)  +\varepsilon_{\lambda,\tilde{\mu}_{\tau}%
}\left(  x\right)  Q_{\lambda,\tilde{\mu}_{\tau}}\left(  x\right)
.\label{122}%
\end{equation}
One would like to pass here to the limit $\tau\rightarrow-\infty.$ According
to $\left(  \ref{102}\right)  ,$ for every $x$ we have $\varepsilon
_{\lambda,\tilde{\mu}_{\tau}}\left(  x\right)  \rightarrow0$ as $\tau
\rightarrow-\infty.$ Moreover, it is not difficult to show that in the same
limit $q_{\lambda,\tilde{\mu}_{\tau},x}\left(  \cdot\right)  \rightarrow
q_{\lambda,x}\left(  \cdot\right)  .$ So the following equation holds for
$\lambda:$%
\begin{equation}
\lambda\left(  x\right)  =\left[  \lambda\ast q_{\lambda,x}\right]  \left(
x\right)  ,\;\;-\infty<x<+\infty.\label{123}%
\end{equation}
By the methods developed below one can show that every bounded solution of
$\left(  \ref{123}\right)  $ is a constant. Since, however, we are proving a
stronger statement, that the dynamical system $\mathcal{T}_{\tau}$ has one
fixed point on each $\mathcal{M}_{q}\left(  \Omega\right)  ,$ which is,
moreover, globally attractive, we will not provide the details.

\section{Self-averaging $\Longrightarrow$ relaxation: a warm-up}

Before presenting the general proof that self-averaging implies relaxation, we
consider the following simpler system: we have infinitely many servers, with
service time $\eta,$ its distribution given by the probability density $p.$ As
the customer comes, he chooses any free server, and is served, leaving the
system afterwards. The inflow is Poissonian, given by the rate function
$f\left(  x\right)  .$ If we impose the condition that the customers are
coming at the rate they are living the system, we get the non-linear Markov
process. The self-averaging relation (\ref{34}) in such a case simplifies to
\[
b\left(  x\right)  =\left[  f\ast p\right]  \left(  x\right)  .
\]

\begin{lemma}
Let $p\left(  x\right)  $ be some smooth probability density, $\mathrm{supp}%
\,p=\left[  0,1\right]  .$ Let $f$ be a positive bounded function on
$\mathbb{R}^{1},$ with $f\left(  x\right)  \leq C$ for $x\in\lbrack1-,0).$
Suppose that
\begin{equation}
f\ast p\left(  x\right)  =f\left(  x\right)  \text{ for all }x\geq
0.\label{100}%
\end{equation}
Then $f\left(  x\right)  \rightarrow c$ as $x\rightarrow\infty,$ for some
$c>0.$
\end{lemma}

\begin{proof}
Let us first show that for every function $\varphi\geq0$ on $\mathbb{R}^{1},$
$\varphi=0$ for $x$ outside the segment $[1-,0)$ there exists a function
$f_{\varphi}$ on $\mathbb{R}^{1},$ satisfying (\ref{100}), which coincides
with $\varphi$ on $[1-,0).$ To construct such a function consider the sequence
$f_{n}$ of functions, defined inductively by
\begin{align*}
f_{0}  & =\varphi,\\
f_{n+1}\left(  x\right)   & =\left\{
\begin{array}
[c]{ll}%
\varphi\left(  x\right)  & \text{ for }x<0,\\
\left[  \;f_{n}\ast p\right]  \left(  x\right)  & \text{ for }x>0.
\end{array}
\right.
\end{align*}
A straightforward check shows that the sequence $f_{n}\left(  x\right)  $ is
non-decreasing for every $x,$ and that $f_{n}\left(  x\right)  \leq C$ for all
$n$ and $x.$ Therefore the function $f_{\varphi}\left(  x\right)
=\lim_{n\rightarrow\infty}f_{n}\left(  x\right)  $ is defined. Clearly, it
satisfies (\ref{100}). This function is given by the formula
\begin{equation}
f_{\varphi}\left(  x\right)  =\left\{
\begin{array}
[c]{ll}%
\left[  \varphi\ast p\right]  \left(  x\right)  +\left[  \;\left(  \varphi\ast
p\Bigm|_{\left\{  x\geq0\right\}  }\right)  \ast\left(  \sum_{n=1}^{\infty
}p^{\ast n}\right)  \right]  \left(  x\right)  & \text{ for }x>0,\\
\varphi\left(  x\right)  & \text{ for }x<0.
\end{array}
\right. \label{101}%
\end{equation}
(Note that for every $x$ the sequence $p^{\ast n}\left(  x\right)  $ decays
exponentially, so the last expression is well-defined.)

Let us show that the function $f,$ satisfying (\ref{100}), is uniquely defined
by its restriction to $[1-,0).$ Indeed, let $g$ be another such function. Then
$h\left(  x\right)  =f\left(  x\right)  -g\left(  x\right)  $ is bounded in
absolute value and satisfies

$h\left(  x\right)  =0$ for $x\leq0,$

$h\ast p\left(  x\right)  =h\left(  x\right)  $ for all $x\geq0.$

\noindent But then \noindent$h\ast p^{\ast n}\left(  x\right)  =h\left(
x\right)  $ for all $x$ and for all $n.$ Since for every $A$ we have $\int
_{0}^{A}p^{\ast n}\left(  x\right)  \,dx\rightarrow0,$ as $n\rightarrow
\infty,$ it follows that $h\equiv0.$

Let us show that the function
\[
s\left(  x\right)  =\sum_{n=1}^{\infty}p^{\ast n}\left(  x\right)
\]
goes to the limit as $x\rightarrow\infty;$ together with (\ref{101}) it would
imply our statement. We will compute that limit, $S\left(  \eta\right)  $. For
that we will use the local limit theorem approximation for the convolutions
$p^{\ast n}\left(  x\right)  .$ Let $m$ be the mean value of $\eta,$ and $v$
be its variance. Denote by $q_{M,V}\left(  \cdot\right)  $ the density of the
normal distribution with the mean $M$ and the variance $V. $ Then easy
calculus computations tell us that
\begin{align*}
S\left(  \eta\right)   & =\lim_{N\rightarrow\infty}\sum_{k\geq N/2}%
q_{km,kv}\left(  Nm\right)  =2\lim_{N\rightarrow\infty}\sum_{k\geq N}%
q_{km,kv}\left(  Nm\right) \\
& =2\lim_{N\rightarrow\infty}\sum_{k\geq N}\frac{1}{\sqrt{2\pi kv}}e^{-\left(
Nm-km\right)  ^{2}/2kv}=\frac{1}{m}.
\end{align*}
Therefore the limit
\[
\lim_{x\rightarrow\infty}f_{\varphi}\left(  x\right)  =\frac{1}{m}\int
_{0}^{+\infty}\left[  \varphi\ast p\right]  \left(  x\right)  \;dx.
\]

\end{proof}

In a special case when
\[
\varphi\left(  x\right)  =1+x\text{ on }\left[  -1,0\right]  ,
\]
and $p\left(  x\right)  $ is the uniform distribution on a segment $\left[
0,1\right]  ,$ there is another formula for $f_{\varphi}:$%
\[
f_{\varphi}\left(  x\right)  =1+x-\sum_{0\leq k<x}\frac{\left(  -1\right)
^{k}}{2k!}\left(  x-k\right)  ^{k}e^{x-k}.
\]
(We got it together with Prof. O. Ogieveckij.) It satisfies the equation:
\begin{equation}
f\left(  x\right)  =\int_{x-1}^{x}f\left(  x\right)  \,dx,\;x\geq0\label{1}%
\end{equation}
with the initial data
\[
f\left(  x\right)  =1+x
\]
on the segment $\left[  -1,0\right]  .$ Note that at $x=0$ it has a jump, from
$1$ to $\frac{1}{2}.$ It becomes more and more smooth; in the non-integer
points it is analytic, but at the integer point $n$ it has $n-1$ derivatives.

It is bounded, and it goes to $\frac{2}{3}$ as $x\rightarrow\infty,$ since
$m=\frac{1}{2}$ and
\[
\int_{0}^{1}\left[  \left(  1+x\right)  \Bigm|_{\left[  -1,0\right]  }\ast
p\right]  \left(  x\right)  \;dx=\frac{1}{3}.
\]
But to see analytically that this series defines a bounded function, and,
moreover,
\[
1+x-\sum_{0\leq k<x}\frac{\left(  -1\right)  ^{k}}{2k!}\left(  x-k\right)
^{k}e^{x-k}\rightarrow\frac{2}{3}\text{ as }x\rightarrow\infty
\]
seems to be quite hard. So, it looks amazing, that the above arguments give
relatively simple proof of this convergence, which proof is probabilistic!

\section{Self-averaging $\Longrightarrow$ relaxation: probabilistic proof?}

As we know, any function $\lambda$, defined for $x<0,$ and vanishing for
$x<-T,$ can be uniquely extended to $x\geq0$ in such a way that the relation
\[
A\left(  \mathbf{0},\lambda\left(  \cdot\right)  ,-T\right)  =b\left(
\cdot\right)
\]
holds with $b\left(  x\right)  =\lambda\left(  x\right)  $ for $x\geq0.$
Therefore for every $x\geq0$ we have
\begin{equation}
\lambda\left(  x\right)  =\left[  \lambda\ast q_{\lambda,x}\right]  \left(
x\right)  ,\label{35}%
\end{equation}
where $q_{\lambda,x}\left(  \cdot\right)  $ is a probability density supported
by the semiaxis $\left\{  y\geq0\right\}  ,$ and which is defined only by the
restriction $\lambda\Bigm|_{\left\{  y\leq x\right\}  }.$ Our goal is to show
that (\ref{35}) implies that $\lambda\left(  x\right)  $ relaxes to some
constant $c$ as $x\rightarrow\infty.$

Since the distributions $q_{\lambda,x}$ depend on $\lambda\left(
\cdot\right)  $ in a very complicated way, we have to treat a more general
statement. Suppose a family of probability densities $q_{x}\left(
\cdot\right)  ,$ supported by the semiaxis $\left\{  y\geq0\right\}  ,$ is
given, where $x\geq0.$ Let $f\left(  x\right)  $ be a non-negative function,
defined on $\mathbb{R}^{1},$ such that
\begin{align}
f\left(  x\right)   & \leq C\text{ for }x<0,\nonumber\\
f\left(  x\right)   & =\left[  f\ast q_{x}\right]  \left(  x\right)  \text{
for }x\geq0.\label{36}%
\end{align}
One would like to show that
\begin{equation}
\lim_{x\rightarrow\infty}f\left(  x\right)  =c,\label{43}%
\end{equation}
for some $c\geq0.$ That will imply the relaxation needed.

Motivated by the analysis of the previous section, we will study the equation
(\ref{36}) by considering the corresponding inhomogeneous Markov random walk.
Unfortunately, the relation (\ref{43}) does not follow from (\ref{36}) in
general, and the reasons are probabilistic! Before explaining it let us
``solve'' (\ref{36}).

So, let the family $\left\{  q_{x},x\geq0\right\}  $ be given; we solve
(\ref{36}) for $f,$ given its restriction $f\Bigm|_{\left\{  x<0\right\}  }.$
We do this in close analogy with the previous section, see (\ref{101}). We
put
\[
f_{0}\left(  x\right)  =\left\{
\begin{array}
[c]{ll}%
f\left(  x\right)  & \text{ for }x<0\\
0 & \text{ for }x\geq0.
\end{array}
\right.
\]
We define
\[
f_{n+1}\left(  x\right)  =\left\{
\begin{array}
[c]{ll}%
f\left(  x\right)  & \text{ for }x<0,\\
\left[  f_{n}\ast q_{x}\right]  \left(  x\right)  & \text{ for }x\geq0.
\end{array}
\right.
\]
Then for every $x$ the sequence $f_{n}\left(  x\right)  $ is increasing, and
the function $f\left(  x\right)  =\lim_{n\rightarrow\infty}f_{n}\left(
x\right)  $ solves (\ref{36}).

To proceed, it is convenient to rewrite the function $f$ in a different way.
We define
\[
g_{1}\left(  x\right)  =\left\{
\begin{array}
[c]{ll}%
\left[  f_{0}\ast q_{x}\right]  \left(  x\right)  & \text{ for }x\geq0,\\
0 & \text{ for }x<0,
\end{array}
\right.
\]
\[
g_{n+1}\left(  x\right)  =\left[  g_{n}\ast q_{x}\right]  \left(  x\right)  .
\]
Then for $x\geq0$ we have
\[
f\left(  x\right)  =\sum_{n\geq1}g_{n}\left(  x\right)  .
\]
Now we will write the formula for $g_{n}$ in terms of convolution. To simplify
the exposition we consider the case $n=5,$ say.
\begin{align}
& g_{5}\left(  x\right)  =\left[  g_{4}\ast q_{x}\right]  \left(  x\right)
\nonumber\\
& =\int g_{4}\left(  x-u\right)  q_{x}\left(  u\right)  \,du\nonumber\\
& =\int g_{3}\left(  x-u-v\right)  q_{x-u}\left(  v\right)  q_{x}\left(
u\right)  \,dudv\label{37}\\
& =\int g_{2}\left(  x-u-v-w\right)  q_{x-u-v}\left(  w\right)  q_{x-u}\left(
v\right)  q_{x}\left(  u\right)  \,dudvdw\nonumber\\
& =\int g_{1}\left(  x-t\right)  q_{x-u-v-w}\left(  t-u-v-w\right)
q_{x-u-v}\left(  w\right)  q_{x-u}\left(  v\right)  q_{x}\left(  u\right)
\,dudvdwdt.\nonumber
\end{align}
Motivated by the last line we will introduce now for every $x>0$ the following
family of d.d.d. (dependent, differently distributed) positive random
variables $\chi_{1},\chi_{2},\chi_{3},...:$

the distribution of $\chi_{1}$ is given by the density $q_{x}\left(
\cdot\right)  ,$

the conditional distribution of $\chi_{2}$ under condition $\chi_{1}$ is given
by the density $q_{x-\chi_{1}}\left(  \cdot\right)  $,

the conditional distribution of $\chi_{3}$ under condition $\chi_{1},\chi_{2}$
is given by the density $q_{x-\chi_{1}-\chi_{2}}\left(  \cdot\right)  $,

etc. ...

\noindent To make this definition complete, we put $q_{x}\left(  \cdot\right)
$ to be the uniform distribution on $\left[  0,1\right]  $ for all negative
$x$-s; note that this extension does not contribute to (\ref{37}), since the
support of all the functions $g_{i}$ is the positive semiaxis.

Hence we are led naturally to consider for every $x$ the sums $\theta_{i}%
=\chi_{1}+\chi_{2}+...+\chi_{i};$ if we denote by $p_{x}^{\left(  i\right)  }$
the probability density of $\theta_{i},$ then we have, by (\ref{37}):
\[
g_{n+1}\left(  x\right)  =\left[  g_{1}\ast p_{x}^{\left(  n\right)  }\right]
\left(  x\right)  .
\]

Summarizing, we have for $x>0:$
\[
f\left(  x\right)  =g_{1}\left(  x\right)  +\sum_{n\geq2}\left[  g_{1}\ast
p_{x}^{\left(  n\right)  }\right]  \left(  x\right)  .
\]

The distributions $p_{\cdot}^{\left(  \cdot\right)  }$ are describing the
following \textbf{non-stationary Markov chain:} at every point $y\in
\mathbb{R}^{1}$ we are given a probability distribution $q_{y}\left(
\cdot\right)  ,$ which has to be interpreted as the transition probability to
make a move once in $y.$ So we start at some $x,$ and we make a (random) move
$-\chi_{1},$ where $\chi_{1}$ is distributed according to $q_{x}\left(
\cdot\right)  .$ Arriving thus to $x-\chi_{1},$ we make a second move
$-\chi_{2}, $ where $\chi_{2}$ is distributed according to $q_{x-\chi_{1}%
}\left(  \cdot\right)  ,$ and so on. Clearly, the local limit theorem for this
chain would imply (\ref{43}).

We have to note, however, that the relation between the validity of the local
limit theorem for this Markov chain and the validity of the relation
(\ref{43}) is more complicated. First of all, the CLT for $\theta_{i}$ might
not hold, notwithstanding that the family $q_{y}\left(  \cdot\right)  $ have
very nice compactness properties. To give one example, consider the family of
probability densities $u_{x}\left(  t\right)  ,$ $x\in\mathbb{R}^{1},$ where
all $u_{x}\left(  \cdot\right)  $ have for their support the segment $\left[
0,1\right]  ,$ and satisfy there $0<c<u_{x}\left(  t\right)  <C<\infty,$
uniformly in $x$ and $t.$ We define now
\[
q_{x}\left(  t\right)  =u_{x}\left(  t-\left\{  x\right\}  \right)  ,
\]
where $\left\{  \cdot\right\}  $ stays for the fractional part. Then all
$q_{y}\left(  \cdot\right)  $-s have their supports within the segment
$\left[  0,2\right]  .$ But the random variables $\theta_{i}$ do not have CLT
behavior! Indeed, the random variable $\theta_{i},$ is localized in the
segment $\left[  \left\lfloor x\right\rfloor -i,\left\lfloor x\right\rfloor
-i+1\right]  ,$ where $\left\lfloor \cdot\right\rfloor $ denotes the integer
part. Nevertheless, for this example it can be shown that the relation
(\ref{43}) still holds, and that involves certain statement of the type of
Perron-Frobenius theorem for our Markov chain. Further modification of this
example, when
\[
q_{x}\left(  t\right)  =u_{x}\left(  t-\left\{  x\right\}  -1\right)  ,
\]
results in the Markov chain with two classes, and in this case both the CLT
and the relation (\ref{43}) fail.

We conjecture here that the CLT theorem for the sums $\theta_{i}$ holds, if
the family $q_{x}\left(  \cdot\right)  $ of transition densities has the
following additional property:

\begin{itemize}
\item For some $k,K,$ $0<k<K<\infty,$
\begin{equation}
k\leq\frac{q_{x_{1}}\left(  t\right)  }{q_{x_{2}}\left(  t\right)  }\leq
K,\label{45}%
\end{equation}
provided at least one of the values $q_{x_{i}}\left(  t\right)  $ is positive.
\end{itemize}

The condition (\ref{45}) is reminiscent of the \textbf{positivity of
ergodicity coefficient }condition, introduced by Dobrushin \cite{D} in his
study of the limit theorems for the non-stationary Markov chains.

In what follows we will take another road, and we get the relaxation property
by analytical methods, which seems in our case to be simpler. But we still use
probability theory, though not the CLT. It would be interesting to obtain the
desired result by proving the corresponding limit theorem.

\section{Self-averaging $\Longrightarrow$ relaxation: finite range case}

In this section we prove the relaxation for the solution of the equation
(\ref{36}) in the finite range case.

\begin{theorem}
Suppose that
\begin{align*}
0  & \leq f\left(  x\right)  \leq C\text{ for }x<0,\\
f\left(  x\right)   & =\left[  f\ast q_{x}\right]  \left(  x\right)  \text{
for }x\geq0,
\end{align*}
while the following conditions on the family $q_{x}$ hold: for some $T$%
\[
\int_{0}^{T}q_{x}\left(  t\right)  \,dt=1
\]
for all $x,$ and
\begin{equation}
C\geq q_{x}\left(  t\right)  \geq\kappa\left(  t\right)  >0\label{016}%
\end{equation}
for $0\leq t\leq T,$ with continuous positive $\kappa\left(  t\right)  .$ Then
the limit exists:
\[
\lim_{x\rightarrow\infty}f\left(  x\right)  =c\geq0.
\]

\end{theorem}

The property $\left(  \ref{016}\right)  $ holds for the NMP, as follows from
the relations $\left(  \ref{301}\right)  $ and $\left(  \ref{302}\right)  .$

\begin{proof}
$i)$ We know that the function $f$ is continuous and bounded, $0\leq f\leq C.
$ So if there exists a value $X$ such that $f$ is monotone for $x\geq X,$ then
the function $f$ has to be constant for $x\geq X+T,$ and we are done. So we
are left with the case when the function $f$ has infinitely many points of
local maxima and local minima, which go to $\infty.$

$ii)$ Given a local maximum, $x_{0},$ we will construct now a sequence $x_{i}
$ of local maximums, $i=0,-1,-2,...,-n=-n\left(  f,x_{0}\right)  $ such that

\begin{itemize}
\item $x_{0}>x_{-1}>x_{-2}>...,$

\item $x_{i}-x_{i-1}<2T,\;\;x_{i}-x_{i-2}\geq T$ for all $i,$

\item $0<x_{-n}<2T,$

\item $f\left(  x_{i-1}\right)  \geq f\left(  x\right)  $ for any $x_{i-1}\leq
x, $ and $f\left(  x_{i-1}\right)  >f\left(  x_{i}\right)  ,$

\item for every $x\in\left[  x_{i-1},x_{i}-T\right]  $ we have $f\left(
x\right)  \geq f\left(  x_{i}\right)  $ (of course if the segment is non-empty).
\end{itemize}

The construction is the following. Let $x_{0}$ be some point of local maxima.
Since
\[
f\left(  x_{0}\right)  =\int_{0}^{T}f\left(  x_{0}-t\right)  q_{x_{0}}\left(
t\right)  \,dt,
\]
we have $f\left(  x_{0}\right)  <F\left(  x_{0}\right)  \equiv\sup\left\{
f\left(  x\right)  :x\in\left[  x_{0}-T,x_{0}\right]  \right\}  ,$ unless $f$
is a constant on $\left[  x_{0}-T,x_{0}\right]  ,$ in which case we are done. Let

\noindent$y=\inf\left\{  x\in\left[  x_{0}-T,x_{0}\right]  :f\left(  x\right)
=F\left(  x_{0}\right)  \equiv\sup\left\{  f\left(  x\right)  :x\in\left[
x_{0}-T,x_{0}\right]  \right\}  \right\}  .$ If $y>x_{0}-T,$ or if $y=x_{0}-T$
and is a local maximum, we define $x_{-1}=y.$ In the opposite case we have
that the point $x_{0}-T$ is not a local maximum of the function $f$ on the
segment $\left[  x_{0}-2T,x_{0}-T\right]  .$ We then consider two cases. In
the first one we suppose that the function $f$ on the segment $\left[
x_{0}-2T,x_{0}-T\right]  $ takes values below $\bar{F}=\frac{f\left(
x_{0}\right)  +f\left(  x_{0}-T\right)  }{2}.$ Let $\left[  y,x_{0}-T\right]
\subset\left[  x_{0}-2T,x_{0}-T\right]  $ be the largest segment for which the
inequality $f\left(  x\right)  \geq\bar{F}$ holds for every $x\in\left[
y,x_{0}-T\right]  .$ We define $x_{-1}$ to be the leftmost point of maximum of
$f$ in $\left[  y,x_{0}-T\right]  .$ In the opposite case we consider the set
$S=\left\{  x\in\left[  x_{0}-2T,x_{0}-T\right]  :f\left(  x\right)  \geq
f\left(  x_{0}-T\right)  \right\}  .$ It contains other points besides
$x_{0}-T. $ However, it can not contain all the segment $\left[
x_{0}-2T,x_{0}-T\right]  .$ Since $f\ $is not a constant on $\left[
x_{0}-2T,x_{0}-T\right]  ,$ $\,\sup_{S}f>f\left(  x_{0}-T\right)  .$ Let
$z\in\left(  x_{0}-2T,x_{0}-T\right)  $ be such that $f\left(  z\right)
<f\left(  x_{0}-T\right)  .$ Let $S_{1}=S\cap\left[  z,x_{0}-T\right]  .$ We
necessarily have that $\sup_{S_{1}}f>f\left(  x_{0}-T\right)  $ as well. We
define $x_{-1}$ to be any point in $S_{1}$ where $f\left(  x_{-1}\right)
=\sup_{S_{1}}f.$ Clearly, $x_{-1}$ is a local maxima of $f,$ while
$x_{0}-x_{-1}<2T.$

We proceed to define the sequences $x_{i}$ by induction, $i=0,-1,-2,...,$
until we arrive to a first value below $2T$, where we stop.

$iii)$ In the same way, starting from a local minima $y_{0},$ we can construct
a sequence $y_{i}$ of local minima, such that

\begin{itemize}
\item $y_{0}>y_{-1}>y_{-2}>...,$

\item $y_{i}-y_{i-1}<2T,\;\;y_{i}-y_{i-2}\geq T$ for all $i,$

\item $0<y_{-n}<2T,$

\item $f\left(  y_{i-1}\right)  \leq f\left(  x\right)  $ for any $y_{i-1}\leq
x, $ and $f\left(  y_{i-1}\right)  <f\left(  y_{i}\right)  ,$

\item for every $x\in\left[  y_{i-1},y_{i}-T\right]  $ we have $f\left(
x\right)  \leq f\left(  y_{i}\right)  $ (if the segment is non-empty).
\end{itemize}

We can suppose additionally that $x_{0}\geq y_{0}\geq x_{-1}.$

$iv)$ Note that the (finite) sequence $x_{i}$ do depend on the initial local
minima $x_{0},$ which was used for the starter. The bigger $x_{0}$ is, the
longer the sequence $x_{i}$ is. So let us introduce the sequence
$x_{0}^{\left(  N\right)  }$ of such starters, and we suppose that
$x_{0}^{\left(  N\right)  }\geq N.$ In that way we will obtain the family
$x_{i}^{\left(  N\right)  }$ of sequences of local maximums of $f,$
$i=0,-1,...,-n\left(  f,x_{0}^{\left(  N\right)  }\right)  ,$ with $n\left(
f,x_{0}^{\left(  N\right)  }\right)  \rightarrow\infty$ as $N\rightarrow
\infty.$ (Of course, in well may happen that for different $N$-s the
corresponding sequences share many common terms.)

Denote by $M$ the limit $\liminf_{N\rightarrow\infty}f\left(  x_{0}^{\left(
N\right)  }\right)  .$ In the same way we can introduce the limit
$m=\limsup_{N\rightarrow\infty}f\left(  y_{0}^{\left(  N\right)  }\right)  .$
Clearly, $M\geq m,$ and if we can show that $M=m,$ then we are done. So we
suppose that $M-m>0,$ and we will bring that to contradiction.

$v)$ Let us fix $\varepsilon>0,$ $\varepsilon<\frac{M-m}{10},$ which is
possible if $M-m>0.$ Then one can choose $N$ so large, that at least 99\% of
terms of the sequence $f\left(  x_{i-1}^{\left(  N\right)  }\right)  -f\left(
x_{i}^{\left(  N\right)  }\right)  $ are less than $\frac{\varepsilon^{2}}%
{2}.$ We will fix that value of $N,$ and we will omit $N$ from our notation.
Therefore without loss of generality we can assume that for some $i$ (in fact,
for many) $\,$we have $f\left(  x\right)  <f\left(  x_{i}\right)
+\varepsilon^{2}$ for all $x\in\left[  x_{i}-T,x_{i}\right]  .$ Therefore for
the set $K\equiv\left\{  x\in\left[  x_{i}-T,x_{i}\right]  :f\left(  x\right)
>f\left(  x_{i}\right)  -\varepsilon\right\}  $ we have:
\begin{equation}
\int_{K-\left(  x_{i}-T\right)  }q_{x_{i}}\left(  t\right)  \,dt>1-\varepsilon
.\label{46}%
\end{equation}
Hence, for its Lebesgue measure we have
\[
\mathrm{mes}\left\{  K\right\}  \geq\frac{1-\varepsilon}{C}.
\]

Consider now the corresponding sequence of minima, $\left\{  y_{k}\right\}  ,$
and the segments $\left[  y_{k}-T,y_{k}\right]  .$ We claim that the set $K$
has to belong to the union of these segments. That would be evident if the
segments in question were covering the corresponding region without any holes.
However, that is not necessarily the case, and there can be holes between the
segments, since in general the differences $y_{i}-y_{i-1}$ can be bigger than
$T.$ Yet, this does not present a problem, since by construction the function
$f$ is smaller than $m$ outside the union of the segments $\left[
y_{k}-T,y_{k}\right]  ,$ which implies that the set $K$ indeed is covered by
these segments. Since $\mathrm{diam\,}\left(  K\right)  \leq T,$ there exists
$k=k\left(  K\right)  ,$ such that $K\subset\left[  y_{k-1}-T,y_{k-1}\right]
\cup\left[  y_{k}-T,y_{k}\right]  \cup\left[  y_{k+1}-T,y_{k+1}\right]  .$
Without loss of generality we can assume the set $K$ ``fits into $\left[
y_{k}-T,y_{k}\right]  $'', in the sense that
\[
\mathrm{mes}\left\{  K\cap\left[  y_{k}-T,y_{k}\right]  \right\}  \geq
\frac{\mathrm{mes}\left\{  K\right\}  }{3}\geq\frac{1-\varepsilon}{3C},
\]
while we have $f\left(  x\right)  >f\left(  y_{k}\right)  -\varepsilon^{2}$
for all $x\in\left[  y_{k}-T,y_{k}\right]  .$ So we have
\begin{equation}
\int_{\left\{  K\cap\left[  y_{k}-T,y_{k}\right]  \right\}  -\left(
y_{k}-T\right)  }q_{y_{k}}\left(  t\right)  \,dt\geq\bar{\kappa}\left(
\frac{1-\varepsilon}{3C}\right)  ,\label{47}%
\end{equation}
where we define the function $\bar{\kappa}\left(  \alpha\right)  $ by
\[
\bar{\kappa}\left(  \alpha\right)  =\inf_{A\subset\left[  0,T\right]
:\,\mathrm{mes}\left\{  A\right\}  \geq\alpha}\int_{A}\kappa\left(  t\right)
\,dt.
\]
By construction, the set $K\cap\left[  y_{k}-T,y_{k}\right]  $ is disjoint
from the set $L\subset\left[  y_{k}-T,y_{k}\right]  ,$ which is defined by
$L=\left\{  x\in\left[  y_{k}-T,y_{k}\right]  :f\left(  x\right)  <f\left(
y_{k}\right)  +\varepsilon\right\}  .$ Since
\[
f\left(  y_{k}\right)  =\int_{0}^{T}f\left(  y_{k}-t\right)  q_{y_{k}}\left(
t\right)  \,dt,
\]
we have similar to (\ref{46}) that
\begin{equation}
\int_{L-\left(  y_{k}-T\right)  }q_{y_{k}}\left(  t\right)  \,dt>1-\varepsilon
.\label{48}%
\end{equation}
But since $q_{y_{k}}\left(  t\right)  \,dt$ is a probability measure, we
should have that
\[
\bar{\kappa}\left(  \frac{1-\varepsilon}{3C}\right)  +1-\varepsilon\leq1,
\]
because of (\ref{47}), (\ref{48}). This, however, fails once $\varepsilon$ is
small enough.
\end{proof}

\section{Self-averaging $\Longrightarrow$ relaxation: infinite range case}

We return to the equation (\ref{36}), $f\left(  x\right)  =\left[  f\ast
q_{x}\right]  \left(  x\right)  .$ Now we will not suppose that the measures
$q_{x}$ have finite support. Instead we suppose that the family $q_{x}$ has
the following compactness property: for every $\varepsilon>0$ there exists a
value $K\left(  \varepsilon\right)  ,$ such that
\begin{equation}
\int_{0}^{K\left(  \varepsilon\right)  }q_{x}\left(  t\right)  \,dt\geq
1-\varepsilon\label{111}%
\end{equation}
uniformly in $x.$ We will also suppose that for every $T$ the (monotone
continuous) function
\begin{equation}
F_{T}\left(  \delta\right)  =\inf_{x}\inf_{\substack{D\subset\left[
0,T\right]  : \\\mathrm{mes}D\geq\delta}}\int_{D}q_{x}\left(  t\right)
\,dt\label{112}%
\end{equation}
is positive once $\delta>0.$ Finally we assume that the family $q_{x}$ is such
that the function $f,$ with solves $\left(  \ref{36}\right)  ,$ is Lipschitz,
with Lipschitz constant $\mathcal{L}=\mathcal{L}\left(  \left\{  q_{\cdot
}\right\}  \right)  .$ As we know from the Sections 4 and 7, these conditions
are indeed satisfied in the specific case of the non-linear Markov process and
the equation $\left(  \ref{35}\right)  .$

\subsection{Approaching stationary point}

\begin{lemma}
\label{T} $i)$ Let $M=\limsup_{x\rightarrow\infty}f\left(  x\right)  .$ Then
for every $T$ and every $\varepsilon$ given there exists some value $K_{1},$
such that
\[
\inf_{x\in\left[  K_{1},K_{1}+T\right]  }f\left(  x\right)  \geq
M-\varepsilon.
\]
$ii)$ Let $m=\liminf_{x\rightarrow\infty}f\left(  x\right)  .$ Then for every
$T$ and every $\varepsilon$ given there exists some value $K_{2},$ such that
\[
\sup_{x\in\left[  K_{2},K_{2}+T\right]  }f\left(  x\right)  \leq
m+\varepsilon.
\]
Moreover, the conclusions of the lemma remains valid if the function $f$
satisfies a weaker equation (see $\left(  \ref{103}\right)  $)
\begin{equation}
f\left(  x\right)  =\left(  1-\varepsilon\left(  x\right)  \right)  \left[
f\ast q_{x}\right]  \left(  x\right)  +\varepsilon\left(  x\right)  Q\left(
x\right)  ,\label{211}%
\end{equation}
with $\varepsilon\left(  x\right)  \rightarrow0$ as $x\rightarrow\infty$ and
$Q\left(  \cdot\right)  \leq C.$
\end{lemma}

\begin{proof}
$i)$ Let $\delta>0.$ Then there exists a value $S>0,$ such that for all $x>S $
we have $f\left(  x\right)  <M+\delta,$ and $\varepsilon\left(  x\right)
Q\left(  x\right)  <\frac{\delta}{2}.$ Further, there exists a value $R>S,$
such that for all $y\geq R$
\[
\int_{R-S}^{\infty}q_{y}\left(  t\right)  dt<\delta,
\]
see (\ref{111}). Finally, there exists a point $y>R+T,$ such that $f\left(
y\right)  >M-\frac{\delta}{2}.$ Due to the equation $\left(  \ref{211}\right)
$ we have
\[
f\left(  y\right)  =\left(  1-\varepsilon\left(  y\right)  \right)  \left[
\int_{0}^{y-S}f\left(  y-t\right)  q_{y}\left(  t\right)  \,dt+\int
_{y-S}^{\infty}f\left(  y-t\right)  q_{y}\left(  t\right)  \,dt\right]
+\varepsilon\left(  y\right)  Q\left(  y\right)  .
\]
Let $\Delta>0,$ and $A=\left\{  x\in\left[  y-T,y\right]  :f\left(  x\right)
<M-\Delta\right\}  ,$ while $a=\int_{A}q_{y}\left(  t\right)  \,dt.$ We want
to show that the measure $a$ has to be small for small $\delta$. Splitting the
first integral into two, according to whether the point $y-t$ is in $A$ or
not, we have
\[
M-\delta<a\left(  M-\Delta\right)  +\left(  1-a-\delta\right)  \left(
M+\delta\right)  +\delta C,
\]
so
\[
a<\delta\frac{C+2-M}{\Delta},
\]
which goes to zero with $\delta,$ provided $\Delta$ is fixed. Therefore
\[
\mathrm{mes}\left\{  A\right\}  \leq F_{T}^{-1}\left(  \delta\frac
{C+2-M}{\Delta}\right)  ,
\]
(see (\ref{112})). Since $F_{T}^{-1}\left(  u\right)  \rightarrow0$ as
$u\rightarrow0,$ that proves that for any given $\Delta$ the Lebesgue measure
$\mathrm{mes}\left\{  A\right\}  \rightarrow0$ as $\delta\rightarrow0$. Since
the function $f$ is Lipschitz, we conclude that $\inf_{x\in\left[
y-T,y\right]  }f\left(  x\right)  \geq M-\Delta-\mathcal{L}\mathrm{mes}%
\left\{  A\right\}  \geq M-2\Delta,$ provided $\delta$ is small enough. Taking
$\Delta=\varepsilon/2$ finishes the proof.

$ii)$ Let $\delta>0.$ Then there exists a value $S>0,$ such that for all $x>S$
we have $f\left(  x\right)  >m-\delta.$ Again, take $R>S,$ such that for all
$y\geq R$
\[
\int_{0}^{R-S}q_{y}\left(  t\right)  dt>1-\delta.
\]
Finally, there exists a point $y>R+T,$ such that $f\left(  y\right)
<m+\delta.$ Due to the equation $\left(  \ref{211}\right)  $ we have
\begin{equation}
f\left(  y\right)  >\left(  1-\varkappa\right)  \left[  \int_{0}^{y-S}f\left(
y-t\right)  q_{y}\left(  t\right)  \,dt+\int_{y-S}^{\infty}f\left(
y-t\right)  q_{y}\left(  t\right)  \,dt\right]  ,\label{113}%
\end{equation}
where $\varkappa$ can be supposed arbitrarily small. Let $\Delta>0,$ and

\noindent$A=\left\{  t\in\left[  0,T\right]  :f\left(  y-t\right)
>m+\Delta\right\}  ,$ while $a=\int_{A}q_{y}\left(  t\right)  \,dt.$ We want
to show that the measure $a$ has to be small for small $\delta$. Splitting the
first integral into two, according to whether the point $y-t$ is in $A$ or
not, and disregarding the second one, we have
\[
m+\delta>\left(  1-\varkappa\right)  \left[  a\left(  m+\Delta\right)
+\left(  1-\delta-a\right)  \left(  m-\delta\right)  \right]  .
\]
For $\varkappa$ so small that $\varkappa\left[  a\left(  m+\Delta\right)
+\left(  1-a-\delta\right)  \left(  m-\delta\right)  \right]  <\delta,$ we
have
\[
m+2\delta>a\left(  m+\Delta\right)  +\left(  1-a-\delta\right)  \left(
m-\delta\right)  ,
\]
so
\[
a<\delta\frac{m+3}{\Delta},
\]
which goes to zero with $\delta,$ provided $\Delta$ is fixed. Therefore
\[
\mathrm{mes}\left\{  A\right\}  \leq F_{T}^{-1}\left(  \delta\frac{m+3}%
{\Delta}\right)  ,
\]
and the rest of the argument coincides with that of the part $i).$
\end{proof}

\subsection{Absorbing by stationary point}

We now will show that if $f$ satisfies $\left(  \ref{36}\right)  ,$ then the
property $\inf_{x\in\left[  K,K+T\right]  }f\left(  x\right)  \geq
M-\varepsilon$ implies that for all $x>K+T$
\begin{equation}
f\left(  x\right)  >M-\varepsilon-c\left(  T\right)  ,\label{203}%
\end{equation}
with $c\left(  T\right)  \rightarrow0$ as $T\rightarrow\infty.$ That clearly
implies relaxation. (Note that we do not claim that $\left(  \ref{203}\right)
$ holds for the solutions of $\left(  \ref{211}\right)  $). We will show it
under the extra assumption that the distribution $p\left(  \cdot\right)  $ has
finite moment of some order above $4.$ This assumption, as well as $\left(
\ref{107}\right)  $ below, will be used only throughout the rest of the
present subsection.

Using the linearity of $\left(  \ref{36}\right)  ,$ we will rewrite our
problem slightly, in order to simplify the notation.

Let the function $f\geq0$ satisfies $f\left(  x\right)  =\left[  f\ast
q_{x}\right]  \left(  x\right)  $ for $x>0,$ and

$i)$ $f\left(  x\right)  >1$ for $x\in\left[  -T,0\right]  ,$

$ii)$ for some $\beta>1$ and $B<\infty$ and for every $x$ we have
\begin{equation}
\int_{0}^{\infty}t^{\beta}q_{x}\left(  t\right)  \,dx\leq B,\label{110}%
\end{equation}
compare with $\left(  \ref{303}\right)  .$ We want to derive from that data
that for some $c\left(  T\right)  >0,$ $c\left(  T\right)  \rightarrow0$ as
$T\rightarrow\infty$
\[
f\left(  x\right)  >1-c\left(  T\right)  \text{ for all }x>0.
\]

Denote by
\[
g_{0}\left(  x\right)  =\left\{
\begin{array}
[c]{cc}%
1 & x\in\left[  -T,0\right] \\
0 & x\notin\left[  -T,0\right]
\end{array}
\right.  .
\]
Since $f\geq g,$ we have $f\left(  x\right)  \geq g_{1}\left(  x\right)
=\left[  g_{0}\ast q_{x}\right]  \left(  x\right)  $ for $x\geq0.$ We define
$g_{1}\left(  x\right)  =g_{0}\left(  x\right)  $ for $x<0.$ Iterating, we
have $f\left(  x\right)  \geq g_{n}\left(  x\right)  ,$ where
\[
g_{n}\left(  x\right)  =\left\{
\begin{array}
[c]{cc}%
g_{0}\left(  x\right)  & x<0\\
\left[  g_{n-1}\ast q_{x}\right]  \left(  x\right)  & x\geq0
\end{array}
\right.  .
\]

Hence, $f\left(  x\right)  \geq g_{\infty}\left(  x\right)  .$ The function
$g_{\infty}\left(  x\right)  $ has the following probabilistic interpretation:
we have a Markov chain on $\mathbb{R}^{1},$ where transition from the point
$x$ is governed by transition densities $q_{x}$ to make the step (to the
left), (and which steps to the left are defined in an arbitrary way for
$x\leq0$); then the value $g_{\infty}\left(  x\right)  $ for $x>0$ is the
probability that starting from $x$ we will visit the interval $\left[
-T,0\right]  .$ The question now is about the lower bound on $g_{\infty
}\left(  x\right)  $ over all possible $q_{x}$ from our class.

So let us take $x>0,$ and let start the Markov chain $X_{n}$ from $x,$ (i.e.
$X_{0}=x$), which goes to the left, and which makes a transition from $y$ to
$y-t$ with the probability $q_{y}\left(  t\right)  dt.$ We need to know the
probability of the event
\[
\mathbb{P}_{x}\left\{  \text{there exists }n\text{ such that }X_{n}\in\left[
-T,0\right]  \right\}  .
\]
In other words, we want to know the probability of $X_{\cdot}$ visiting
$\left[  -T,0\right]  .$ We would like to show that
\begin{equation}
\mathbb{P}_{x}\left\{  X_{\cdot\text{ }}\text{visits }\left[  -T,0\right]
\right\}  \geq\gamma\left(  \beta,B,T\right) \label{106}%
\end{equation}
with
\[
\gamma\left(  \beta,B,T\right)  \rightarrow1\text{ as }T\rightarrow\infty
\]
uniformly over the families $q_{x}$ from our class.

Note, however, that in general such an estimate does not hold. For example,
the process $X_{\cdot}$ can well stay positive for all times. The more
interesting example where the process goes to $-\infty$, follows, so we will
need further restrictions on the family $q_{x}$.

\textbf{Example. }Let $T$ be given. We will construct the family $q_{x}^{T}$
from our class (\ref{110}), such that for the corresponding process
$X_{\cdot\text{ }}^{T}$
\[
\mathbb{P}_{x}\left\{  X_{\cdot\text{ }}^{T}\,\text{visits }\left[
0,T\right]  \right\}  =0.
\]
We define $q_{x}^{T}\left(  t\right)  $ for $x\in(k,k+1]$ with integer
$k\neq0$ to be any distribution localized in the segment $\left[
k-1,k\right]  $ (the uniform distribution on $\left[  k-1,k\right]  $ is OK).
For $x\in(\frac{1}{2^{k}},\frac{1}{2^{k-1}}],$ $k=1,2,...,$ it is defined by
\[
q_{x}^{T}\left(  t\right)  =\left\{
\begin{array}
[c]{ll}%
e^{-t} & \text{ if }t>T+1\\
2^{k+1}\left(  1-\int_{T+1}^{\infty}e^{-t}dt\right)  & \text{ if }t\in\left[
x-\frac{1}{2^{k}},x-\frac{1}{2^{k+1}}\right]  \text{ }\\
0 & \text{ otherwice.}%
\end{array}
\right.
\]
For $x\leq0$ it is defined in an arbitrary way. $\blacksquare$

The mechanism of violating the relation (\ref{106}) is that the time the
process $X_{\cdot\text{ }}^{T}$ can spend in the segment $\left[  0,1\right]
$ is unbounded in $T$. As the following theorem shows, this feature is the
only obstruction for the statement desired to hold.

\begin{theorem}
Consider the Markov chain $X_{\cdot\text{ }}$ defined above via the transition
densities $q_{x}\left(  t\right)  .$ Suppose that condition (\ref{110}) holds,
and that in addition these densities are uniformly bounded in the vicinity of
the origin: for all real $x$ and all $t$ in the segment $\left[  0,1\right]
$, say,
\begin{equation}
q_{x}\left(  t\right)  \leq C.\label{107}%
\end{equation}
Then for some $\gamma\left(  \beta,B,C,T\right)  \rightarrow1$ as
$T\rightarrow\infty$ we have:
\[
\mathbb{P}_{x}\left\{  X_{\cdot\text{ }}\text{visits }\left[  -T,0\right]
\right\}  \geq\gamma\left(  \beta,B,C,T\right)  .
\]

\end{theorem}

The condition $\left(  \ref{107}\right)  $ in the case of NMP follows easily
from the estimate $\left(  \ref{152}\right)  ,$ see Lemma \ref{oc}.

\begin{proof}
We will estimate the probability of the complementary event:
\begin{align*}
& \mathbb{P}_{x}\left\{  X_{\cdot\text{ }}\text{misses }\left[  -T,0\right]
\right\} \\
& =\sum_{k=0}^{\infty}\int_{0}^{x}\left[  \int_{y+T}^{\infty}q_{y}\left(
t\right)  \,dt\right]  P_{k}\left(  x,dy\right)  .
\end{align*}
Here $P_{k}\left(  x,dy\right)  $ is the probability distribution of the chain
$X_{\cdot\text{ }}$ after $k$ steps, and the expression $\left[  \int
_{y+T}^{\infty}q_{y}\left(  t\right)  \,dt\right]  P_{k}\left(  x,dy\right)  $
is the probability that the chain $X_{\cdot\text{ }}$ arrives after $k$ steps
to the location $y,$ and then makes a jump over the segment $\left[
-T,0\right]  .$ So we have
\begin{align*}
& \mathbb{P}_{x}\left\{  X_{\cdot\text{ }}\text{misses }\left[  -T,0\right]
\right\} \\
& \leq\int_{0}^{x}B\left(  y+T\right)  ^{-\beta}\sum_{k=0}^{\infty}%
P_{k}\left(  x,dy\right) \\
& \leq\sum_{n=0}^{\left[  x\right]  +1}B\left(  n+T\right)  ^{-\beta}%
\sum_{k=0}^{\infty}P_{k}\left(  x,\left[  n,n+1\right]  \right)  ,
\end{align*}
where $P_{k}\left(  x,\left[  n,n+1\right]  \right)  $ is the probability of
the event $X_{k}\in\left[  n,n+1\right]  ,$ and where in the second line we
are using the following simple estimate:
\[
\int_{r}^{\infty}q_{y}\left(  t\right)  \,dt=r^{-\beta}\int_{r}^{\infty
}r^{\beta}q_{y}\left(  t\right)  \,dt\leq r^{-\beta}\int_{0}^{\infty}t^{\beta
}q_{y}\left(  t\right)  \,dt.
\]
Now,
\begin{align}
& \sum_{k=0}^{\infty}\mathbb{P}_{x}\left\{  X_{k}\in\left[  n,n+1\right]
\right\} \label{108}\\
& =\sum_{k=0}^{\infty}\sum_{l<k}\mathbb{P}_{x}\left\{  X_{k}\in\left[
n,n+1\right]  ,X_{l}>n+1,X_{l+1}\in\left[  n,n+1\right]  \right\} \nonumber\\
& =\sum_{l=0}^{\infty}\mathbb{P}_{x}\left\{  X_{l}>n+1,X_{l+1}\in\left[
n,n+1\right]  \right\} \nonumber\\
& \times\sum_{k>0}\mathbb{P}_{x}\left\{  X_{l+k}\in\left[  n,n+1\right]
\Bigm|X_{l}>n+1,X_{l+1}\in\left[  n,n+1\right]  \right\}  .\nonumber
\end{align}
Let now the random variables $\zeta_{i}$ be i.i.d., uniformly distributed in
the segment $\left[  0,\frac{1}{C}\right]  ,$ where $C$ is the same as in
(\ref{106}). Then is easy to see that
\[
\mathbb{P}_{x}\left\{  X_{l+k}\in\left[  n,n+1\right]  \Bigm|X_{l}%
>n+1,X_{l+1}\in\left[  n,n+1\right]  \right\}  \leq\mathbf{\Pr}\left\{
\zeta_{1}+...+\zeta_{k}\leq1\right\}  .
\]
Since the last probability decays exponentially in $k,$ while

\noindent$\sum_{l=0}^{\infty}\mathbb{P}_{x}\left\{  X_{l}>n+1,X_{l+1}%
\in\left[  n,n+1\right]  \right\}  =1,$ we conclude that
\[
\sum_{k=0}^{\infty}\mathbb{P}_{x}\left\{  X_{k}\in\left[  n,n+1\right]
\right\}  \leq K\left(  C\right)  .
\]
Since the series $\sum n^{-\beta}$ converges for $\beta>1,$ the proof follows.
\end{proof}

\section{Self-averaging $\Longrightarrow$ relaxation: noisy case}

In this section we prove the relaxation for the NMP with general initial
condition, i.e. for the solution of the equation
\[
\lambda\left(  x\right)  =\left(  1-\varepsilon_{\lambda,\mu}\left(  x\right)
\right)  \left[  \lambda\ast q_{\lambda,\mu,x}\right]  \left(  x\right)
+\varepsilon_{\lambda,\mu}\left(  x\right)  Q_{\lambda,\mu}\left(  x\right)
\]
see $\left(  \ref{103}\right)  .$ We are not able to prove it in the
generality of the previous Sections. Below we will use all the specific
features of the NMP, and in particular we will use the comparison between
different NMP-s and GMP-s, corresponding to various initial states and input
rates. The comparison mentioned is based on the coupling arguments.

\subsection{Coupling}

\begin{definition}
Let $\mu_{1},\mu_{2}$ be two states on $\Omega.$ We call the state $\mu_{1}$
to be \textbf{higher} than $\mu_{2},$ $\mu_{1}\succcurlyeq\mu_{2},$ if there
exists a coupling $P\left[  d\omega_{1},d\omega_{2}\right]  $ between the
states $\mu_{1},\mu_{2},$ with the property:
\[
P\left[  \left(  \Omega\times\Omega\right)  ^{>}\right]  =1,
\]
where
\[
\left(  \Omega\times\Omega\right)  ^{>}=\left\{  \left[  \left(  n_{1}%
,\tau_{1}\right)  ,\left(  n_{2},\tau_{2}\right)  \right]  \in\Omega
\times\Omega:n_{1}\geq n_{2}\right\}  .
\]

\end{definition}

Clearly, if $\mu_{1}\succcurlyeq\mu_{2},$ then $N\left(  \mu_{1}\right)  \geq
N\left(  \mu_{2}\right)  .$

\begin{definition}
Let $\mu_{1},\mu_{2}$ be two states on $\Omega.$ We call the state $\mu_{1}$
to be \textbf{taller} than $\mu_{2},$ $\mu_{1}\curlyeqsucc\mu_{2},$ if there
exists a coupling $P\left[  d\omega_{1},d\omega_{2}\right]  $ between the
states $\mu_{1},\mu_{2},$ with the property:
\[
P\left[  \left(  \Omega\times\Omega\right)  ^{\gg}\right]  =1,
\]
where
\[
\left(  \Omega\times\Omega\right)  ^{\gg}=\left\{  \left[  \left(  n_{1}%
,\tau_{1}\right)  ,\left(  n_{2},\tau_{2}\right)  \right]  \in\Omega
\times\Omega:\tau_{1}=\tau_{2},n_{1}\geq n_{2}\text{ or }\omega_{2}%
=\mathbf{0}\right\}  .
\]

\end{definition}

\begin{lemma}
Let $\mu_{1}\left(  0\right)  ,\,\mu_{2}\left(  0\right)  $ be two initial
states on $\Omega$ at $t=0,$ and $\lambda_{1}\left(  t\right)  ,$
$\,\lambda_{2}\left(  t\right)  ,$ $t\geq0$ be two Poisson densities of the
input flows. The service time distribution is the same $\eta$ as before. Let
$\mu_{i}\left(  t\right)  $ be two corresponding GFP-s, with $\mu_{i}\left(
0\right)  =\mu\left(  0\right)  $. Suppose that $\mu_{1}\left(  0\right)
\curlyeqsucc\mu_{2}\left(  0\right)  ,$ and that $\lambda_{1}\left(  t\right)
\geq\lambda_{2}\left(  t\right)  .$ Then $\mu_{1}\left(  t\right)
\succcurlyeq\mu_{2}\left(  t\right)  ,$ so in particular
\[
N\left(  \mu_{1}\left(  t\right)  \right)  \geq N\left(  \mu_{2}\left(
t\right)  \right)  .
\]

\end{lemma}

\begin{proof}
To see this let us construct the coupling between the processes $\mu
_{i}\left(  t\right)  .$ Let us color the customers arriving according to the
$\lambda_{2}\left(  t\right)  $ flow as red. We also assign the red color to
the customers which were present at time $t=0$ from the initial state $\mu
_{2}\left(  0\right)  $. Let $\gamma\left(  t\right)  =\lambda_{1}\left(
t\right)  -\lambda_{2}\left(  t\right)  ,$ and consider $\gamma\left(
t\right)  $ as the extra input flow of blue customers (with independent
service times). We also add blue customers at time $t=0,$ which are needed to
complete the state $\mu_{2}\left(  0\right)  $ up to $\mu_{1}\left(  0\right)
.$ Then the total (color blind) flow coincides with $\lambda_{1}$ flow, while
the total (color blind) process coincides with $\mu_{1}\left(  t\right)  .$

The service rule for the two-colored process is color blind: all the customers
are served in order of their arrival time. We claim now that along every
coupled trajectory $\left(  \omega_{1}\left(  t\right)  ,\omega_{2}\left(
t\right)  \right)  $ we have $R\left(  \omega_{1}\left(  t\right)  \right)
\geq R\left(  \omega_{2}\left(  t\right)  \right)  ,$ where $R\left(
\cdot\right)  $ is the number of red customers at the moment $t,$ waiting to
be served. That evidently will prove our statement.

Clearly, the number $R\left(  \omega\left(  t\right)  \right)  $ is the
difference,
\[
R\left(  \omega\left(  t\right)  \right)  =\mathcal{A}\left(  \omega\left(
t\right)  \right)  -\mathcal{S}\left(  \omega\left(  t\right)  \right)  ,
\]
where $\mathcal{A}\left(  \omega\left(  t\right)  \right)  $ is the total
number of red customers, having arrived before $t,$ while $\mathcal{S}\left(
\omega\left(  t\right)  \right)  $ is the total number of red customers, who
left the system before $t.$ Clearly, $\mathcal{A}\left(  \omega_{1}\left(
t\right)  \right)  =\mathcal{A}\left(  \omega_{2}\left(  t\right)  \right)  .$
Let us show that $\mathcal{S}\left(  \omega_{1}\left(  t\right)  \right)
\leq\mathcal{S}\left(  \omega_{2}\left(  t\right)  \right)  .$

This is easy to see once one visualizes the procedure of resolving the rod
conflicts, which corresponds to our service rule, for the two-colored rod
case. Namely, one has first to put all the red rods, and resolve all their
conflicts by shifting some of them to the right accordingly. The number of
thus obtained rods to the left of the point $t$ is the number $\mathcal{S}%
\left(  \omega_{2}\left(  t\right)  \right)  .$ Clearly, if one adds some blue
rods to the red ones, then each red rod would be shifted to the right by at
least the same amount as without the blue rods. As a result, every red rod
would either stay where it was, or move to the right, so indeed $\mathcal{S}%
\left(  \omega_{1}\left(  t\right)  \right)  \leq\mathcal{S}\left(  \omega
_{2}\left(  t\right)  \right)  .$
\end{proof}

\subsection{Compactness}

Consider a General Flow Process $\mu\left(  t\right)  $ with initial state
$\mu\left(  0\right)  =\nu$ at $T=0$ and the input rate $\lambda\left(
t\right)  \equiv c<1.$ This is an ergodic process, so the weak limit
\[
\lim_{t\rightarrow\infty}\mu_{\nu,c}\left(  t\right)  =\nu_{c}%
\]
exists and does not depend on the initial state $\nu.$ We would like to show
that if $N\left(  \nu\right)  <\infty,$ then also
\begin{equation}
\lim_{t\rightarrow\infty}N\left(  \mu_{\nu,c}\left(  t\right)  \right)
=N\left(  \nu_{c}\right) \label{50}%
\end{equation}
(see $\left(  \ref{03}\right)  $). This turns out to be somewhat delicate
problem, because the speed of the convergence $\mu_{\nu,c}\left(  t\right)
\rightarrow\nu_{c}$ is only linear in time, and it can happen that for every
$\delta>0$ the moment $\mathbb{E}_{\nu}\left(  n\left(  \omega\right)
^{1+\delta}\right)  $ does not exist, which property persists in time, and the
moments $\mathbb{E}_{\mu_{\nu,c}\left(  t\right)  }\left(  n\left(
\omega\right)  ^{1+\delta}\right)  $ are infinite for every $t.$

\subsubsection{Compactification}

With every point $\omega=\left(  n,\tau\right)  \in\Omega,\,n>0,$ there is
associated the random variable $\eta\Bigm|_{\tau}=\left(  \eta-\tau
\Bigm|\eta>\tau\right)  .$ Consider now the following queueing problem: the
customers are arriving at positive Poissonian times with the rate
$\lambda\left(  t\right)  \equiv c,$ while service times are independent and
$\eta$-distributed. In addition, at moment $0$ there is a customer with
service time distributed according to $\eta\Bigm|_{\tau},$ and $n-1$ $\eta
$-distributed customers. Then the expected size of queue at the moment $t$ is
precisely $N\left(  \mu_{\delta_{\omega},c}\left(  t\right)  \right)  .$ In
general, $N\left(  \mu_{\nu,c}\left(  t\right)  \right)  =\mathbb{E}_{\nu
}\left(  N\left(  \mu_{\delta_{\omega},c}\left(  t\right)  \right)  \right)
.$ We want to study the dependence of $N\left(  \mu_{\nu,c}\left(  t\right)
\right)  $ on $\nu.$ We abbreviate it by $N\left(  \nu,t\right)  .$

In general, the family $\eta\Bigm|_{\tau}$ is not weakly compact. In order to
prove our statement we have to generalize it, including $\left\{
\eta\Bigm|_{\tau},\tau\geq0\right\}  $ into a compact family. The
generalization is as follows. Note that the random variables $\eta
\Bigm|_{\tau}$ have the property that $\mathbb{E}\left(  \eta\Bigm|_{\tau
}\right)  ^{2+\delta}\leq M_{\delta}.$ Consider now the set $\mathcal{\bar{N}%
}$ of all positive random variables with this property: $\zeta\in
\mathcal{\bar{N}\Longleftrightarrow}\mathbb{E}\left(  \zeta\right)
^{2+\delta}\leq M_{\delta}.$ This set $\mathcal{\bar{N}}$ already is weakly
compact, due to Prokhorov theorem. We denote by $\mathcal{N\subset\bar{N}}$
the closure of the family $\left\{  \eta\Bigm|_{\tau},\tau\geq0\right\}  $ in
$\mathcal{\bar{N}}.$ We now extend our configuration space $\Omega,$ to
consist of pairs $\omega=\left(  n,\zeta\right)  ,$ $\zeta\in\mathcal{N}.$ The
initial state will then be a measure $\nu$ on $\Omega,$ i.e. on the set of
pairs $\left(  n,\zeta\right)  ,\,n\geq1,$ plus the single point $n=0.$ We
will use the old notation for all the extended objects.

Unfortunately, the function $N\left(  \nu,t\right)  $ is not continuous on
$\mathcal{M}\left(  \Omega\right)  ,$ due to the fact that $\Omega$ is (still)
not compact, in the $n$-direction. This obstruction would be removed once one
is contented to restrict the function $N\left(  \nu,t\right)  $ to
$\mathcal{M}\left(  \Omega_{N}\right)  \subset\mathcal{M}\left(
\Omega\right)  ,$ where $\Omega_{N}=\left\{  \omega=\left(  n,\zeta\right)
\in\Omega:n\leq N\right\}  .$ Then it is enough to check that $N\left(
\nu,t\right)  $ is continuous on $\Omega\subset\mathcal{M}\left(
\Omega\right)  ,$ where the imbedding $\Omega\subset\mathcal{M}\left(
\Omega\right)  $ is via $\left(  n,\zeta\right)  \rightsquigarrow
\delta_{\left(  n,\zeta\right)  }.$ The function we are dealing with is then
the following:

Let $x>0,$ and $N\left(  \left(  n,x\right)  ,t\right)  $ be the expected size
of the queue at the moment $t,$ if

\begin{itemize}
\item the customers are arriving at positive Poissonian times with the rate
$\lambda\left(  t\right)  \equiv c,$

\item the service times are independent and $\eta$-distributed,

\item at moment $0$ there is a customer with non-random service time, which
equals $x,$ together with $n-1$ $\eta$-distributed customers, waiting in the queue.
\end{itemize}

Now,
\[
N\left(  \left(  n,\zeta\right)  ,t\right)  \equiv N\left(  \delta_{\left(
n,\zeta\right)  },t\right)  =\mathbb{E}\left(  N\left(  \left(  n,\zeta
\right)  ,t\right)  \right)  .
\]
Since the function $N\left(  \left(  n,x\right)  ,t\right)  ,$ though
continuous, has infinite support in the $x$-variable for $n,t$ fixed, the
continuity of $N\left(  \left(  n,\zeta\right)  ,t\right)  $ in $\zeta$ (in
weak topology) needs to be checked. However, $N\left(  \left(  n,x\right)
,t\right)  =ct+n$ for all $x>t$, and that makes the check trivial.

In general case
\[
N\left(  \nu,t\right)  =\mathbb{E}_{\nu}\left(  \mathbb{E}_{\zeta}\left(
N\left(  \left(  n,\zeta\right)  ,t\right)  \right)  \right)  .
\]
Now, since $\mathbb{E}_{\zeta}\left(  N\left(  \left(  n,\zeta\right)
,t\right)  \right)  $ is a continuous function on $\left\{  1,...,N\right\}
\times\mathcal{N\times}\mathbb{R}_{+},$ the continuity of $N\left(
\nu,t\right)  $ on $\mathcal{M}\left(  \Omega_{N}\right)  $ follows from
compactness of $\left\{  1,...,N\right\}  \times\mathcal{N}.$

Therefore we have obtained the following conditional statement:

\begin{lemma}
\label{lim} Suppose for some $\ell\geq0$ the convergence $N\left(
\nu,t\right)  \rightarrow\ell$ holds for every $\nu\in\mathcal{M}\left(
\Omega\right)  ,$ as $t\rightarrow\infty.$ Then the convergence is uniform on
every $\mathcal{M}\left(  \Omega_{N}\right)  .$
\end{lemma}

\subsubsection{Convergence}

\begin{lemma}
For every $n,\tau$ we have
\begin{equation}
\lim_{t\rightarrow\infty}N\left(  \mu_{\delta_{\left(  n,\tau\right)  }%
,c}\left(  t\right)  \right)  =N\left(  \nu_{c}\right) \label{131}%
\end{equation}
(though, of course, not uniform in $n,\tau$).
\end{lemma}

\begin{proof}
Since $\mu_{\delta_{\left(  n,\tau\right)  },c}\left(  t\right)
\rightarrow\nu_{c}$ weakly, $\limsup_{t\rightarrow\infty}N\left(  \mu
_{\delta_{\left(  n,\tau\right)  },c}\left(  t\right)  \right)  \geq N\left(
\nu_{c}\right)  .$

To prove the opposite inequality we need to produce a uniform upper bound on
the family $\mu_{\delta_{\left(  n,\tau\right)  },c}\left(  t\right)  ,$ in
order to have its uniform integrability. By this we mean the following
property: for every $\varkappa>0$ there exists a value $N_{\varkappa}$ such
that for all $t$%
\[
\mathbb{E}_{\mu_{\delta_{\left(  n,\tau\right)  },c}\left(  t\right)  }\left(
N\left(  \omega\right)  \mathbf{I}_{N\left(  \omega\right)  \geq N_{\varkappa
}}\right)  <\varkappa,
\]
where $\mathbf{I}$ stands for the indicator. If it were possible to find an
$\varepsilon$ such that $\nu_{c}\curlyeqsucc\varepsilon\delta_{\left(
n,\tau\right)  }+\left(  1-\varepsilon\right)  \delta_{\mathbf{0}}$, then we
would be done, since we then have that $N\left(  \mu_{\varepsilon
\delta_{\left(  n,\tau\right)  }+\left(  1-\varepsilon\right)  \delta
_{\mathbf{0}},c}\left(  t\right)  \right)  \leq N\left(  \nu_{c}\right)  $ for
all $t.$ However this is not the case, since the measure $\nu_{c}$ has no
atoms. Therefore we have to pass to the imbedded Markov chain, as it is done
in \cite{S}, sect. 5.1.

Consider the service process started in the configuration $\left(
n,\tau\right)  ,$ with the customer arrival rate $\lambda\equiv c.$ Let
$\xi_{i}$ be the number of customers in the system right after the (random)
moment $t_{i},$ when the $i$-th customer was served. We put $\xi_{0}=n$,
$t_{0}=0.$ Then
\[
\xi_{i+1}=\max\left\{  0,\xi_{i}-1+\theta_{i}\right\}  ,
\]
where $\theta_{i}$ is the number of customers which came to the system during
the $i$-th service session. Then the random variables $\theta_{0},\theta
_{1},\theta_{2},...$ are independent, with $\theta_{1},\theta_{2},...$
identically distributed. The Markov chain $\xi_{i}$ is ergodic. It is
stationary, except for the first step. We denote by $\pi$ its stationary
distribution, and by $\pi^{\left(  n,\tau\right)  }$ the distribution of the
variable $\xi_{1}$. We claim that it is enough for our purposes to study this
chain. Indeed, if we define the process
\[
\bar{\omega}\left(  t\right)  =\xi_{i}+1\text{ for }t\in(t_{i-1},t_{i}],
\]
and $\bar{\mu}\left(  t\right)  $ be its distribution, then we clearly have
$\bar{\mu}\left(  t\right)  \succcurlyeq\mu\left(  t\right)  .$

Let $\pi^{1},\pi^{2}$ be two probability distributions on $\mathbb{N}=\left\{
0,1,2,...\right\}  .$ As above, we will say that $\pi^{1}\succcurlyeq\pi^{2}$
if there is a coupling $\Pi$ of $\pi^{1},\pi^{2},$ supported by $\left(
\mathbb{N\times N}\right)  ^{>}=\left\{  \left(  n_{1},n_{2}\right)
\in\mathbb{N\times N}:n_{1}\geq n_{2}\right\}  .$ If $\xi_{i}^{1},\xi_{i}^{2}$
are two stationary Markov chains, corresponding to two initial distributions
$\pi^{1},\pi^{2}$ at the moment $i=1,$ and if $\pi^{1}\succcurlyeq\pi^{2},$
then also $\xi_{i}^{1}\succcurlyeq\xi_{i}^{2}.$

Let us show now that for some $\varepsilon>0$ we have
\begin{equation}
\pi\succcurlyeq\varepsilon\pi^{\left(  n,\tau\right)  }+\left(  1-\varepsilon
\right)  \delta_{0}.\label{204}%
\end{equation}
This is almost evident. Indeed, in case $t_{i}-t_{i-1}>\tau$ let us consider
the random number $\theta_{i}^{\tau}$ of customers arriving to the server
during the initial portion $\tau$ of time of the $i$-th service session. Then
the event $\left\{  t_{i}-t_{i-1}>\tau,\theta_{i}^{\tau}\geq n\right\}  $
happens with positive probability, while conditioning by this event we have
that the conditional distribution of $\xi_{i}$ is higher than $\pi^{\left(
n,\tau\right)  }.$ The relation $\left(  \ref{204}\right)  $ then follows,
since the stationary measure can be computed by averaging over the trajectories.
\end{proof}

From Lemma \ref{lim} we then know that the convergence in $\left(
\ref{131}\right)  $ is uniform over $\left\{  \left(  n,\tau\right)  :n\leq
N\right\}  $ for every $N.$

The next statement will allow us to treat the unbounded component.

\begin{lemma}
\label{N0} There exist $N_{0}=N_{0}\left(  c\right)  $ and $T=T\left(
c\right)  ,$ such that the following holds:

Let $\nu=\delta_{\left(  n,\tau\right)  }$ denote the initial state,
concentrated on the configuration with $n\geq N_{0}$ customers, with the first
one being already served for a time $\tau.$ Then
\begin{equation}
N\left(  \mu_{\delta_{\left(  n,\tau\right)  },c}\left(  t\right)  \right)
\leq n,\label{08}%
\end{equation}
for every $\tau\geq0$ and every $t\geq T.$
\end{lemma}

\begin{proof}
We start with presenting our choice of $N_{0}$ and $T.$ Namely, we take
$N_{0}$ to be any integer bigger than $N\left(  \nu_{c}\right)  ,$ while $T$
is defined by the property that for every $\tau$ and every $t\geq T$
\[
N\left(  \mu_{\delta_{\left(  N_{0},\tau\right)  },c}\left(  t\right)
\right)  <N_{0}.
\]
(The existence of $T$ follows from the uniformity of the convergence $N\left(
\mu_{\delta_{\left(  N_{0},\tau\right)  },c}\left(  t\right)  \right)
\rightarrow N\left(  \nu_{c}\right)  $ in $\tau.$)

We claim now that for any $n>N_{0},$ any $\tau$ and any $t\geq T$%
\[
N\left(  \mu_{\delta_{\left(  n,\tau\right)  },c}\left(  t\right)  \right)
<n.
\]
To see this let us consider the following auxiliary service discipline: we
start in the state $\left(  n,\tau\right)  ,$ the input rate is $\lambda\equiv
c,$ but the server serves the customers only if the queue has more than
$n-N_{0}$ clients; otherwise the server remains idle. Let us denote the
resulting states of the corresponding process by $\tilde{\mu}_{\delta_{\left(
n,\tau\right)  },c}\left(  t\right)  .$ Evidently, $N\left(  \tilde{\mu
}_{\delta_{\left(  n,\tau\right)  },c}\left(  t\right)  \right)  =N\left(
\mu_{\delta_{\left(  N_{0},\tau\right)  },c}\left(  t\right)  \right)
+n-N_{0}.$ Therefore, for any $\tau$ and any $t\geq T$%
\[
N\left(  \tilde{\mu}_{\delta_{\left(  n,\tau\right)  },c}\left(  t\right)
\right)  <n.
\]
On the other hand, the processes $\tilde{\mu}_{\delta_{\left(  n,\tau\right)
},c}\left(  \cdot\right)  $ and $\mu_{\delta_{\left(  n,\tau\right)  }%
,c}\left(  \cdot\right)  $ can be coupled in such a way that with probability
one $N\left(  \tilde{\omega}\right)  \geq N\left(  \omega\right)  $ at all
times, so
\[
N\left(  \mu_{\delta_{\left(  n,\tau\right)  },c}\left(  t\right)  \right)
\leq N\left(  \tilde{\mu}_{\delta_{\left(  n,\tau\right)  },c}\left(
t\right)  \right)  .
\]

\end{proof}

Together, the three last lemmas imply the relation $\left(  \ref{50}\right)
:$

\begin{lemma}
For all initial states $\nu$ with $N\left(  \nu\right)  <\infty$
\[
\lim_{t\rightarrow\infty}N\left(  \mu_{\nu,c}\left(  t\right)  \right)
=N\left(  \nu_{c}\right)  .
\]

\end{lemma}

\begin{proof}
Let $N_{\varepsilon}$ be the smallest integer $k$, satisfying

\begin{itemize}
\item $k>N_{0}$ (see Lemma \ref{N0}),

\item $N\left(  \nu^{\uparrow}\right)  <\varepsilon,$ where we denote by
$\nu^{\uparrow}$ the measure obtained from $\nu$ by restricting it to the set
$\Omega^{\uparrow}=\left\{  \left(  n,\tau\right)  :n>k\right\}  $ $.$

\noindent Likewise, we define the measure $\nu^{\downarrow}$ by $\nu
^{\downarrow}=\nu-\nu^{\uparrow}.$
\end{itemize}

Let us write
\[
\mu_{\nu,c}\left(  t\right)  =\int d\nu\left(  n,\tau\right)  \mu
_{\delta_{\left(  n,\tau\right)  },c}\left(  t\right)  .
\]
Then
\[
N\left(  \mu_{\nu,c}\left(  t\right)  \right)  =N\left(  \int d\nu
^{\downarrow}\left(  n,\tau\right)  \mu_{\delta_{\left(  n,\tau\right)  }%
,c}\left(  t\right)  \right)  +N\left(  \int d\nu^{\uparrow}\left(
n,\tau\right)  \mu_{\delta_{\left(  n,\tau\right)  },c}\left(  t\right)
\right)  .
\]
From compactness we know that
\[
N\left(  \int d\nu^{\downarrow}\left(  n,\tau\right)  \mu_{\delta_{\left(
n,\tau\right)  },c}\left(  t\right)  \right)  \rightarrow\nu^{\downarrow
}\left(  \Omega\right)  N\left(  \nu_{c}\right)  \text{ as }t\rightarrow
\infty.
\]
And the relation $\left(  \ref{08}\right)  $ tells us that for $t>T\left(
c\right)  $%
\[
N\left(  \int d\nu^{\uparrow}\left(  n,\tau\right)  \mu_{\delta_{\left(
n,\tau\right)  },c}\left(  t\right)  \right)  \leq N\left(  \nu^{\uparrow
}\right)  .
\]
Hence for all $t$ large enough
\[
\left(  1-2\varepsilon\right)  N\left(  \nu_{c}\right)  \leq N\left(  \mu
_{\nu,c}\left(  t\right)  \right)  \leq N\left(  \nu_{c}\right)  +\varepsilon.
\]

\end{proof}

\subsection{End of the proof in noisy case}

Let $\mu_{\nu,\lambda_{\nu}\left(  \cdot\right)  }\left(  t\right)  $ be the
non-linear Markov process\textit{\ }with the initial state $\nu,$ having
finite mean queue, $N\left(  \nu\right)  <\infty.$ We will show that the
function $\lambda\left(  t\right)  \equiv\lambda_{\nu}\left(  t\right)  $ goes
to a limit as $t\rightarrow\infty.$ The idea is the following:

Suppose $m=\liminf_{t\rightarrow\infty}\lambda\left(  t\right)  <\limsup
_{t\rightarrow\infty}\lambda\left(  t\right)  =M.$ As we already know, for
every $T$ and every $\varepsilon>0$ there exist some values $K_{1},K_{2}$ such
that
\begin{equation}
\sup_{x\in\left[  K_{1},K_{1}+T\right]  }\lambda\left(  x\right)  \leq
m+\varepsilon,\label{161}%
\end{equation}
while
\begin{equation}
\inf_{x\in\left[  K_{2},K_{2}+T\right]  }\lambda\left(  x\right)  \geq
M-\varepsilon.\label{162}%
\end{equation}
We want to bring this to contradiction, arguing as follows:

\begin{itemize}
\item First of all, we note that the mean queue, $N\left(  \mu_{\nu
,\lambda_{\nu}\left(  \cdot\right)  }\left(  t\right)  \right)  $ does not
change in time, staying equal to the initial value $N\left(  \nu\right)  .$ On
the other hand

\item We can compare the state $\mu_{\nu,\lambda_{\nu}\left(  \cdot\right)
}\left(  K_{1}+T\right)  $ with the state $\mu_{\mu_{\nu,\lambda_{\nu}\left(
\cdot\right)  }\left(  K_{1}\right)  ,m+\varepsilon}\left(  T\right)  .$ Due
to $\left(  \ref{161}\right)  ,$ the latter is higher, so
\begin{equation}
N\left(  \mu_{\mu_{\nu,\lambda_{\nu}\left(  \cdot\right)  }\left(
K_{1}\right)  ,m+\varepsilon}\left(  T\right)  \right)  \geq N\left(
\nu\right)  .\label{163}%
\end{equation}
By the same reasoning,
\begin{equation}
N\left(  \mu_{\mu_{\nu,\lambda_{\nu}\left(  \cdot\right)  }\left(
K_{2}\right)  ,M-\varepsilon}\left(  T\right)  \right)  \leq N\left(
\nu\right)  .\label{164}%
\end{equation}

\item We then claim that once $T$ is large enough, the state $\mu_{\mu
_{\nu,\lambda_{\nu}\left(  \cdot\right)  }\left(  K_{1}\right)  ,m+\varepsilon
}\left(  T\right)  $ is close to the equilibrium $\nu_{m+\varepsilon},$ and
moreover
\begin{equation}
N\left(  \mu_{\mu_{\nu,\lambda_{\nu}\left(  \cdot\right)  }\left(
K_{1}\right)  ,m+\varepsilon}\left(  T\right)  \right)  \leq N\left(
\nu_{m+\varepsilon}\right)  +\varepsilon^{\prime}.\label{165}%
\end{equation}
By the same reasoning,
\begin{equation}
N\left(  \mu_{\mu_{\nu,\lambda_{\nu}\left(  \cdot\right)  }\left(
K_{2}\right)  ,M-\varepsilon}\left(  T\right)  \right)  \geq N\left(
\nu_{M-\varepsilon}\right)  -\varepsilon^{\prime\prime}.\label{166}%
\end{equation}

\item Since $N\left(  \nu_{M-\varepsilon}\right)  >N\left(  \nu_{m+\varepsilon
}\right)  $ once $\varepsilon$ is small, the relations $\left(  \ref{163}%
\right)  $-$\left(  \ref{166}\right)  $ are inconsistent once $\varepsilon
^{\prime}$ and $\varepsilon^{\prime\prime}$ are also small enough.
\end{itemize}

We need to prove the relations $\left(  \ref{165}\right)  $ and $\left(
\ref{166}\right)  .$ It turns out that the relation $\left(  \ref{166}\right)
$ is much easier. Indeed, to show it, we can compare the state $\mu_{\mu
_{\nu,\lambda_{\nu}\left(  \cdot\right)  }\left(  K_{2}\right)  ,M-\varepsilon
}\left(  T\right)  $ with the state $\mu_{\mathbf{0},M-\varepsilon}\left(
T\right)  .$ The latter is evidently lower --
\[
N\left(  \mu_{\mu_{\nu,\lambda_{\nu}\left(  \cdot\right)  }\left(
K_{2}\right)  ,M-\varepsilon}\left(  T\right)  \right)  \geq N\left(
\mu_{\mathbf{0},M-\varepsilon}\left(  T\right)  \right)  ,
\]
-- and as soon as $T$ is large enough, $\mu_{\mathbf{0},M-\varepsilon}\left(
T\right)  $ is close to $\nu_{M-\varepsilon}.$ Since $\mu_{\mathbf{0}%
,M-\varepsilon}\left(  T\right)  $ is also lower than $\nu_{M-\varepsilon},$
\begin{equation}
N\left(  \mu_{\mathbf{0},M-\varepsilon}\left(  T\right)  \right)  \leq
N\left(  \nu_{M-\varepsilon}\right)  .\label{167}%
\end{equation}
Since $\mu_{\mathbf{0},M-\varepsilon}\left(  T\right)  \rightarrow
\nu_{M-\varepsilon}$ as $T\rightarrow\infty,$ $\left(  \ref{167}\right)  $
implies that $N\left(  \mu_{\mathbf{0},M-\varepsilon}\left(  T\right)
\right)  \rightarrow N\left(  \nu_{M-\varepsilon}\right)  ,$ which proves
$\left(  \ref{166}\right)  .$

In the above proof the important step was to replace the state $\mu
_{\nu,\lambda_{\nu}\left(  \cdot\right)  }\left(  K_{2}\right)  $ with a lower
state $\mathbf{0,}$ which is in fact the lowest. Turning to $\left(
\ref{165}\right)  ,$ we see that this step can not be mimicked there, since
there is no highest state! So, to proceed, we need some apriori upper bound on
the state $\mu_{\nu,\lambda_{\nu}\left(  \cdot\right)  }\left(  K_{1}\right)
.$

\begin{lemma}
Let $\nu$ be an arbitrary initial state, with $N\left(  \nu\right)  <\infty. $
Then there exist $\bar{c}\left(  \nu\right)  <1$ and $T<\infty,$ such that for
every $t>T$%
\[
\lambda_{\nu}\left(  t\right)  <\bar{c}\left(  \nu\right)  .
\]

\end{lemma}

\begin{proof}
The statement of the lemma is equivalent to the fact that $M=\limsup
_{t\rightarrow\infty}\lambda\left(  t\right)  <1.$ So suppose the opposite,
that $M\geq1.$ As we then know from Lemma \ref{T}, for every $T$ and every
$\varepsilon>0$ we can find a segment $\left[  K,K+T\right]  ,$ such that
$\lambda_{\nu}\left(  t\right)  >1-\varepsilon$ for all $t\in\left[
K,K+T\right]  .$ This, however, contradicts to the statement $\left(
\ref{01}\right)  $ of Lemma \ref{la}.
\end{proof}

So without loss of generality we can assume that the initial state $\nu$ is
such that $N\left(  \nu\right)  <\infty,$ while $\lambda_{\nu}\left(
t\right)  <\bar{c}<1$ for all $t>0.$ Clearly, the state $\mu_{\nu,\lambda
_{\nu}}\left(  t\right)  $ is dominated by $\mu_{\nu,\bar{c}}\left(  t\right)
.$ From the previous section we know that $N\left(  \mu_{\nu,\bar{c}}\left(
t\right)  \right)  \rightarrow N\left(  \nu_{\bar{c}}\right)  $ as
$t\rightarrow\infty.$ Moreover, for any $\varepsilon>0$ fixed we know from
Lemma \ref{N0} that there exist a level $N\left(  \bar{c},\varepsilon\right)
$ and a time $T\left(  \bar{c}\right)  ,$ such that for all $t>$ $T\left(
\bar{c}\right)  $ in the state $\mu_{\nu,\bar{c}}\left(  t\right)  $ we have:
\begin{equation}
\sum_{n>N\left(  \bar{c},\varepsilon\right)  }n\,\mathbf{\Pr}\left\{  N\left(
\omega\right)  =n\right\}  <\varepsilon.\label{170}%
\end{equation}
Again we may assume that $T\left(  \bar{c}\right)  =0.$

Define now the time duration $\tilde{T}=\tilde{T}\left(  N\left(  \bar
{c},\varepsilon\right)  ,m+\varepsilon\right)  $ as such that for every state
$\tilde{\nu}$ on $\Omega,$ supported by configurations $\left\{  \left(
n,\tau\right)  :n\leq N\left(  \bar{c},\varepsilon\right)  \right\}  ,$ and
every $t>\tilde{T}$%
\begin{equation}
N\left(  \mu_{\tilde{\nu},m+\varepsilon}\left(  t\right)  \right)  \leq
N\left(  \nu_{m+\varepsilon}\right)  +\varepsilon.\label{171}%
\end{equation}
(The existence of $\tilde{T}$ follows from the compactness, as was explained
in the preceding section.) As we know from Lemma \ref{T}, there exist a moment
$K=K\left(  \tilde{T}\right)  ,$ such that $\sup_{t\in\left[  K,K+\tilde
{T}\right]  }\lambda_{\nu}\left(  t\right)  \leq m+\varepsilon.$ We claim that
at the moment $K+\tilde{T}$ the state $\mu_{\nu,\lambda_{\nu}}\left(
K+\tilde{T}\right)  $ is not much higher than $\nu_{m+\varepsilon}, $ so in
particular
\[
N\left(  \mu_{\nu,\lambda_{\nu}}\left(  K+\tilde{T}\right)  \right)  \leq
N\left(  \nu_{m+\varepsilon}\right)  +2\varepsilon.
\]
Indeed, let us write
\begin{align*}
& \mu_{\nu,\lambda_{\nu}}\left(  K\right) \\
& =\mu_{\nu,\lambda_{\nu}}\left(  K\right)  \Bigm|_{\left\{  \left(
n,\tau\right)  :n\leq N\left(  \bar{c},\varepsilon\right)  \right\}  }%
+\mu_{\nu,\lambda_{\nu}}\left(  K\right)  \Bigm|_{\left\{  \left(
n,\tau\right)  :n>N\left(  \bar{c},\varepsilon\right)  \right\}  }\\
& \equiv\varkappa_{1}+\varkappa_{2}.
\end{align*}
Then
\[
\mu_{\nu,\lambda_{\nu}\left(  \cdot\right)  }\left(  K+\tilde{T}\right)
=\mu_{\varkappa_{1},\tilde{\lambda}\left(  \cdot\right)  }\left(  \tilde
{T}\right)  +\mu_{\varkappa_{2},\tilde{\lambda}\left(  \cdot\right)  }\left(
\tilde{T}\right)  ,
\]
where $\tilde{\lambda}\left(  t\right)  =\lambda_{\nu}\left(  K+t\right)  .$
Then for the first summand the relation $\left(  \ref{171}\right)  $ holds,
since the state $\varkappa_{1}$ relaxes after time $\tilde{T}$ under ``higher
dynamics'' with the rate $m+\varepsilon\left(  \geq\tilde{\lambda}\right)  ,$
so is very close to $\varkappa_{1}\left(  \Omega\right)  \nu_{m+\varepsilon}.$
For the second summand the relation $\left(  \ref{170}\right)  $ holds, since
$\tilde{\lambda}\left(  t\right)  <\bar{c}.$ That proves the relation $\left(
\ref{165}\right)  .$

\begin{acknowledgement}
We would like to thank our colleagues -- in particular, Yu. Golubev, T.
Liggett, O. Ogievetsky, G. Olshansky, S. Pirogov, A. Vladimirov -- for
valuable discussions and remarks, concerning this paper. We are grateful also
to the Institute for Pure and Applied Mathematics (IPAM) as UCLA, for the
uplifting atmosphere and support during the Spring 2002 program on Large Scale
Communication Networks, where part of this work was done.
\end{acknowledgement}

\end{document}